\documentclass[12pt]{article}
\usepackage{amssymb}
\usepackage{a4}
\evensidemargin \oddsidemargin
\def\1{{\bf 1}}
\def\id{\mbox{id\,}}
\def\ot{\!\otimes\!}

\def\F{\mbox{$\cal F$}}
\def\Fu{{\cal F}^{(1)}}
\def\Fd{{\cal F}^{(2)}}
\def\bF{\mbox{$\overline{\cal F}$}}

\def\f{{\scriptscriptstyle {\cal F}}}

\def\ra{\rangle}
\def\la{\langle}
\def \A{{\cal A}}
\def \B {{\cal B}}

\def \D {{\cal D}}
\def \En {{\cal E}}

\def \H {{\cal H}}
\def \bH {\overline{\cal H}}
\def \I {{\cal I}}
\def \O {{\cal O}}

\def \P {{\cal P}}
\def \W {{\cal W}}

\def \L {{\cal L} }
\def \M {{\cal M}}
\def \NN {{\cal N}}
\def \X {{\cal X} }
\def \Hau{{\sf H}^{\scriptscriptstyle (1)}}
\def \Han{{\sf H}^{\scriptscriptstyle (n)}}
\def \Haus{{\sf H}^{\scriptscriptstyle (1)}_{\scriptscriptstyle \star}}
\def \Hans{{\sf H}^{\scriptscriptstyle (n)}_{\scriptscriptstyle \star}}
\def \Has{{\sf H}_{\scriptscriptstyle \star}}
\def \varphihs{\varphi^{\scriptscriptstyle H}_{\scriptscriptstyle \star}}

\def \psin{\psi^{\scriptscriptstyle (n)}}

\def\R{\mbox{$\cal R$}}
\def\bR{\mbox{$\overline{\cal R}$}}

\def\hA{\mbox{$\widehat{\cal A}$}}
\def\hB{\mbox{$\widehat{\cal B}$}}

\def\Ha{{\sf H}}
\def\hH{\mbox{$\hat{H}$}}

\newcommand{\tr}{\hat\triangleright}
\newcommand{\trc}{\triangleright}
\def\g{\mbox{\bf g\,}}
\def\h{\mbox{\bf h}}

\def\b#1{{\mathbb #1}}
\def\nn{\nonumber \\}

\newcommand{\be}{\begin{equation}}
\newcommand{\ee}{\end{equation}}
\newcommand{\bea}{\begin{eqnarray}}
\newcommand{\eea}{\end{eqnarray}}
\newcommand{\ba}{\begin{array}}
\newcommand{\ea}{\end{array}}
\newtheorem{lemma}{Lemma}

\def\sq{\mbox{\rlap{$\sqcap$}$\sqcup$}}
\newenvironment{proof}[1]{\vspace{5pt}\noindent{\bf Proof #1}\hspace{6pt}}%
{\hfill\sq}
\newcommand{\bp}{\begin{proof}}
\newcommand{\ep}{\end{proof}\par\vspace{10pt}\noindent}
\gdef\Feynmanlength{\setlength{\unitlength}{0.01pt}}
\global\newcount\LINETYPE
\global\newcount\LINEDIRECTION
\global\newcount\LINECONFIGURATION
\newcommand{\LTYPE}{\LINETYPE}
\newcommand{\LDIR}{\LINEDIRECTION}
\newcommand{\LCONFIG}{\LINECONFIGURATION}
\global\LINETYPE=1  \global\LINEDIRECTION=0
\global\LINECONFIGURATION=0
\global\newcount\fermion    \fermion=1
\global\newcount\scalar     \scalar=2
\global\newcount\photon     \photon=3
\global\newcount\gluon      \gluon=4
\global\newcount\SPECIAL    \SPECIAL=5
\gdef\N{0}  \gdef\NE{1}  \gdef\E{2}   \gdef\SE{3}
\gdef\S{4}  \gdef\SW{5}  \gdef\W{6}   \gdef\NW{7}
\global\newcount\REG            \global\REG=0
\global\newcount\FLIPPED        \global\FLIPPED=1
\global\newcount\CURLY          \global\CURLY=2
\global\newcount\FLIPPEDCURLY   \global\FLIPPEDCURLY=3
\global\newcount\FLAT           \global\FLAT=4
\global\newcount\FLIPPEDFLAT    \global\FLIPPEDFLAT=5
\global\newcount\CENTRAL        \global\CENTRAL=6
\global\newcount\FLIPPEDCENTRAL \global\FLIPPEDCENTRAL=7
             
\global\newcount\SQUASHEDGLUON  \global\SQUASHEDGLUON=8

\newcount\adjx \adjx=0
\newcount\adjy \adjy=0
\global\newdimen\BIGPHOTONS     \BIGPHOTONS=0pt  
\global\newdimen\THICKPHOTONS     \THICKPHOTONS=0pt
\global\newdimen\THICKPHOTONSWITCH    \THICKPHOTONSWITCH=0pt
\gdef\THICKPHOTONTEST{
\THICKPHOTONSWITCH=0pt
\ifdim\THICKPHOTONS=0pt \relax
  \else \ifnum\LTYPE=3
           \ifnum\LDIR=2 \THICKPHOTONSWITCH=1pt \fi 
           \ifnum\LDIR=6 \THICKPHOTONSWITCH=1pt \fi 
        \fi
\fi
}  

\global\newcount\phantomswitch   \global\phantomswitch=0
\global\newcount\stemlength   \global\stemlength=275
\global\newcount\absstemlength        
\global\newcount\stemlengthx          
\global\newcount\stemlengthy          
\newdimen\FRONTSTEM  \FRONTSTEM=0pt   
\newdimen\BACKSTEM   \BACKSTEM=0pt    
\newdimen\EITHERSTEM \EITHERSTEM=0pt  
\global\newcount\arrowlength                
\global\newdimen\ATTIP   \global\ATTIP=0pt  
\global\newdimen\ATBASE  \global\ATBASE=1pt 
\global\newcount\unitboxnumber
\global\newcount\unitboxnumberpo  
\global\newcount\particlelengthx
\gdef\plengthx{\particlelengthx}
\global\newcount\particlelengthy
\gdef\plengthy{\particlelengthy}
\global\newcount\boxlengthx
\global\newcount\boxlengthy
\global\newcount\particleadjustx
\global\newcount\particleadjusty
\global\newcount\particlelength   
\global\newcount\particlefrontx
\gdef\pfrontx{\particlefrontx}
\global\newcount\PFRONTx
\global\newcount\particlefronty
\gdef\pfronty{\particlefronty}
\global\newcount\PFRONTy
\global\newcount\particlebackx
\gdef\pbackx{\particlebackx}
\global\newcount\particlebacky
\gdef\pbacky{\particlebacky}
\global\newcount\particlemidx
\gdef\pmidx{\particlemidx}
\global\newcount\particlemidy
\gdef\pmidy{\particlemidy}
\global\newcount\seglength  \global\newcount\gaplength
\global\gaplength=850  
\global\seglength=1416  
\global\newcount\Xone    \global\newcount\Yone
\global\newcount\Xtwo    \global\newcount\Ytwo
\global\newcount\Xthree  \global\newcount\Ythree
\global\newcount\Xfour   \global\newcount\Yfour   
\global\newcount\Xfive   \global\newcount\Yfive   
\global\newcount\Xsix    \global\newcount\Ysix
\global\newcount\Xseven  \global\newcount\Yseven
\global\newcount\Xeight  \global\newcount\Yeight
\newsavebox{\lastline}  
\global\newcount\numlineparts
\global\newcount\upperlineadjx  \upperlineadjx=0  
\global\newcount\upperlineadjy  \upperlineadjy=0  
\global\newcount\lowerlineadjx  \lowerlineadjx=0  
\global\newcount\lowerlineadjy  \lowerlineadjy=0  
\global\newcount\thirdlineadjx  \thirdlineadjx=0  
\global\newcount\thirdlineadjy  \thirdlineadjy=0  
\global\newcount\fourthlineadjx \fourthlineadjx=0  
\global\newcount\fourthlineadjy \fourthlineadjy=0  
\global\newcount\unitboxwidth   \unitboxwidth=1000
\global\newcount\unitboxheight  \unitboxheight=0  
\global\newcount\numupperunits  \numupperunits=8  
\global\newcount\numlowerunits  \numlowerunits=8  
\global\newcount\numthirdunits  \numthirdunits=8  
\global\newcount\numfourthunits \numfourthunits=8  
\global\newcount\fermioncount   \global\fermioncount=0
\global\newcount\scalarcount    \global\scalarcount=0
\global\newcount\photoncount    \global\photoncount=0
\global\newcount\gluoncount     \global\gluoncount=0
\global\newcount\SPECIALcount   \global\SPECIALcount=0
\global\newcount\vertexcount    \global\vertexcount=-1
\global\newcount\XDIR
\global\newcount\YDIR
\gdef\SETDIR{  
\ifcase\LDIR
     \global\XDIR=0  \global\YDIR=1   
\or  \global\XDIR=1  \global\YDIR=1   
\or  \global\XDIR=1  \global\YDIR=0   
\or  \global\XDIR=1  \global\YDIR=-1  
\or  \global\XDIR=0  \global\YDIR=-1  
\or  \global\XDIR=-1 \global\YDIR=-1  
\or  \global\XDIR=-1 \global\YDIR=0   
\or  \global\XDIR=-1 \global\YDIR=1   
\else\DIRECTERROR
\fi}  
\gdef\moduloeight#1{
\ifnum#1>7 \global\advance #1 by -8
\relax
\moduloeight#1
\relax
\else \relax
\fi}
\gdef\multroothalf#1{\global\multiply #1 by 7071 \global\divide #1 by 10000}
\gdef\negate#1{\global\multiply #1 by -1}

\gdef\slanttest(#1,#2){
\ifodd\LDIR
\multiply #1 by 7071  \divide #1 by 10000
\multiply #2 by 7071  \divide #2 by 10000
\fi
}
\gdef\gslanttest(#1,#2){
\ifodd\LDIR
\multroothalf#1
\multroothalf#2
\fi
}
\gdef\setplength{ 
\global\particlelengthx=\unitboxwidth
\global\particlelengthy=\unitboxheight
\global\multiply \particlelengthx by \unitboxnumber
\global\multiply \particlelengthy by \unitboxnumber
\global\advance \particlelengthx by \particleadjustx
\global\advance \particlelengthy by \particleadjusty
}
\gdef\boxlengthdefault{  
\global\boxlengthx=\plengthx
\global\boxlengthy=\plengthy
\ifnum\plengthx<0 \global\multiply\boxlengthx by -1 \fi
\ifnum\plengthy<0 \global\multiply\boxlengthy by -1 \fi
}
\gdef\rearcoords{
\global\particlebacky=\particlefronty
\global\particlebackx=\particlefrontx
\global\advance \particlebackx by \particlelengthx
\global\advance \particlebacky by \particlelengthy
}
\gdef\midcoords{
\global\particlemidy=\particlefronty
\global\particlemidx=\particlefrontx
\global\stemlengthx=\particlelengthx
\global\stemlengthy=\particlelengthy
\global\divide\stemlengthx by 2
\global\divide\stemlengthy by 2
\global\advance \particlemidx by \stemlengthx
\global\advance \particlemidy by \stemlengthy
}
\gdef\setparticle{\setplength\rearcoords\midcoords\boxlengthdefault}
\gdef\setcoords(#1,#2,#3)(#4,#5,#6)[#7,#8]{
\global\upperlineadjx=#1
\global\lowerlineadjx=#2
\global\thirdlineadjx=#3
\global\upperlineadjy=#4
\global\lowerlineadjy=#5
\global\thirdlineadjy=#6
\global\unitboxwidth=#7
\global\unitboxheight=#8
}
\gdef\drawoldpic#1(#2,#3){  
\global\particlefrontx=#2
\global\particlefronty=#3
\rearcoords
\midcoords
\put(#2,#3){\usebox{#1}}
}
\gdef\drawsavedline`#1' as #2[#3#4](#5,#6)[#7]{
\global\LINETYPE=#2
\global\LINEDIRECTION=#3
\global\LINECONFIGURATION=#4
\global\particlefrontx=#5
\global\particlefronty=#6
\global\unitboxnumber=#7
\selectcase
\rearcoords
\midcoords
\ifnum\phantomswitch=0 \drawas{#1}\fi
}

\gdef\drawas#1{
\global\savebox{#1}(\boxlengthx,\boxlengthy){
\setlength{\unitlength}{0.01pt}
\begin{picture}(\boxlengthx,\boxlengthy)
\multiput(\upperlineadjx,\upperlineadjy)(\unitboxwidth,\unitboxheight)
{\numupperunits}{\upperunitbox}
\ifnum\numlineparts > 1  
\multiput(\lowerlineadjx,\lowerlineadjy)(\unitboxwidth,\unitboxheight)
{\numlowerunits}{\lowerunitbox}
\fi
\ifnum\numlineparts > 2  
\multiput(\thirdlineadjx,\thirdlineadjy)(\unitboxwidth,\unitboxheight)
{\numthirdunits}{\thirdunitbox}
\fi
\ifnum\numlineparts > 3  
\multiput(\fourthlineadjx,\fourthlineadjy)(\unitboxwidth,\unitboxheight)
{\numfourthunits}{\lowerunitbox}
\fi
\end{picture} }
\global\PFRONTx=\pfrontx  \global\PFRONTy=\pfronty   
\SETFRONTSTEM
\THICKPHOTONTEST
\ifdim\THICKPHOTONSWITCH=1pt\global\advance\PFRONTy by 20  \fi
\put(\PFRONTx,\PFRONTy) {\usebox{#1}}   
\ifdim\THICKPHOTONSWITCH=1pt
\global\advance\PFRONTy by -40
\put(\PFRONTx,\PFRONTy) {\usebox{#1}}
\global\advance \PFRONTy by 20
\fi  
\SETBACKSTEM
\seglength=1416   \gaplength=850   
}
\gdef\drawandsaveline`#1' as #2[#3#4](#5,#6)[#7]{
\global\newsavebox{#1}
\drawsavedline`#1' as #2[#3#4](#5,#6)[#7]
}

\gdef\drawline#1[#2#3](#4,#5)[#6]{   
\drawsavedline`\lastline' as #1[#2#3](#4,#5)[#6]}

\gdef\TYPEERROR{\message{*** ERROR IN PARTICLE TYPE
SELECTION ***}
\message{+++ Try with line type \fermion,\scalar,\photon,\gluon
(see manual) +++}\SETERR}
\gdef\DIRECTERROR{\SETERR\message{*** ERROR IN PARTICLE
DIRECTION SELECTION
***}
\message{+++ Try again with direction N, NE, E, SE  etc. or see
manual +++}}
\gdef\UNIMPERROR{\message{*** ERROR IN PARTICLE OPTIONS
SELECTION ***}
\message{
+++ The requested options combination has not yet been implemented
+++}\SETERR}
\gdef\SETERR{\gdef\upperunitbox{{\tiny Error}}
\gdef\lowerunitbox{\relax}
\gdef\thirdunitbox{\relax}
}
\gdef\neglengthcheck{\ifnum\unitboxnumber < 1
\message{   *** ERROR:  PARTICLE OF NEGATIVE OR ZERO
LENGTH REQUESTED. ***   }
\message{   ***         TAKING ABSOLUTE VALUE. ***   }\negate
\unitboxnumber\fi}
\gdef\selectcase{
\neglengthcheck   
\SETDIR
\ifcase\LINETYPE
\TYPEERROR  
\or \selectfermion  
\or \selectscalar   
\or \selectphoton   
\or \selectgluon    
\or \selectspecial  
\else \TYPEERROR \fi  }
\gdef\selectfermion{
\ifnum\fermioncount=0

\global\newcount\fermionlength  
\global\newcount\fermionlengthx
\global\newcount\fermionlengthy
\global\newcount\fermionfrontx  
\global\newcount\fermionfronty  
\global\newcount\fermionbackx
\global\newcount\fermionbacky
\gdef\ALLfermion{  
\global\fermionfrontx=\particlefrontx \global\fermionfronty=\particlefronty
\ifnum\unitboxnumber > 50000
\message{   *** WARNING *** Fermion of length
\the\unitboxnumber\space requested ***   }
\ifnum\unitboxnumber > 80000
\message{   *** Reducing fermion length to 30000 (max 80000) ***   }
\global\unitboxnumber=30000 \fi \fi  
\global\fermionlength=\unitboxnumber 
\global\particleadjustx=0   \global\particleadjusty=0 
\global\numlineparts = 1    \global\numupperunits=1
\global\upperlineadjx=-200  \global\upperlineadjy=0
\global\fermionlengthx=\fermionlength
\global\fermionlengthy=\fermionlength
\gslanttest(\fermionlengthx,\fermionlengthy)
\global\multiply\fermionlengthx by \XDIR  
\global\multiply\fermionlengthy by \YDIR  
\global\unitboxheight=\fermionlengthy
\global\unitboxwidth=\fermionlengthx
\global\advance \fermionlengthx by \particleadjustx
\global\advance \fermionlengthy by \particleadjusty
\global\particlelengthx=\fermionlengthx
\global\particlelengthy=\fermionlengthy
\boxlengthdefault    \rearcoords    \midcoords
\global\fermionbackx=\particlebackx
\global\fermionbacky=\particlebacky
\ifcase\LINECONFIGURATION  
\ifnum\XDIR=0
\gdef\upperunitbox{\line(\XDIR,\YDIR){\boxlengthy}} 
\else
\gdef\upperunitbox{\line(\XDIR,\YDIR){\boxlengthx}}
\fi
\else \UNIMPERROR
\fi
}

\fi
\global\advance\fermioncount by 1  
\ALLfermion
}
\gdef\selectscalar{
\ifnum\scalarcount=0

\newcount\scalarlength
\newcount\scalarlengthx
\newcount\scalarlengthy
\newcount\scalarfrontx  
\newcount\scalarfronty  
\newcount\scalarbackx
\newcount\scalarbacky
\gdef\ALLscalar{
\global\scalarfrontx=\particlefrontx
\global\scalarfronty=\particlefronty
\numlineparts = 1      \numupperunits=\unitboxnumber
\ifcase\LINECONFIGURATION
\global\upperlineadjx=-200     \global\upperlineadjy=0
\slanttest(\seglength,\gaplength)   
\gdef\upperunitbox{\line(\XDIR,\YDIR){\seglength}}
\else \UNIMPERROR 
\fi
\global\unitboxwidth=\seglength  \global\advance\unitboxwidth by
\gaplength
\global\multiply \unitboxwidth by \XDIR
\global\unitboxheight=\seglength  \global\advance\unitboxheight
by \gaplength
\global\multiply \unitboxheight by \YDIR
\global\particleadjustx=\gaplength \global\multiply\particleadjustx by \XDIR
\global\particleadjusty=\gaplength \global\multiply\particleadjusty by \YDIR
\negate\particleadjustx   \negate\particleadjusty
\setparticle  
\global\scalarlengthx=\particlelengthx  
\global\scalarlengthy=\particlelengthy  
\ifnum\boxlengthx > 50000
\message{   *** WARNING *** Scalar of length in excess of 50000cp
requested!}\fi
\ifnum\boxlengthy > 50000
\message{   *** WARNING *** Scalar of length in excess of 50000cp
requested!}\fi
\global\scalarbackx=\pbackx      \global\scalarbacky=\pbacky   
}

*****************************************************************************

\fi
\global\advance\scalarcount by 1  
\ALLscalar
}
\gdef\selectphoton{   
\ifnum\photoncount=0

\newcount\numwiggles    \newcount\numwigglespo
\global\newcount\photonlengthx
\global\newcount\photonlengthy
\global\newcount\photonfrontx  
\global\newcount\photonfronty  
\global\newcount\photonbackx
\global\newcount\photonbacky
\newcount\halfwigglelength
\global\font\Twelverom=cmr12
\global\font\Tenrom=cmr10
\gdef\Lbr{{\Twelverom(}}   \gdef\Rbr{{\Twelverom)}}
\gdef\SLbr{{\Tenrom(}}     \gdef\SRbr{{\Tenrom)}}
\gdef\Smile{{\large$\smile$}}  
\gdef\Frown{{\large$\frown$}}  
\ifdim\BIGPHOTONS>0pt  \gdef\Smile{$\smile$} \gdef\Frown{$\frown$} \fi
\gdef\selectphoton{   
\global\advance\photoncount by 1  
\global\photonfrontx=\particlefrontx
\global\photonfronty=\particlefronty
\ifnum\unitboxnumber > 50
\message{   *** WARNING *** Photon with
\the\unitboxnumber\space half-wiggles requested ***   }
\ifnum\unitboxnumber > 150
\message{   *** Reducing photon length to 10 half-wiggles (max 150) ***   }
\ifnum\unitboxnumber > 1000
\message{   *** Probable Cause:  Photon selected instead of Fermion ***   }
\fi \global\unitboxnumber=10 \fi \fi  
\numwiggles=\unitboxnumber
\divide\numwiggles by 2
\global\unitboxnumberpo=\numwiggles
\global\multiply \unitboxnumberpo by -1
\numwigglespo=\unitboxnumber
\advance\numwigglespo by \unitboxnumberpo
\global\numlineparts = 2  
\global\numupperunits=\numwigglespo  
\global\numlowerunits=\numwiggles  
\particleadjustx=0  
\particleadjusty=0  
\ifcase\LINEDIRECTION
     \Nphoton    
\or  \NEphoton   
\or  \Ephoton    
\or  \SEphoton   
\or  \Sphoton    
\or  \SWphoton   
\or  \Wphoton    
\or  \NWphoton   
\else\DIRECTERROR \fi
\setplength
\global\divide\plengthx by 2  \global\divide\plengthy by 2
\rearcoords  \boxlengthdefault   \midcoords
\global\photonbackx=\pbackx  
\global\photonbacky=\pbacky  
\global\photonlengthx=\plengthx  
\global\photonlengthy=\plengthy  
}
\gdef\SETUNITBOX(##1)[##2][##3]{ 
\gdef\upperunitbox{\oval(##1,##1)[##2]}
\gdef\lowerunitbox{\oval(##1,##1)[##3]}
}
\gdef\Nphoton{  
\ifcase\LINECONFIGURATION  
\setcoords(-490,-250,0)(260,1250,0)[0,2000]
\gdef\upperunitbox{\SLbr}   \gdef\lowerunitbox{\SRbr}
\particleadjusty=10
\or 
\setcoords(-271,-501,0)(250,1250,0)[0,2000]
\gdef\upperunitbox{\SRbr}   \gdef\lowerunitbox{\SLbr}
\or 
\particleadjusty=0
\setcoords(-501,-351,0)(300,1400,0)[0,2200]
\gdef\upperunitbox{\Lbr}   \gdef\lowerunitbox{\Rbr}
\or 
\setcoords(-353,-499,0)(300,1400,0)[0,2200]
\gdef\upperunitbox{\Rbr}   \gdef\lowerunitbox{\Lbr}
\or 
\setcoords(-481,-371,0)(280,1300,0)[0,2000]
\gdef\upperunitbox{\Lbr}   \gdef\lowerunitbox{\Rbr}
\particleadjusty=150
\ifnum\numwiggles=\number\numwigglespo \particleadjustx=-50 \fi
\or 
\setcoords(-321,-391,0)(280,1300,0)[0,2000]
\gdef\upperunitbox{\Rbr}   \gdef\lowerunitbox{\Lbr}
\particleadjusty=150
\ifnum\numwiggles=\number\numwigglespo \particleadjustx=80 \fi
\or 
\setcoords(-490,-260,0)(300,1500,0)[0,2400]
\gdef\upperunitbox{\Lbr}   \gdef\lowerunitbox{\Rbr}
\or 
\setcoords(-301,-531,0)(300,1500,0)[0,2400]
\gdef\upperunitbox{\Rbr}   \gdef\lowerunitbox{\Lbr}
\else \UNIMPERROR
\fi
}
\gdef\NEphoton{
\ifcase\LINECONFIGURATION  
\setcoords(425,425,0)(1250,0,0)[1250,1250]       \SETUNITBOX(1250)[br][tl]
\ifnum\numwigglespo > \number \numwiggles \particleadjustx=15 \fi
\or 
\setcoords(1050,-200,0)(625,625,0)[1250,1250]    \SETUNITBOX(1250)[tl][br]
\ifnum\numwigglespo > \number \numwiggles \particleadjustx=25 \fi
\or 
\setcoords(500,500,0)(1400,0,0)[1400,1400]       \SETUNITBOX(1400)[br][tl]
\or 
\setcoords(1200,-200,0)(700,700,0)[1400,1400]    \SETUNITBOX(1400)[tl][br]
\or 
\setcoords(400,400,0)(1200,0,0)[1200,1200]       \SETUNITBOX(1200)[br][tl]
\or 
\setcoords(1000,-200,0)(600,600,0)[1200,1200]    \SETUNITBOX(1200)[tl][br]
\else \UNIMPERROR
\fi
\numupperunits=\numwiggles   \numlowerunits=\numwigglespo
}
\gdef\Ephoton{    
\ifcase\LINECONFIGURATION  
\setcoords(-285,715,0)(-150,-400,0)[2005,0]
\gdef\upperunitbox{\Frown}   \gdef\lowerunitbox{\Smile}
\or  
\setcoords(-285,715,0)(-420,-170,0)[2005,0]
\gdef\upperunitbox{\Smile}   \gdef\lowerunitbox{\Frown}
\else \UNIMPERROR
\fi
\particleadjustx=-15 
}
\gdef\SEphoton{
\ifcase\LINECONFIGURATION  
\setcoords(-200,1050,0)(-625,-625,0)[1250,-1250]
\SETUNITBOX(1250)[tr][bl]
\ifnum\numwigglespo > \number \numwiggles \particleadjustx=25 \fi
\or 
\setcoords(425,425,0)(0,-1250,0)[1250,-1250]
\SETUNITBOX(1250)[bl][tr]
\ifnum\numwigglespo > \number \numwiggles \particleadjustx=15 \fi
\or 
\setcoords(-200,1200,0)(-700,-700,0)[1400,-1400]
\SETUNITBOX(1400)[tr][bl]
\or 
\setcoords(500,500,0)(0,-1400,0)[1400,-1400]
\SETUNITBOX(1400)[bl][tr]
\or 
\setcoords(-200,1000,0)(-600,-600,0)[1200,-1200]
\SETUNITBOX(1200)[tr][bl]
\particleadjustx=-20
\or 
\setcoords(420,420,0)(0,-1200,0)[1200,-1200]
\SETUNITBOX(1200)[bl][tr]
\particleadjustx=40
\else \UNIMPERROR
\fi
}
\gdef\Sphoton{  
\ifcase\LINECONFIGURATION  
\setcoords(-252,-490,0)(-740,-1740,0)[0,-2000]
\gdef\upperunitbox{\SRbr}   \gdef\lowerunitbox{\SLbr}
\or 
\setcoords(-490,-260,0)(-740,-1740,0)[0,-2002]
\gdef\upperunitbox{\SLbr}   \gdef\lowerunitbox{\SRbr}
\or 
\setcoords(-299,-449,0)(-870,-1970,0)[0,-2200]
\gdef\upperunitbox{\Rbr}    \gdef\lowerunitbox{\Lbr}
\particleadjusty=-95
\or 
\setcoords(-517,-371,0)(-900,-2000,0)[0,-2200]
\gdef\upperunitbox{\Lbr}    \gdef\lowerunitbox{\Rbr}
\particleadjusty=-165
\or 
\setcoords(-299,-409,0)(-885,-1905,0)[0,-2000]
\gdef\upperunitbox{\Rbr}   \gdef\lowerunitbox{\Lbr}
\particleadjustx=50     \particleadjusty=-380
\ifodd\unitboxnumber\relax\else\particleadjustx=250
\particleadjusty=-400 \fi
\or 
\setcoords(-519,-449,0)(-900,-1920,0)[0,-2000]
\gdef\upperunitbox{\Lbr}   \gdef\lowerunitbox{\Rbr}
\particleadjusty=-370
\ifodd\unitboxnumber\relax\else\particleadjustx=-240
\particleadjusty=-400 \fi
\or 
\gdef\upperunitbox{\Rbr}   \gdef\lowerunitbox{\Lbr}
\setcoords(-325,-555,0)(-900,-2100,0)[0,-2400]
\particleadjusty=-40
\or 
\setcoords(-505,-275,0)(-900,-2100,0)[0,-2400]
\gdef\upperunitbox{\Lbr}   \gdef\lowerunitbox{\Rbr}
\particleadjusty=-30  
\else \UNIMPERROR
\fi
}
\gdef\SWphoton{
\ifcase\LINECONFIGURATION  
\setcoords(-825,-825,0)(0,-1250,0)[-1250,-1250]
\SETUNITBOX(1250)[br][tl]
\or 
\setcoords(-175,-1425,0)(-625,-625,0)[-1250,-1250]
\SETUNITBOX(1250)[tl][br]
\or 
\setcoords(-900,-900,0)(0,-1410,0)[-1400,-1400]
\SETUNITBOX(1400)[br][tl]
\or 
\setcoords(-200,-1600,0)(-700,-700,0)[-1400,-1400]
\SETUNITBOX(1400)[tl][br]
\or 
\setcoords(-800,-800,0)(0,-1200,0)[-1200,-1200]
\SETUNITBOX(1200)[br][tl]
\or 
\setcoords(-200,-1400,0)(-600,-600,0)[-1200,-1200]
\SETUNITBOX(1200)[tl][br]
\else \UNIMPERROR
\fi
}
\gdef\Wphoton{
\ifcase\LINECONFIGURATION 
\setcoords(-2245,-1245,0)(-150,-400,0)[-2005,0]
\gdef\upperunitbox{\Frown}   \gdef\lowerunitbox{\Smile}
\or 
\setcoords(-2245,-1245,0)(-400,-150,0)[-2005,0]
\gdef\upperunitbox{\Smile}   \gdef\lowerunitbox{\Frown}
\else \UNIMPERROR
\fi
\particleadjustx=57 
\ifnum\numwigglespo=\number\numwiggles \particleadjustx=0  \fi
\numlowerunits=\numwigglespo   \numupperunits=\numwiggles
}
\gdef\NWphoton{
\ifcase\LINECONFIGURATION  
\setcoords(-200,-1425,0)(625,625,0)[-1250,1250]
\SETUNITBOX(1250)[bl][tr]
\or 
\setcoords(-825,-825,0)(0,1250,0)[-1250,1250]
\SETUNITBOX(1250)[tr][bl]
\ifnum\numwigglespo > \number \numwiggles \particleadjusty=-15 \fi
\or 
\setcoords(-200,-1600,0)(700,700,0)[-1400,1400]
\SETUNITBOX(1400)[bl][tr]
\or 
\setcoords(-900,-900,0)(0,1400,0)[-1400,1400]
\SETUNITBOX(1400)[tr][bl]
\or 
\setcoords(-200,-1400,0)(600,600,0)[-1200,1200]
\SETUNITBOX(1200)[bl][tr]
\or 
\setcoords(-800,-800,0)(0,1200,0)[-1200,1200]
 \SETUNITBOX(1200)[tr][bl]
\else \UNIMPERROR
\fi
}

\fi
\selectphoton
}
\gdef\selectgluon{
\ifnum\gluoncount=0
\fi
\selectgluon
}
\gdef\selectspecial{\UNIMPERROR}
\gdef\checkvertex{ 
\ifnum\vertexcount=-1
\fi}
\gdef\drawvertex#1[#2#3](#4,#5)[#6]{\checkvertex\drawvertex#1
[#2#3](#4,#5)[#6]}
\gdef\vertexcap#1{\checkvertex\vertexcap#1}
\gdef\vertexcaps{\checkvertex\vertexcaps}
\gdef\vertexlink#1{\checkvertex\vertexlink#1}
\gdef\vertexlinks{\checkvertex\vertexlinks}
\gdef\stemvertex#1{\checkvertex\stemvertex#1}
\gdef\stemvertices{\checkvertex\stemvertices}
\gdef\flipvertex{\checkvertex\flipvertex}
\global\arrowlength=349  
\gdef\drawarrow[#1#2](#3,#4){
\global\LDIR=#1
\SETDIR
\global\boxlengthx=#3
\global\boxlengthy=#4  
\ifdim#2=1pt
\adjx=\arrowlength      \adjy=\arrowlength
\multiply\adjx by \XDIR \multiply\adjy by \YDIR  
\slanttest(\adjx,\adjy)
\global\advance\boxlengthx by \adjx    \global\advance\boxlengthy by \adjy
\fi
\ifnum\phantomswitch=0\put(\boxlengthx,\boxlengthy)
{\vector(\XDIR,\YDIR){0}}\fi
}  
\gdef\SETFRONTSTEM{
\EITHERSTEM=\FRONTSTEM   \advance\EITHERSTEM by \BACKSTEM
\ifdim\EITHERSTEM>0pt
\global\stemlengthx=\stemlength   \global\stemlengthy=\stemlength
\global\absstemlength=\stemlength
\SETDIR
\gslanttest(\stemlengthx,\stemlengthy)
\gslanttest(\absstemlength,\REG)
\ifnum\XDIR=0 \stemlengthx=0 \fi
\ifnum\YDIR=0 \stemlengthy=0 \fi
\global\multiply\stemlengthx by \XDIR
\global\multiply\stemlengthy by \YDIR
\ifdim\FRONTSTEM=1pt
\ifnum\phantomswitch=0
          \put(\pfrontx,\pfronty){\line(\XDIR,\YDIR){\absstemlength}}\fi
\global\advance\plengthx by \stemlengthx
\global\advance\plengthy by \stemlengthy
\global\advance\PFRONTx by \stemlengthx
\global\advance\PFRONTy by \stemlengthy
\global\advance\pmidx by \stemlengthx
\global\advance\pmidy by \stemlengthy
\global\advance\pbackx by \stemlengthx
\global\advance\pbacky by \stemlengthy
\ifnum\LTYPE=3
\global\photonfrontx=\PFRONTx  \global\photonfronty=\PFRONTy
\global\photonbackx=\pbackx    \global\photonbacky=\pbacky
\fi  
\ifnum\LTYPE=4
\global\gluonfrontx=\PFRONTx  \global\gluonfronty=\PFRONTy
\global\gluonbackx=\pbackx    \global\gluonbacky=\pbacky
\fi  
\fi  
\fi  
}    
\gdef\SETBACKSTEM{
\ifdim\BACKSTEM=1pt
\ifnum\phantomswitch=0
       \put(\pbackx,\pbacky){\line(\XDIR,\YDIR){\absstemlength}}\fi
\global\advance\plengthx by \stemlengthx
\global\advance\plengthy by \stemlengthy
\global\advance\pbackx by \stemlengthx
\global\advance\pbacky by \stemlengthy
\fi  
\global\stemlength=275  \FRONTSTEM=0pt  \BACKSTEM=0pt
}    
\gdef\drawloop#1[#2#3](#4,#5){  

\global\newcount\loopfrontx    \global\newcount\loopfronty
\global\newcount\loopbackx    \global\newcount\loopbacky
\global\newcount\loopmidx    \global\newcount\loopmidy
\global\newdimen\CENTRALLOOP
\gdef\drawloop#1[#2#3](#4,#5){
\global\CENTRALLOOP=0pt  
\global\LINETYPE=#1
\ifnum\LTYPE=\gluon\relax\else\UNIMPERROR\LTYPE=1
\message{Reverting to Gluons}
\fi
\global\LINEDIRECTION=#2  
\global\fourthlineadjx=#3 
\ifnum\fourthlineadjx=0 
  \global\CENTRALLOOP=1pt  
  \global\fourthlineadjx=8
  \global\LDIR=0
\fi
\global\fourthlineadjy=\fourthlineadjx  
\global\advance\fourthlineadjy by -4
\global\loopfrontx=#4   \global\loopfronty=#5
\ifdim\CENTRALLOOP=1pt
  \global\advance\loopfrontx by -2413  \global\advance\loopfronty by -425
\fi                          
\global\unitboxnumber=1  
\ifnum\LINETYPE=\photon \unitboxnumber=2 \fi
\checkdir
\drawline\LINETYPE[\LDIR\LCONFIG](\loopfrontx,\loopfronty)
[\unitboxnumber]
\DRAWLOOP
\ifnum\fourthlineadjy>-1 
\global\loopmidx=\loopfrontx   \global\loopmidy=\loopfronty
\global\advance\loopmidx by \loopbackx  \global\advance\loopmidy by
\loopbacky
\divide\loopmidx  by 2 \divide\loopmidy by 2  
\ifdim\CENTRALLOOP=1pt
  \global\advance\loopfrontx by 200    \global\advance\loopfronty by 425
  \global\advance\loopbackx by -200    \global\advance\loopbacky by -425
\fi
\fi 
}
\gdef\DRAWLOOP{
\global\advance\fourthlineadjx by -1
\ifnum\fourthlineadjx=0\relax  
\else
\ifnum\fourthlineadjx=\fourthlineadjy 
   \global\loopbackx=\pbackx   \global\loopbacky=\pbacky
\fi
\global\advance\LDIR by 1
\moduloeight\LDIR
\checkdir
\drawline\LINETYPE[\LDIR\LCONFIG](\pbackx,\pbacky)[\unitboxnumber]
\fi 
\ifnum\fourthlineadjx>1 \DRAWLOOP  \fi  
}
\gdef\checkdir{
\ifnum\LTYPE=\gluon
\ifodd\LDIR \global\LCONFIG=0 \else \global\LCONFIG=2 \fi
\fi 
}

\drawloop#1[#2#3](#4,#5)}
\Feynmanlength  

\begin{document}

\title{On second quantization on noncommutative spaces with twisted
symmetries}

\author{   Gaetano Fiore,  \\\\
$^{1}$ Dip. di Matematica e Applicazioni, Universit\`a ``Federico II''\\
   V. Claudio 21, 80125 Napoli, Italy\\         
$^{2}$         I.N.F.N., Sez. di Napoli,
        Complesso MSA, V. Cintia, 80126 Napoli, Italy}
\date{}

\maketitle
\abstract{By application of the general
twist-induced $\star$-deformation procedure
we translate second quantization of a system of bosons/fermions
on a symmetric spacetime in a non-commutative language.
The procedure deforms in a coordinated way the spacetime algebra and its
symmetries, the wave-mechanical description of a
system of $n$ bosons/fermions, the algebra of
creation and annihilation operators and also the commutation relations
of the latter with functions of spacetime; our key requirement
is the mode-decomposition independence of the quantum field.
In a minimalistic view, the use of noncommutative coordinates can be seen just as a way to better express non-local interactions of a special kind.
In a non-conservative one, we obtain a closed, covariant framework for QFT
on the corresponding noncommutative spacetime consistent with quantum
mechanical axioms and Bose-Fermi statistics. One distinguishing
feature is that the field commutation relations remain of the type
``field (anti)commutator=a distribution''. We illustrate the results
by choosing as examples interacting non-relativistic and free relativistic
QFT  on Moyal space(time)s.

\vskip1cm

\newpage


\section{Introduction}

The idea of Quantum Field Theory (QFT) on noncommutative spacetime
goes back to Heisenberg\footnote{Heisenberg proposed it in a letter
to Peierls \cite{Hei30} to solve the problem of divergent integrals
in relativistic QFT. The idea propagated via Pauli to Oppenheimer.
In 1947 Snyder, a student of Oppenheimer, published the first
concrete proposal of a quantum theory built on a noncommutative
space \cite{Sny47}. The idea was put aside after the first successes
of renormalization in QED.}, but has been investigated more
intensively only in the last fifteen years. Motivations range from
its appearance in string theory (an effective description in the
presence of a $D3$-brane with a large $B$-field is provided
\cite{SeiWit99} by a non-local Yang-Mills action obtained replacing
the pointwise product by a $\star$-product), to more fundamental
ones, like the search \cite{DopFreRob95} for a proper framework
where the principles of quantum mechanics and of general relativity
can be conciliated; such a framework could yield as a bonus also an
inthrinsic regularization mechanism of ultraviolet divergences in
QFT - another important motivation (Heisenberg's original one).
Trying to mimic QFT on commutative spacetime,
it is desirable to reproduce as many equivalent approaches as possible.
However, we believe their legitimation should come from the equivalence to some
operator approach admitting a proper quantum mechanical interpretation.
In fact, on Moyal spaces the non-equivalence \cite{Bah04} of
the naive Euclidean and Minkowski formulations of relativistic
QFT, or of the path-integral  \cite{Fil96} and operator
approaches to quantization, may be the source \cite{BahDopFrePia}
of complications like non-unitarity after naive Wick back-rotation
\cite{GomMeh00}, violation
of causality \cite{SeiSusTou00,BozFisGroPitPutSchWul03} even at
large distance, mixing of UV and IR divergences \cite{MinRaaSei00}
and subsequent non-renormalizability \cite{CheRoi00}, the need for
translation non-invariant counterterms to recover renormalizability
\cite{GroWul03}.

Historically, Second Quantization played a crucial role in the
foundation of QFT as a bottom-up approach from the wave-mechanical
description of a system of $n$ identical quantum particles. For
instance, the nonrelativistic field operator of a zero spin particle
(in the Schr\"odinger picture) and its hermitean conjugate are
introduced by
\be
\varphi({\rm x}):= \varphi_i({\rm x})a^i,
\qquad\quad \varphi^*({\rm x})\!=\! \overline{\varphi_i({\rm x})} a^+_i,
\qquad\qquad\mbox{(infinite sum over $i$),}         \label{Schfield}
\ee
where $\{ e_i\}_{i\!\in\!\b{N}}$ is an orthonormal basis of a suitable subspace
$\H$ dense in the 1-particle Hilbert space and $\varphi_i$, $a^+_i,a^i$ the
wavefunction, creation, annihilation operators associated to $e_i$.
It is tempting to {\it adopt the Second Quantization approach also
on noncommutative spaces}. The main motivation is the wish to start
from the particle interpretation of quantum fields and their
Bose/Fermi statistics\footnote{Changing the statistics, i.e. the
rule to compute the number of allowed states of $n$-particle
systems, would have dramatic physical consequences: even a very tiny
violation of Pauli exclusion principle would e.g. appreciably modify
the behaviour of multi-fermion systems and affect the properties of
matter, first of all its stability, especially in extreme conditions
(e.g. in neutron stars).} in generic (in particular, 3+1)
dimensions. In this paper we do, using a {\it twist} \cite{Dri83} to
deform in a coordinated way space(time), its symmetries and all
objects transforming under space(time) transformations; this has the
advantage of restoring all the undeformed symmetries in terms of
noncocommutative Hopf algebras.

Let us summarize the main conceptual steps. A rather general way to
deform an algebra is by deformation quantization
\cite{BayFlaFroLicSte}. This means keeping the vector space, but
deforming the original product $\cdot$ into a new one $\star$. On
the space of functions on a manifold $X$  $f\star h$  can be defined
applying to $f\ot h$ first a suitable bi-pseudodifferential operator
$\bF$ (reducing to 1 when the deformation parameter $\lambda$
vanishes) and then the pointwise multiplication $\cdot$. If one replaces
all $\cdot$ by $\star$'s in an equation of motion, e.g. in the
Schr\"odinger equation on $X=\b{R}^3$ of a particle with electrical
charge $q$
\be
\ba{c} \Haus\psi({\rm x})=i\hbar  \partial_t
\psi({\rm x}),\quad \qquad
\Haus:=\big[\!\frac{-\hbar^2}{2m}D^a\!\star\!
D_a\!+\!V\big]\star,\quad \qquad D_a\!=\!\partial_a\!+\! i qA_a,
\ea                  \label{1Schr}
\ee
one obtains a pseudodifferential equation and therefore introduces a
(very special) amount of non-locality in the interactions.
A very interesting situation is when $\bF$ is related to the
symmetries of $X$.
If for simplicity $X\!=\!\b{R}^m$,  $G\!=\!ISO(m)$ is the isometry group of
 $X$, $\g$ its Lie algebra,
$U\g$ its Universal Enveloping Algebra (UEA), and we choose  $\bF$
as the inverse of a unitary  twist
$\F\!=\!\1\!\ot\!\1\!+\!\lambda\F^\alpha\!\ot\!
\F_\alpha\!+\!...\!\in\! (U\g\ot U\g)[[\lambda]]$
\cite{Dri83} (see also \cite{Tak90,ChaPre94}), then the action
of $U\g$ on a $\star$-product $f\star h$ obeys (section
\ref{TwistMod}) the deformed Leibniz rule corresponding to the
coproduct of the triangular noncocommutative Hopf algebra
$\widehat{U\g}$ obtained twisting $U\g$ with $\F$ (section \ref{TwistSym}).
In a similar way, inverting the $\lambda$-power
expansion defining the $\star$-product in all
$U\g$-module algebras,
$$
f \cdot h=f\star
h+\lambda(\F^\alpha\trc f)\star(\F\!_\alpha\trc
h)+...,
$$
we can express  also all other commutative notions [wavefunctions
$\psi$, differential operators (Hamiltonian, etc) and integration
(section \ref{twistdifcal}), $U\g$-covariant $a^i,a^+_i$ (section
\ref{Hei-Cli}), the wave-mechanical description of $n$
bosons/fermions (section \ref{Config}),...]
purely in terms of their $\star$-analogs
and thus translate (sections \ref{Schrpic} ,\ref{Heispic}) nonrelativistic second
quantization on $X$ in a ``noncommutative language'',
which we finally express by the use of ``hatted'' objects
$\hat x^a,\hat\psi...$ only; in this final translation, summarized
in formulae (\ref{hatDn}), (\ref{hatintprop}), (\ref{hatqfccr}),  all
formal $\lambda$-power series arising from $\F$
either disappear or  are expressed through the
triangular structure $\R\!:=\!\F_{21}\F^{-1}$, which has
much better representation properties than $\F$.
In a sense, the philosophy is like the one by J. Wess and
coworkers in formulating \cite{AscBloDimMeySchWes05,AscDimMeyWes06}
noncommutative diffeomorphisms and related notions
(metric, connections, tensors, etc).

In a minimalistic view one can  see the replacement of
all $\cdot$ by $\star$'s in (\ref{1Schr})
just as a way to introduce non-locality
in the interactions, and the use of noncommutative coordinates
just as a help to solve this equation,  but keep
considering spacetime {\it commutative}, in the sense of describing
the measurement processes of the space coordinates of an event
still by the (commuting)  multiplication operators ${\rm x}^a\cdot$.
In a more open-minded view, one can
re-interpret the results as the construction of
a noncommutative space(time) and  on it of a  `closed',
$\widehat{U\g}$-covariant candidate framework for QFT
consistent with the basic principles of quantum mechanics and
Bose/Fermi statistics, what is the main purpose of this paper.
Its observational consequences (e.g.
of adopting noncommutative coordinates $\hat x^a$ for describing
the measurement processes of the space coordinates of an event, or
the meaning of twisted
spacetime symmetries) deserve separate investigations.

We anticipate the key features of our deformed nonrelativistic
QFT framework on  $X\!=\!\b{R}^3$ as
follows. A basic property of $\varphi,\varphi^*$ of (\ref{Schfield})
is their basis-independence, i.e. invariance under the group
$U(\infty)$ of unitary transformations of $\H$;
$\varphi_i,a^+_i$ transform according to the same representation
$\rho$, $\varphi_i^*,a^i$ according the contragredient $\rho^\vee$.
The group $G\!=\! ISO(3)$ of (active) space-symmetry
transformations (combined translations and
rotations of the system) is a subgroup of $U(\infty)$ and of the
Galilei covariance group of the theory, $G'$. Deforming the setting
through a twist $\F\!\in\! (U\g\ot U\g)[[\lambda]]$ (section
\ref{N2ndQ}) must leave  the field invariant under $\widehat{U\g}$
(and $\widehat{Uu(\infty)}$). Consequently, deformed $a^+_i,a^i$ no
longer commute with deformed
functions, but the deformed field does; moreover, the commutation
relations of the deformed field appear as the undeformed (section \ref{Schrpic}). The same occurs with Heisenberg fields (defined as
usual with a commutative time), and the deformed nonrelativistic
theory is $\widehat{U\g'}$ covariant (section \ref{Heispic}).

In section \ref{R2ndQ} we
extend the Second Quantization procedure to construct relativistic
free fields on a deformed Minkowski spacetime covariant under the
associated deformed Poincar\'e symmetry. We then concentrate the
attention on the simplest examples, Moyal-Minkowski spaces and the
corresponding twisted Poincar\'e Hopf algebra $\widehat{U\P}$
\cite{ChaKulNisTur04,Wes04,Oec00}.  It has been debated
\cite{ChaPreTur05,Tur06,BalManPinVai05,AkoBalJos08,
BuKimLeeVacYee06,Zah06,LizVaiVit06,Abe06,RicSza07,AscLizVit07} how to implement
$\widehat{U\P}$ covariance even for free fields, three main issues
being whether one should deform also the commutation relations {\it
a)} among coordinates of different spacetime points, {\it b)} among
creation and annihilation operators, {\it c)} of the latter with
spacetime coordinates. Our procedure leads to a peculiar combination
of {\it a)}, {\it b)}, {\it c)} [respectively (\ref{hatDn}) \& (\ref{hqccr}) \& (\ref{earel})  in the general deformed Minkowski case,
(\ref{hatDmoyal}) \& (\ref{aa+cr})
in the Moyal-Minkowski one], but again functions $\star$-commute
with quantum fields. It is encouraging that the same combination
arises [eq. (46) of \cite{FioWes07}] also from consistency with the
Wightman axioms for relativistic QFT, see also
\cite{Fio08Proc}\footnote{We put aside the other combination
proposed in eq. (44) of \cite{FioWes07} for the reasons presented in
\cite{Fio08Proc}.}. Ref. \cite{FioWes07} found also quite disappointingly that
the $n$-point functions of a scalar theory, both free and
self-interacting, depend on the differences of coordinates at
independent spacetime points as in the undeformed theory, i.e. the
effect of $\star$-products disappears. This is due to the
translation invariance of the interaction $\int\varphi^{\star n}$
considered in \cite{FioWes07} and should change in the presence
of gauge interactions as in sections \ref{N2ndQ}, \ref{moyalE}.


Our framework is consistent with Bose/Fermi statistics, in contrast
with claims often made in the literature \cite{BalManPinVai05} and
despite changes
appearing at two levels. At the level of the deformed
wave-mechanical description (sections \ref{Config}, \ref{R2ndQ}),
the realization of the permutation group $S_n$ on noncommutative
$n$-particle wavefunctions differs from the usual one by a unitary
transformation $\wedge^n$ related to the twist; as $\wedge^n$ is not
completely symmetric, the noncommutative wavefunctions of $n$
bosons/fermions are not completely (anti)symmetric, but are so up to
$\wedge^n$, and hence still singlets under this realization of
$S_n$. Such a mechanism, which actually makes Bose/Fermi statistics
compatible also with transformations under {\it quasitriangular}
Hopf algebras, was already proposed in \cite{FioSch96} on the
abstract Hilbert space rather than on its realization on the space
of square-integrable wavefunctions, as done here. At the level
(section \ref{Hei-Cli}) of the Heisenberg/Clifford algebra, the
Fock-type representation of the deformed algebra is on the {\it
ordinary} (i.e. undeformed) Fock space, which describes the states
of a system of bosons/fermions; the deformed $a^i,a^+_i$ act
on the Fock space as ``dressed'' operators (i.e. composite in the
undeformed ones), but do not map the Fock space out of itself.

Whether our framework is implementable beyond the level of formal
$\lambda$-power series can be studied case by case
using ``noncommutative mathematics'' only.
On Moyal deformations of space(time) - probably the simplest and
best known noncommutative spaces - this seems the case
because the $\star$-products admit (section \ref{moyal})
non-perturbative (in $\lambda$) definitions
in terms of Fourier transforms.

Sections  \ref{TwistSym}, \ref{TwistMod} give mathematical preliminaries.
Section \ref{TwistSym} is a basic introduction to twisting
of Hopf algebras $H$, in particular UEA's, and the associated notation.
Section  \ref{TwistMod} is a systematic presentation of
known and new results regarding the twisting of $H$-module algebras and
its applications;
one can focus on the main results in a first reading and come back to the
details when they are referred to in the following sections. 
As said, sections \ref{N2ndQ}, \ref{R2ndQ} treat second quantization
respectively of non-relativistic interacting and relativistic non-interacting
fields.
Section \ref{moyalE} is a detour from field theory to quantum
mechanics on the Moyal deformation of $\b{R}^3$.  
It first shows how the (anti)symmetry of 2-particle wavefunctions is 
translated in terms of noncommutative coordinates; then it illustrates 
how the occurrence of the $\star$ in (\ref{1Schr})
modifies the 1-particle Schr\"odinger equation in two
simple models (a constant magnetic field and a plane-wave
electromagnetic field) and how the use of noncommutative coordinates
can help to solve it (in the first model).

We stick to vector spaces $V$ and algebras $\A$ over
$\b{C}$. In the sequel $V(\A)$ stands for the vector space
underlying $\A$. $V[[\lambda]]$, $\A[[\lambda]]$ respectively
stand for the
topological vector space and algebra (over $\b{C}[[\lambda]]$)
of power series in $\lambda$ with coefficients in $V,\A$;
their tensor products are meant as completed in the
$\lambda$-adic topology.

\tableofcontents

\newpage
\section{Twisting Hopf algebras $H$}
\label{TwistSym}

Consider a cocommutative Hopf $*$-algebra $(H,*,\Delta ,
\varepsilon,S)$; $H$ stands for the algebra,
$*,\Delta, \varepsilon,S$ stand for the
$*$-structure, coproduct, counit and antipode respectively.
For readers not so familiar with Hopf algebras, we recall
that $\varepsilon$ gives the trivial representation,
$\Delta,S$ are the abstract operations by which one constructs
the tensor product of any two representations and the
contragredient of any representation, respectively; $S$
is uniquely determined by $\Delta$.
We are especially interested in the UEA examples $H=U\g$,
with the $*$-structure
determined by a real form of the Lie algebra $\g$. Then
$$
\ba{lll}
\varepsilon(\1)=1,\qquad \quad &\Delta(\1)=\1\ot\1,\qquad \quad & S(\1)=\1,\\[8pt]
\varepsilon(g)=0,\qquad \quad & \Delta(g)=g\ot\1+\1\ot g,\qquad \quad &  S(g)=-g,\qquad \qquad \mbox{if }g\in\g;
\ea
$$
$\varepsilon,\Delta$
are extended to all of $H\!=\!U\g$ as $*$-algebra maps,
$S$ as  a $*$-antialgebra map:
\be
\ba{lll}
\varepsilon:H\mapsto\b{C},\quad\qquad  & \varepsilon(ab)=\varepsilon(a)\varepsilon(b),\quad\qquad  & \varepsilon(a^*)=[\varepsilon(a)]^*,\\[8pt]
\Delta:H\mapsto
H\ot H,\quad \qquad  & \Delta(ab)=\Delta(a)\Delta(b), \qquad
 & \Delta(a^*)=[\Delta(a)]^{*\ot *},\\[8pt]
S:H\mapsto H,\quad\qquad  & S(ab)=S(b)S(a),\quad\qquad  &
S\left\{\left[S(a^*)\right]^*\right\}=a.
\ea
\label{deltaprop}
\ee
The extensions of $\Delta,S$ are unambiguous, as
$\Delta\big([g,g']\big)=\big[\Delta(g),\Delta(g')\big]$,
$S\big([g,g']\big)=\big[S(g'),S(g)\big]$ if
$g,g'\in\g$. We shall often abbreviate
$(H,*,\Delta,\varepsilon,S)$ just as $H$.
Clearly $(H[[\lambda]],*,\Delta ,\varepsilon,S)$ is a (topological)
cocommutative Hopf $*$-algebra if we extend the product, $\Delta,\varepsilon,S$
$\b{C}[[\lambda]]$-linearly   and $*$ $\b{C}[[\lambda]]$-antilinearly.

Consider a {\it twist} \cite{Dri83}
(see also \cite{Tak90,ChaPre94}), i.e.
an element $\F\!\in\! (H\ot H)[[\lambda]]$ fulfilling\footnote{By
definition $\F=\sum_{k=0}^\infty f_k\lambda^k$
with  $f_k\!\in\! H\ot H$; $f_0=\1\ot\1$
by the second equality in (\ref{twistcond})$_1$.
Fixed a basis $\{h_\mu\}$ of $H$, $f_k$ can be decomposed
as a finite combination $f_k\!=\!\sum_{\mu,\nu} f_k^{\mu\nu}h_\mu\ot h_\nu$,
with $f_k^{\mu\nu}\!\in\!\b{C}$.
The generic term $\F^{(1)}_I \ot\F^{(2)}_I$ in the decomposition of $\F$ is
$\lambda^kf_k^{\mu\nu}h_\mu\ot h_\nu$, and
$\sum_I$ means $\sum_{k,\mu,\nu}$.}
\bea
&&\F\!\equiv\!\sum_I \F^{(1)}_I \ot\F^{(2)}_I
=\1\ot\1+ O(\lambda),\qquad \qquad (\epsilon\ot\id)\F=
(\id\ot\epsilon)\F=\1, \qquad    \qquad          \label{twistcond}\\[8pt]
&&(\F\ot\1)[(\Delta\ot\id)(\F)]=(\1\ot\F)[(\id\ot\Delta)(\F)]=:\F^3
\quad\mbox{(cocycle condition)}.
                                           \label{cocycle}
\eea
Let $H_s\!\subseteq\!H$ the smallest Hopf $*$-subalgebra such that
$\F\!\in\! (H_s\ot H_s)[[\lambda]]$.  Let also
\be
\beta:=\sum_I \F^{(1)}_I S\left(\F^{(2)}_I \right)\in H_s[[\lambda]].    \label{defbeta}
\ee
For our purposes $\F$ is {\it unitary} ($\F^{*\ot *}=\F^{-1}$),
whence  \ 
$\beta^*=S\!\left(\beta^{-1}\right)$; \
%
%
without loss of generality
$\lambda$ can be assumed real. Let $\hH\!:=\!H[[\lambda]]$
and for any $g\!\in\!\hH$\footnote{In (\ref{inter-2}) one could
replace $\beta^{-1}$ by
$S(\beta)$, as $S(\beta)\beta\in\mbox{Centre}(H)[[\lambda]]$.
In terms of the decomposition $\bF\!\equiv\!\F^{-1}\!=\!\sum_I \bF^{(1)}_I \ot \bF^{(2)}_I$
one can show that
\be
 \beta^{-1}=\sum_I S\left(\bF^{(1)}_I \right)\bF^{(2)}_I , \qquad \quad
\qquad S\big(\beta^{-1}\big)=\sum_I S\left(\bF^{(2)}_I\right)
\bF^{(1)}_I. \label{defbeta'} \ee }
\be
g^{\hat *}\!:=\!g^*, \qquad\hat\Delta(g) \!:=\!\F \Delta(g) \F^{-1},
\qquad\hat\varepsilon(g)\!:=\!\varepsilon(g), \qquad
\hat S(g)\!:=\!\beta S(g)\beta^{-1}              ,\label{inter-2}
 \ee
one finds that the analogs of conditions (\ref{deltaprop}) are satisfied,
and therefore $(\hH,\hat *,\hat\Delta , \hat\varepsilon,\hat S)$ is a Hopf $*$-algebra (which we shall often abbreviate as $\hH$) with triangular structure
$\R\!:=\!\F_{21}\F^{-1}$, deformation of the initial one. $(\hH_s,\hat *,\hat\Delta , \hat\varepsilon,\hat S)$, with  $\hH_s:=H_s[[\lambda]]$, is a Hopf $*$-subalgebra. Drinfel'd has shown \cite{Dri83} that
any triangular deformation of the initial  Hopf algebra
can be obtained in this way
(up to isomorphisms). Given
$\hat\Delta,\Delta$, the twist $\F$ is determined up to
multiplication $\F\to\F'=\F T$ by a unitary and $\g$-invariant [i.e.
commuting with $\Delta(\g)$] element $T\!\in\! (H\ot H)[[\lambda]]$
fulfilling (\ref{twistcond}), (\ref{cocycle})\footnote{On the other hand,
if we know a
non-unitary twist for the Hopf $*$-algebra $\hH$, a transformation
of this kind allows to construct \cite{Dri89,Jur94}
a unitary one: it suffices to choose $T=(\F^{*\ot *}\F)^{- 1/2}$.}.
$\R$ is independent of $T$ if the latter is symmetric. Suitable additional
conditions may restrict the choice or uniquely determine $T$ and hence the twist.

Eq. (\ref{cocycle}), (\ref{inter-2}) imply the
generalized intertwining relation
$\hat\Delta^{(n)}(g)\!=\!\F^n\Delta^{(n)}(g)(\F^n)^{-1}$
 for the iterated coproduct. By definition
$$
\F^n\in (\hat H_s)^{\ot n},
\qquad\Delta^{(n)}:  H\mapsto H^{\ot n},
\qquad\hat\Delta^{(n)}: \hH\mapsto \hH^{\ot n},
$$
reduce to $\F,\Delta,\hat\Delta$ for $n=2$, whereas for $n>2$ they
can be defined recursively as \be \ba{l}
\F^{m\!+\!1}=(\1^{\ot{(m\!-\!1)}}\ot\F)[(\id^{\ot{(m\!-\!1)}}
\ot\Delta)\F^m],\\[8pt]
\Delta^{(m\!+\!1)}=(\id^{\ot{(m\!-\!1)}}\ot\Delta)\circ\Delta^{(m)},
\qquad\hat\Delta^{(m\!+\!1)}=(\id^{\ot^{m\!-\!1}}\ot\hat\Delta)
\circ\hat\Delta^{(m)}.                \label{iter-n} \ea
\ee
The results for
$\F^n,\Delta^{(n)},\hat\Delta^{(n)}$ are the same if at any step $m$
we respectively apply $\F(\cdot\ot\cdot)\Delta,\Delta,\hat\Delta $ not
to the last but to a different tensor factor;
for $n\!=\!3$ this means that (\ref{cocycle}) and the equalities
$\Delta^{(3)}\!=\!(\Delta\ot\id)\!\circ\!\Delta$,
$\hat\Delta^{(3)}\!=\!(\hat\Delta\ot\id)\!\circ\!\hat\Delta$ hold
(coassociativity of $\Delta,\hat\Delta$).
For any $g\in \hH=H[[\lambda]]$ we shall use the following Sweedler notations
for the decompositions of $\Delta^{(n)}(g),\hat\Delta^{(n)}(g)$ in
$\hH^{\ot n}$\footnote{This is the analog of (\ref{twistcond})$_1$,
i.e. $\sum_I$ means $\sum_{k,\mu,\nu,...}$, etc.}:
$$
\Delta^{(n)}(g)=\sum_I  g^I_{(1)} \otimes g^I_{(2)} \otimes ...
\otimes g^I_{(n)},\qquad\qquad
\hat\Delta^{(n)}(g)=\sum_I  g^I_{(\hat 1)} \otimes g^I_{(\hat 2)} \otimes ...
\otimes g^I_{(\hat n)}.
$$

\section{Twisting $H$-modules and $H$-module algebras}
\label{TwistMod}

We recall that, given  a Hopf $*$-algebra $H$ over $\b{C}$,
a left $H$-module $(\M,\trc)$ is a vector space $\M$ over $\b{C}$
equipped with a left action, i.e. a $\b{C}$-bilinear map
$(g,a)\!\in\! H\!\times\!\M\mapsto g\trc a\!\in\!\M$
such that (\ref{leibniz})$_1$ holds. Equipped also with
an antilinear involution $*$  fulfilling (\ref{leibniz})$_2$
$(\M,\trc,*)$ is a left $H$-$*$-module. Finally, a
left  $H$-module ($*$-)algebra is a $*$-algebra $\A$ over $\b{C}$
equipped with a left $H$-($*$-)module structure
$\big(V(\A),\trc,*\big)$ such that (\ref{leibniz})$_3$ holds:
\be(gg')\!\trc\! a=g
\!\trc\! (g'\! \!\trc\! a)\!,\qquad (g\!\trc\! a)^*\!=[S(g)]^*\!\trc\! a^*\!,
\qquad g\!\trc\! (ab)=\!\sum_I\!  \left(g^I_{(1)}\!\trc\! a\right)\!
 \left(g^I_{(2)}\!\trc\! b\right).\qquad        \label{leibniz}
\ee
If $g\!\in\!\g$ formula (\ref{leibniz})$_3$ becomes the Leibniz rule.
Given  a Hopf  ($*$-)algebra $\hH$ the analogous objects (over
$\b{C}[[\lambda]]$) and maps are defined putting a $\hat{}$ over the previous
symbols. Hereby, (\ref{leibniz}) is replaced by
\be
(gg')\tr \hat
a=g \tr (g'\tr \hat a)\!,\qquad(g\tr \hat a)^{\hat *}\!=[\hat
S(g)]^{\hat *}\tr \hat a^{\hat *}\!, \qquad g\tr (\hat a\hat
b)=\!\sum_I\!  \left(g^I _{(\hat 1)}\tr \hat a\right)\!\! \left(g^I
_{(\hat 2)}\tr \hat b\right)\!.         \quad \label{defleibniz}
\ee
(\ref{defleibniz})$_3$ gives the deformation of the Leibniz rule.
Extending the action $\trc$ $\b{C}[[\lambda]]$-bilinearly one can trivially extend
any $H$-module $(\M,\trc)$ into a $\hH$-module $(\M[[\lambda]],\trc)$.
If $(\M,\trc,*)$ is a $H$-$*$-module and $\F$ is unitary then
$(\M[[\lambda]],\trc,*_\star)$ with
\be
 a^{*_\star}:=S(\beta)\trc a^*.       \label{star'}
\ee
is a $\hH$-$*$-module\footnote{See the appendix.
It is not possible to keep $*_\star=*$ as in \cite{AscDimMeyWes06},
section 8, since that was based on the condition
 $\F^{*\ot *}=(S\ot S)(\F_{21})$ rather than
$\F^{*\ot *}=\F^{-1}$.}. Given a $H$-module ($*$-)algebra $\A$ and choosing
$\M=V(\A)$, the twist gives also a systematic way to make $V(\A)[[\lambda]]$
into a $\hH$-module ($*$-)algebra $\A_\star$ by endowing it with
a new product, the  $\star$-product, defined by
\be
a\star b:=\sum_I  \left(\bF^{(1)}_I  \trc a\right) \left(\bF^{(2)}_I
\trc b\right).                                   \label{starprod}
\ee
The  $\b{C}[[\lambda]]$-bilinearity of $\star$ is manifest, the
associativity follows from (\ref{cocycle}), whereas
\bea
g\trc (a\star b)\!&\stackrel{(\ref{leibniz})_3}{=}&\!\sum_{I,I'}\!  \left(g^I_{(1)}\bF^{(1')}_{I'}  \trc a\right) \!
\left(g^I_{(2)}\bF^{(2')}_{I'}\trc b\right)\!\stackrel{(\ref{inter-2})}{=}\!
\sum_{I,I'}\!  \left(\bF^{(1')}_{I'}g^I_{(\hat 1)}  \trc a\right) \!
\left(\bF^{(2')}_{I'}g^I_{(\hat 2)}\trc b\right)\nn
\!&=&\!\sum_I\! \left(g^I_{(\hat 1)}\trc  a\right)
\!\star\! \left(g^I _{(\hat 2)}\trc b\right)\nonumber
\eea
proves the property (\ref{defleibniz})$_3$.
In the appendix we prove the compatibility with $*_\star$,
\be
(a\star b)^{*_\star}=b^{*_\star}\star a^{*_\star}.\label{*comp}
\ee
By (\ref{twistcond}), the $\star$-product coincides with the original one
 if $a$ or $b$ is $H_s$-invariant:
\be
g\trc a=\epsilon(g) a\quad \mbox{or}\quad g\trc b=\epsilon(g) b\quad
\forall g\in H_s
\qquad\qquad \Rightarrow\qquad \qquad a\star b=ab.\qquad \label{Trivstar}
\ee

\medskip
In the literature $\star$-products are mostly introduced to deform
abelian algebras (e.g. the algebra of functions on a manifold) into
non-abelian ones. We stress that the above construction works also
if $\A$ is non-abelian, e.g. if $\A=H$ \cite{Dri83,AscDimMeyWes06}.

\medskip
Given two $H$-modules $(\M,\trc)$, $(\NN,\trc)$,
the tensor product $(\M\!\ot\!\NN,\trc)$
also is if we define $g\trc (a\ot b):= \sum_I
\left(g^I_{(1)}\trc a\right) \ot\left( g^I_{(2)}\trc b\right)$.
As above, this is extended to a $\hH$-($*$-)module
$(\M\!\ot\!\NN[[\lambda]],\trc)$.
Introducing the ``$\star$-tensor product''
\cite{AscBloDimMeySchWes05,AscDimMeyWes06} \be \ba{l} (a\ot_\star
b):=\F^{-1} \trc^{\ot 2}(a\ot b)\equiv \sum_I(\bF^{(1)}_I\trc a)\ot
(\bF^{(2)}_I\trc b) \ea \label{startensor}\ee (an invertible
endomorphism, i.e. a change of bais, of $\M\!\ot\!\NN[[\lambda]]$), we find
\be
g\trc(a\ot_\star b)= \sum_I \left(g^I_{(\hat 1)}\trc a\right)
\ot_\star \left(g^I_{(\hat 2)}\trc b\right). \label{startransf}
\ee
Note that
this has the same form as the law $g\tr(a\ot b)= \sum_I
\left(g^I_{(\hat 1)}\tr a\right)\ot \left(g^I_{(\hat 2)}\tr
b\right)$ used to build
a $\hH$-module  $(\hat\M\!\ot\!\hat\NN,\tr)$ out of two
$(\hat\M,\tr)$, $(\hat\NN,\tr)$.
Given two $H$-module
($*$-)algebras $\A,\B$, this applies in particular to $\M=V(\A)$,
$\NN=V(\B)$. The tensor ($*$-)algebra $\A\ot\B$ [whose product is
defined by $(a\ot b)(a'\ot b')=(aa'\ot bb')$] also is a $H$-module
($*$-)algebra under the action $\trc$. By introducing the
$\star$-product (\ref{starprod}) $\A\ot\B$  is deformed into a
$\hH$-module ($*$-)algebra $(\A\ot\B)_\star$. To clarify the
relation with the products (and $*$-structures) in
$\A_\star,\B_\star$ we compute (see the appendix)
\be
\ba{c}
(a\ot_\star b)\star (a'\ot_\star b')=\sum_I a\star (\R^{(2)}_I\trc
a') \ot_\star (\R^{(1)}_I\trc b) \star b',\\[8pt]
(a\ot_\star b)^{*_\star}=\sum_I \big(\R^{(2)}_I\trc
a\big)\ot_\star \big(\R^{(1)}_I\trc b\big),
\ea\label{braid}
\ee
where $\sum_I\!\R^{(1)}_I\!\ot\! \R^{(2)}_I$ is the
decomposition of $\R$ in
$\hH\!\ot\!\hH$. Note that if $\A,\B$ are unital\footnote{This is no real
restriction: a non-unital algebra can be always extended unitally.} then
$a\ot_\star b=a_1\star b_2$ and
$(a\ot_\star b)^{*_\star}=b_2^{*_\star}\star a_1^{*_\star}$,
with the short-hand
notation $a_1\!:=\! a\ot\1_{\scriptscriptstyle \B}$,
$b_2\!:=\!\1_{\scriptscriptstyle \A}\ot b$.
 From (\ref{braid}) we recognize that
$(\A\ot\B)_{\star}$ is isomorphic to
the {\it braided  tensor product (algebra)}
\cite{Maj95,ChaPre94} of $\A_\star$ with $\B_\star$; here the
braiding is involutive and therefore spurious, as
$\R\R_{21}=\1\!\ot\!\1$ (triangularity of $\R$). So $(\A\!\ot\!\B)_{\star}$ encodes both the
usual $\star$-product within  $\A,\B$ and the $\star$-tensor product
(or braided tensor product) between the two. By (\ref{cocycle}) the
$\star$-tensor product is associative, and the previous results hold
also for iterated $\star$-tensor products.

\subsection{Module algebras defined by generators and relations}

Here we show that if $\A$ is defined by generators and relations,
then also $\A_\star$ is, with the same generators and
Poincar\'e-Birkhoff-Witt series, and with related relations; and similarly
for tensor products.

Given a generic $H$-($*$-)module $\M$ and fixed a
(for simplicity discrete) basis $\{a_i\}_{i\in{\cal I}}$
of $\M$, consider the free algebra $\A^f$ (over $\b{C}$) with
$\{a_i\}_{i\in{\cal I}}$ as a set of generators.
$\A^f$  is automatically also a $H$-module
($*$-)algebra under the action
$$
g\trc(a_{i_1} a_{i_2}...a_{i_k})=\sum_I  \left(g^I_{(1)}\trc a_{i_1}\right)
\left(g^I_{(2)}\trc a_{i_2}\right)...\left(g^I _{(k)}\trc a_{i_k}\right)
$$
and has the spaces $\M^k$ of homogeneous polynomials of degree $k$ as
$H$-($*$-)submodules. Endowing $V(\A^f)[[\lambda]]$
 with the $\star$-product (\ref{starprod}) one
deforms $\A^f$ into a $\hH$-module ($*$-)algebra $\A^f_\star$
 having the spaces $\M^k_\star$ of homogeneous
$\star$-polynomials of degree $k$ as $\hH$-($*$-)submodules.
Inverting the definition of the generic $\star$-monomial of degree
$k$ one can express the generic monomial of degree $k$ as a
homogeneous $\star$-polynomial of degree $k$:
\be
a_{i_1}
a_{i_2}...a_{i_k}=F^k{}_{i_1i_2...i_k}^{j_1j_2...j_k} a_{j_1}\star
a_{j_2}\star ...\star a_{j_k}        \label{starmon}
\ee
[here $F^k$ is
the $\b{C}[[\lambda]]$-valued matrix  defined by
$F^k=\tau^{\ot k}(\F^k)$, $\tau$ being
the representation defined by  $g\trc a_i=\tau_i^j(g)a_j$].
Formula (\ref{starmon}) is an identity in $V(\A^f)[[\lambda]]=V(\A^f_\star)$
which can be used to establish a one-to-one correspondence
$\wedge:f\!\in\!\A^f[[\lambda]]\!\mapsto\!
\wedge(f)\equiv\hat f\!\in\!\A^f_\star$ by the
requirement that the latter reduces to the identity
on $V(\A^f)[[\lambda]]=V(\A^f_\star)$: the polynomials $f,\hat f$
$$
f=\sum_{k=0}^m f_k^{j_1...j_k}a_{j_1}...a_{j_k},\qquad
\hat f=\sum_{k=0}^{n} \hat f_k^{j_1...j_k}a_{j_1}\star ...
\star a_{j_k}
\qquad\qquad f_k^{j_1...j_k},\hat f_k^{j_1...j_k}\in\b{C}[[\lambda]],
$$
(sum over repated indices $j_h$ is understood) are such that the equality
\be
f=\hat f
\label{defhat}
\ee
holds in $V(\A^f)[[\lambda]]=V(\A^f_\star)$
(then clearly their degrees $m,n$ must coincide);
$\wedge$ is $\b{C}[[\lambda]]$-linear by construction.
If e.g. $f$  is the monomial at
the lhs(\ref{starmon}), then $\hat f$ is the rhs(\ref{starmon}).
Using (\ref{cocycle})
it is easy to show that $\wedge,\wedge^{-1}$ fulfill
\be
\ba{l}
\wedge(f f')=\sum_I \wedge\!\!\left[\Fu_I\!\trc\! f \right]\,\star\:
\wedge\!\!\left[\Fd_I\!\trc\!f'\right],\\[8pt]
\wedge^{-1}(\hat f \!\star\!\hat f')=\sum_I
\left[\bF^{(1)}_I\!\trc\!\wedge^{-1}(\hat f )\right]
\left[\bF^{(2)}_I\!\trc\! \wedge^{-1}(\hat f')\right].
\ea  \label{hatprop}
\ee
Assume that $\A=\A^f/\I$, where ${\cal I}$ is a
$H$-invariant ($*$-)ideal  generated by a set of
polynomial relations\footnote{This assumption is satisfied
in most cases of interest. For instance,
if $\A$ has unit this must be one of the $a_i$, say $a_0=\1$, must span
a 1-dimensional submodule, and
among these relations there must be $\1 a_i-a_i=a_i\1-a_i=0$ for all $i$.
If $\A$
is a UEA of a Lie algebra the remaining relations are of the form
$a_ia_j-a_ja_i=c_{ij}^ka_k$.
If instead $\A$ is abelian among these relations there must be $a_ia_j-a_ja_i=0$,
for all $i,j$. Imposing further polynomial relations one can obtain
the algebra of polynomial functions on an algebraic manifold. The algebra of
differential operators on the latter is obtained adding as generators
a basis of vector fields (first order differential operators) and as additional relations the Lie algebra ones fulfilled by the latter together
with the Leibniz-rule-type commutation relations with the functions
on the manifold. Similarly, superalgebras can be defined replacing
some of the above commutation relations by anticommutation relations.
And so on.}
\be
f^J
=0,\qquad \qquad\qquad J\in{\cal J}.    \label{genrel}
\ee
Imposing (\ref{genrel}), in particular those of degree $k$,
will make $\M^k$ into a $H$-($*$-)submodule $\M'{}^k$
consisting of (no more necessarily homogeneous) polynomials of degree $k$.
The $\star$-polynomial relations
\be
\hat f^J
=0,\qquad \qquad\qquad J\in{\cal J} \label{genstarrel}
\ee
will generate a $\hH$-invariant ideal ${\cal I}_\star$,
therefore $\A_\star:=\A^f_\star/{\cal I}_\star$ is a
$\hH$-module ($*$-)algebra  with generators $a_i$ and
relations (\ref{genstarrel}). Imposing the latter $\M^k_\star$
become $\hH$-($*$-)submodules $\M'{}^k_\star$ consisting of
corresponding suitable $\star$-polynomials of degree $k$. By
construction the Poincar\'e-Birkhoff-Witt series of $\A,\A_\star$
coincide. Given a polynomial $f\!\in\!\A[[\lambda]]$, we shall denote still
by $\hat f\!\in\!\A_\star$ the polynomial such that the equality (\ref{defhat})
holds now in $V(\A_\star)=V(\A)[[\lambda]]$, and by
$\wedge:f\!\in\!\A[[\lambda]]\!\mapsto\!\hat f\!\in\!\A_\star$ the
corresponding linear map; for the latter (\ref{hatprop}) remains
valid. Summing up, {\bf by $\wedge,\wedge^{-1}$ one respectively expresses
polynomials in the $a_i$ as $\star$-polynomials in the $a_i$, and viceversa}.

Denoting by $\{a_i\}_{i\in{\cal I}}$, $\{b_i'\}_{i'\in{\cal I}'}$ sets
of generators of unital $\A,\B$, a set of generators of
both $\A\!\ot\!\B$ and $(\A\!\ot\!\B)_\star$ will consist of
$\{a_{i1},b_{i'2}\}_{i\in{\cal I},i'\in{\cal I}'}$. As generators of
$\A\ot\B$ [resp. $(\A\ot\B)_\star$] the $a_{i1}$ fulfill
(\ref{genrel}) [resp. (\ref{genstarrel})], the $b_{i'2}$ the
analogous relations for $\B$, and
\be a_{i1} b_{i'2}=
b_{i'2}a_{i1} \qquad \qquad \mbox{[resp. } a_{i1}\star b_{i'2}=
\sum_I  (\R^{(2)}_I\trc b_{i'2}) \star
(\R^{(2)}_I\trc a_{i1})\mbox{ ].} \label{braid'}
\ee
The $a_{i1}$  generate
a $H$- (resp. $\hH$-) module (*-)subalgebra, which we
call $\A_1$ (resp. $\A_{1\star}$). As a
$H$- (resp. $\hH$-) module (*-)algebra, this is
isomorphic to $\A$ (resp. $\A_\star$). Similarly for $\B_2$
(resp. $\B_{2\star}$).

Finally note that, as in (\ref{genstarrel}), (\ref{braid'})$_2$
the original product of $\A$ does not appear any more, one can introduce
$\A_\star$, $\B_\star$, $(\A\ot\B)_\star$ just in terms
of these generators and relations. For convenience we shall denote
them as $\hA$, $\hB$, $\widehat{\A\!\ot\!\B}$, the generators as
$\hat a_i,\hat b_{i'}$, $\hat a_{i1},\hat b_{i'2}$, $*_\star$ as $\hat *$,
and omit the
$\star$-product symbols; e.g. $\hA$ can be abstractly defined as the
algebra generated by $\hat a_i$ fulfilling the relations
obtained from (\ref{genstarrel}) replacing $a_i\to\hat a_i$
and dropping $\star$,
\be
\hat f^J(\hat a_1,\hat a_2,...)=0,\qquad \qquad\qquad J\in{\cal J},
                                     \label{genhatrel}
\ee
with $*$-structure, if any,  defined by
$\hat a_i{}^{\hat *}=\tau^h_j\big[S(\beta)\big]K^j_i a_h$, where
$K^j_i\!\in\!\b{C}$ are defined by $a_i^*=K^j_ia_j$ ($K^j_i=\delta^j_i$
if $a_i$ are real). On $\hA$ the action $\trc$ fulfills (\ref{defleibniz})$_3$.

\subsection{Deforming maps}
\label{defmaps}

Here we describe how a $\hH$-module ($*$)-algebra $\A_\star$
can be realized within $\A[[\lambda]]$ (i.e. how its elements
can be realized as power series in $\lambda$ with coefficients in $\A$)
through a socalled {\it deforming map}, provided
the left $H$-module ($*$)-algebra $\A$
admits a ($*$)-algebra map $\sigma:H\mapsto\A$ such
that the $H$-action on $\A$ can be
expressed in the (cocommutative) left ``adjoint-like'' form
\be
g\trc a =
\sum_I \sigma\left(g^I_{(1)}\right) a \,\sigma\left(S g^I_{(2)}\right).
\label{gag}
\ee
This will hold of course also for the corresponding
linear extensions $\sigma\!:\! \hH\!=\! H[[\lambda]]\rightarrow
\A[[\lambda]]$ and
 $\trc :H[[\lambda]]\times\A[[\lambda]]\mapsto\A[[\lambda]]$, and
suggests a way to make $\A[[\lambda]]$ into a $\hH$-module $*$-algebra
by defining the corresponding action $\tr$ in the (noncocommutative)
``adjoint-like'' form:
\be
g\,\tr\, a := \sum_I  \sigma\left(g^I _{(\hat 1)}\right) a \,
\sigma\left(\hat S g^I _{(\hat 2)}\right).
\label{gog}
\ee
In the case $\A\!=\!H$ then $\sigma\!=\!\id$, (\ref{gag}) is the adjoint action
of $H$, and the action defined by
(\ref{gog}) makes $H[[\lambda]]$ into a $\hH$-module $*$-algebra - see the end
of this subsection.
It is easy to check that the {\it deforming map} \cite{Fio96}
$D_{\f}^{\sigma}:a\!\in\!V(\A)[[\lambda]]\mapsto \check
a\!\in\!V(\A)[[\lambda]]$ defined by
\be
\check a\equiv
D_{\f}^{\sigma}(a):=\sum_I \sigma\!\left(\Fu_I \right)\,
a\:\sigma\!\left[S \left(\Fd_I\right) \beta^{-1}\right]
\label{defDf}
\ee
has the interesting property of intertwining between $\trc,\tr$
\be
g\,\tr[D_{\f}^{\sigma}(a)]=D_{\f}^{\sigma}\left(g\trc a\right).
                                       \label{intertw}
\ee
An immediate consequence is that if $\M\subseteq V(\A)$ is a $H$-$*$-submodule,
then $D_{\f}^{\sigma}(\M)$ is  a $\hH$-$*$-submodule.
An alternative expression for $D_{\f}^{\sigma}$ is (cf. with
\cite{GurMaj94,GroMadSte02,AscDimMeyWes06})
\be
\check a\equiv D_{\f}^{\sigma}(a)=\sum_I \left(\bF^{(1)}_I \trc a\right)\,
\sigma\left(\bF^{(2)}_I \right).           \label{altdefDf}
\ee
Moreover
\be
[D_{\f}^{\sigma}(a)]^*=D_{\f}^{\sigma}\big[a^{*_\star}\big]\label{*inter}
\ee
(we emphasize: at the lhs is the {\it undeformed} $*$), implying [with the help of (\ref{star'})]
$$
\big(g\tr \check a)^*=[\hat S(g)]^*\tr(\check a)^*.
$$
Finally, $D_{\f}^{\sigma}$ satisfies also
\be
D_{\f}^{\sigma}(a\star a')=D_{\f}^{\sigma}(a)D_{\f}^{\sigma}(a')\label{prodinter}
\ee
[eq. (\ref{altdefDf}), (\ref{*inter}), (\ref{prodinter}) are proved in the appendix]. Summing up, {\bf the deforming map  is a $\hH$-module $*$-algebra isomorphism
$D_{\f}^{\sigma}:\A_\star\leftrightarrow\A[[\lambda]]$}, or, changing notation,
$\widehat{D}_{\f}^{\sigma}:\hA\leftrightarrow\A[[\lambda]]$.

If $\A^s\subseteq\A$ is a $H$-module ($*$-)subalgebra
then $\check{\A}^s:=D_{\f}^{\sigma}(\A^s)\subseteq \A[[\lambda]]$ is
a $\hH$-module ($*$-)subalgebra isomorphic to $\A^s_{\star}$,
with $\hH$-module $*$-algebra isomorphism
$D_{\f}^{\sigma}:\A^s_{\star}\mapsto \check{\A}^s$.
Note also that, by (\ref{twistcond}), (\ref{altdefDf}),
$\check a=a$ if $a$ is $H_s$-invariant:
\be
g\trc a=\epsilon(g) a\quad \forall g\in H_s
\qquad\qquad \Rightarrow\qquad \qquad D_{\f}^{\sigma}(a)=a.\qquad
                                               \label{trivDf}
\ee
From (\ref{defDf}) the inverse of $D_{\f}^{\sigma}$ is readily seen to be
\be
\big(D_{\f}^{\sigma}\big)^{-1}(a):=\sum_I \sigma\!\left(\bF^{(1)}_I \right)\,
a\:\sigma\!\left[\beta S \left(\bF^{(2)}_I\right)\right].  \label{defDfinv}
\ee

If  $\A,\B$ are $H$-module $*$-algebras admitting $*$-algebra maps
$\sigma_{\scriptscriptstyle \A}:H\mapsto\A$, $\sigma_{\scriptscriptstyle
\B}:H\mapsto\B$ fulfilling (\ref{gag}), then
$$
\sigma_{\scriptscriptstyle \A\ot\B}:=(\sigma_{\scriptscriptstyle \A}\ot\sigma_{\scriptscriptstyle \B})\circ\Delta,
\qquad \qquad
\hat\sigma_{\scriptscriptstyle \A\ot\B}:= (\sigma_{\scriptscriptstyle \A}\ot\sigma_{\scriptscriptstyle \B}) \circ \hat\Delta
$$
define $*$-algebra maps $\sigma_{\scriptscriptstyle \A\ot\B}:H\mapsto\A\ot\B$,
$\hat\sigma_{\scriptscriptstyle \A\ot\B}:\hH\mapsto (\A\ot\B)[[\lambda]]$.
The former  fulfills the analog of (\ref{gag}), whereas replacing in
(\ref{gog}) $\sigma$ by the latter and $a$ by
$c\!\in\!(\A\ot\B)[[\lambda]]$,
$$
g \tr c  := \sum_I \hat \sigma_{\scriptscriptstyle \A\ot\B}
\!\left(g^I _{(\hat 1)}\right)\: c\, \: \hat \sigma_{\scriptscriptstyle \A\ot\B}\!\left[\hat S
\left(g^I _{(\hat 2)}\right)\right],
$$
defines the action $\tr$
making $(\A\ot\B)[[\lambda]]$ into a $\hH$-module ($*$-)algebra;
incidentally, note that
$g\tr(a\ot b)\neq \sum_I \big( g^I _{(\hat 1)}\trc a\big)\ot
\big( g^I _{(\hat 2)}\trc b\big)$.
In the appendix we prove that  the corresponding deforming map
$D_{\f}^{\sigma_{\scriptscriptstyle \A\ot\B}}:(\A\ot\B)_\star\mapsto (\A\ot\B)[[\lambda]]$ is defined by
\be
 D_{\f}^{\sigma_{\scriptscriptstyle \A\ot\B}}(c):= \F\!_\sigma\sum_I
\left(\bF^{(1)}_I \trc c\right)\, \sigma_{\scriptscriptstyle
\A\ot\B}\!\left(\bF^{(2)}_I \right)\F\!_{\sigma}{}^{-1},\qquad\qquad
\F\!_\sigma\!:=\!(\sigma_{\scriptscriptstyle
\A}\ot\sigma_{\scriptscriptstyle \B}) (\F).   \label{defDfsAB}
\ee
Using $(\id\!\ot\!\hat\Delta)(\bR)\!=\!\bR_{12}\bR_{13}$ and
(\ref{gog}) and setting $\R_\sigma\!=\!(\sigma_{\scriptscriptstyle
\A}\ot\sigma_{\scriptscriptstyle \B})(\R)$,  we also prove that
\be
\check \alpha_1 =
\check \alpha\ot\1,\qquad \quad \check b_2
=\bR_\sigma(\1\ot \check b)\R_\sigma=
\sum_I\sigma_{\scriptscriptstyle \A}\! \left(\bR^{(1)}_I\right)\ot
\bR^{(2)}_I\tr \check b \label{utile}
\ee
(here $\bR\!:=\!\R^{-1}\!\equiv\!\R_{21}$).
If $\A$ is generated by a set of $H$-equivariant $\{a_i\}_{i\in{\cal
I}}$ fulfilling (\ref{genrel}) we find that the $\check
a_i:=D_{\f}^{\sigma}(a_i)$, which  make up an alternative set of
generators of $\A[[\lambda]]$, in fact span a $\hH$-submodule and
close the polynomial relations (\ref{genhatrel}), so they provide an
explicit realization of $\hA\sim \A_\star$ within $\A[[\lambda]]$.
Therefore $D_{\f}^{\sigma}$ can be seen as a change from a set of
$H$-equivariant to a set of $\hH$-equivariant generators of
$\A[[\lambda]]$.  Applying
$D_{\f}^{\sigma}$ to (\ref{defhat}) [with the constraints
(\ref{genrel}), (\ref{genstarrel})] one finds
\be
D_{\f}^{\sigma}\left[f(a_1,a_2,...)\right]= \hat f(\check a_1,\check
a_2,...)].              \label{Df-hat}
\ee
Similarly, defining
$$
\check a_{i1} :=D_{\f}^{\sigma_{\scriptscriptstyle \A\ot\B}}(a_{i1})
\qquad \qquad \check b_{i'2}:=
D_{\f}^{\sigma_{\scriptscriptstyle \A\ot\B}}(b_{i'2})
$$
 we find that
 $\{\check a_{i1},\check b_{i'2}\}_{(i,i')\!\in\!{\cal I}\!\times\!{\cal I}'}$ is
an alternative set of generators of $(\A\ot\B)[[\lambda]]$.
The $\{\check a_{i1}\}_{i\!\in\!{\cal I}}$
 fulfill  (\ref{genhatrel}) and generate a $\hH$-module ($*$-)subalgebra
$\check{\A_1}:=D_{\f}^{\sigma_{\scriptscriptstyle \A\ot\B}}(\A_1)$,
isomorphic to $\A_\star$ as a $\hH$-module ($*$-)algebra.
Similarly, the  $\{\check b_{i'2}\}_{i'\!\in\!{\cal I}'}$ fulfill
the analogous relations and generate a $\hH$-module ($*$-)subalgebra
$\check{\B_2}:=D_{\f}^{\sigma_{\scriptscriptstyle \A\ot\B}}(\B_2)$
isomorphic to $\B_\star$.
Moreover they fulfill the analog of (\ref{braid'})$_2$.

Given a $H$-($*$-)module $\M$,
$(\bF \!\trc^{\ot 2}\!\M\ot\M[[\lambda]])_{\pm}=(\M\ot\M)_{\pm}[[\lambda]]$ as vector spaces; both are also $\hH$-($*$-)submodules of
$\M\ot\M[[\lambda]]$  [with involution (\ref{star'})].
Given a $H$-module ($*$-)algebra $\A$, since $(\A\ot\A)_+$ is a $H$-module
($*$-)subalgebra of $\A\ot\A$, the {\it twisted completely symmetric tensor
product algebra} $(\A\ot\A)_{+\star}$ is a $\hH$-module
($*$-)subalgebra of $(\A\ot\A)_{\star}$. It is easy to see that
for any $b,c\in\A[[\lambda]]$
\be
\ba{l} b_1\!\star\! c_2 +
b_2\!\star\! c_1 =\sum_I \!\left[\bF^{(1)}_I\!\trc\! b\,\ot\,\bF^{(2)}_I\!\trc
c +\bF^{(2)}_I\!\trc\! c \,\ot\, \bF^{(1)}_I\!\trc\!
b\right]\in\!V[(\A\ot\A)_{+}][[\lambda]], \label{bcsym}
\ea
\ee
and actually the underlying
vector space $V[(\A\ot\A)_{+\star}]=V[(\A\ot\A)_{+}][[\lambda]]$ is the
linear span of elements of this form. If $\A$ can be defined in terms of
generators $a_i$ and polynomial relations, we
obtain a basis of $V[(\A\ot\A)_{+\star}]$
letting $b,c$ run over a basis $\{\hat P_J\}_{J\!\in\! {\cal J}}$
of $V(\A_\star)=V(\A)[[\lambda]]$
consisting of $\star$-polynomials $\hat P_J(a_1\star,a_2\star,...)$.
If a $*$-algebra map $\sigma:H\mapsto\A$ of the type (\ref{gag}) exists,
applying $D_{\f}^{\sigma_{\scriptscriptstyle \A\ot\A}}$ to
(\ref{bcsym}) we find $\check b_1\check c_2+\check b_2\check c_1$, which
gives the realization of the elements of $(\A\ot\A)_{+\star}$
within $(\A\ot\A)[[\lambda]]$ in terms of polynomials in the $\check
a_{i1},\check a_{i2}$, since $\check b_1=\hat P_J(\check a_{11},\check
a_{21},...)$, $\check b_2=\hat P_J(\check a_{12},\check
a_{22},...)$, etc.
The generalization to $n$-fold tensor products is straightforward.

\medskip
We conclude with some link to the literature.
As said, a first example of a deforming map is on $\A=H$ itself with $\sigma=\id$  (we shall abbreviate $D_{\f}\!\equiv\!D_{\f}^{\mbox{\footnotesize id}} $);
this was introduced  in \cite{GurMaj94}. In
\cite{AscDimMeyWes06} it has been applied to $H=U\Xi$, where
$\Xi$ stands for the Lie algebra of
infinitesimal diffeomorphisms of a manifold $X$.
By (\ref{twistcond}), (\ref{*inter}), (\ref{prodinter}),  $D_{\f}$
is an Hopf $*$-algebra isomorphism between
$(\hH,*,\hat\Delta,\epsilon,\hat S,\R)$ and
$(H_\star,*_\star,\Delta_\star,\epsilon,S_\star,\R_\star)$
if one defines, as in \cite{AscDimMeyWes06},
$$
\Delta_\star:=(D_{\f}^{-1}\!\ot\! D_{\f}^{-1})\!\circ\! \hat \Delta \!\circ\! D_{\f},\qquad\quad S_\star:=D_{\f}^{-1}\!\circ\! \hat S \!\circ\! D_{\f},\qquad
\R_\star=(D_{\f}^{-1}\!\ot\! D_{\f}^{-1})(\R).
$$
So the two are essentially the same Hopf $*$-algebra; $D_{\f}^{-1}$
can be seen just as a change of generators. $*$-modules and module
$*$-algebras of the former are also of the latter, if one defines
$g\trc_\star a\!:=\! D_{\f}(g)\trc a$ for any
$g\!\in\!V(\hH)\!=\!V(H_\star)$\footnote{In fact,
$g\trc_\star (g'\trc_\star a)\!=\! D_{\f}(g)\trc\big[
D_{\f}(g')\trc a\big]\!=\! D_{\f}(g)D_{\f}(g')\trc a
\!=\! D_{\f}(g \star g')\trc a\!=\! (g \star g')\trc_\star a$, as
required.}. If both $g,h\!\in\!\g$ then also
$g\trc_\star h\!\in\!\g$; this defines a ``$\star$-Lie bracket''
$[g,h]_\star$ \cite{AscDimMeyWes06} making $\g$ a {\it Lie
$\star$-algebra} (in the parlance of \cite{AscDimMeyWes06}), a {\it generalized}
(in the parlance of \cite{Gur,LyuSud98}), or
a  {\it quantum Lie algebra} (in the parlance of many authors, e.g.
\cite{Ber90,SchWatZum93}, based on the results of \cite{Wor89}). We point out an
important difference w.r.t. \cite{AscDimMeyWes06}. There the twist
was assumed to fulfill $\F^{*\ot*}\!=\!(S\!\ot\!S)(\F_{21})$, rather
than $\F^{*\ot*}\!=\!\F^{-1}$; then the $*$-structures of the two
isomorphic deformed Hopf algebras had to be taken respectively as
$*_{\f},*$, where $g^{*_{\f}}\!:=\! \beta g^*\beta^{-1}$, rather
than $*,*_\star$ [there the two Hopf $*$-algebras were denoted as
$(U\g\!^{\f}\!,\cdot,*_{\f},\Delta^{\f},\epsilon,S^{\f},\R)$ and
$(U\g_\star,\star,*,\Delta_\star,\epsilon,S_\star,\R_\star)$].

In \cite{Fio96} we introduced suitable $\sigma$ and $D_{\f}^{\sigma}$
for general $H$-covariant Heisenberg or Clifford algebras,
what we shall recall in subsection \ref{Hei-Cli}. In next subsection
we introduce $\sigma$ and $D_{\f}^{\sigma}$ on the algebra of
differential operators on  $\b{R}^m$.

\subsection{Twisting functional, differential, integral calculi on $\b{R}^m$}
\label{twistdifcal}

Here we apply the $\star$-deformation procedure to the
algebras of functions and of differential operators on $\b{R}^m$,
as well as to integration over $\b{R}^m$.

 $X=\b{R}^m$ is invariant under the Lie group
$IGL(m)$ of real inhomogeneous general linear
transformations. We call $igl(m)$ the Lie algebra of $IGL(m)$.
Any set of cartesian coordinates $x^1,...,x^m$ of
$\b{R}^m$ together with the unit $\1$ spans a $Uigl(m)$-$*$-module
$\M$, with $g\trc\1=\varepsilon(g)\1$;
the derivatives $\partial_1,...,\partial_m$ span the contragredient one
$\M'$.
We shall denote as $\D_p$ the Heisenberg algebra on
$\b{R}^m$ (sometimes this is called Weyl algebra), i.e. the
$*$-algebra with generators the unit $\1$ and
$x^1,...,x^m,\partial_1,...,\partial_m$, relations given
by trivial commutators but the ones $[\partial_j,x^i]=\1\delta^i_j$,
$*$-structure given by $x^i{}^*=x^i$, $\partial_j{}^*=-\partial_j$.
$\1,x^1,...,x^m$ generate the abelian subalgebra $\X_p$
of polynomials in $x^i$. $\D_p$ is a $Uigl(m)$-module $*$-algebra with
a $Uigl(m)$-module $*$-subalgebra $\X_p$. Setting for brevity
$x^{m\!+\!1}\!\equiv\!\1$, on the generators the action can be expressed in
the form\footnote{Relation (\ref{deftau})$_1$ is the definition of
the representation $\tau$ of $Uigl(m)$ on $\M$; (\ref{deftau})$_3$ follows from
$g\trc x^{m\!+\!1}\!=\!\tau_{m\!+\!1}^k(g)x^k
=\varepsilon(g)x^{m\!+\!1}$. From
$\varepsilon(g)\!=\!\sum_I S\left(g^I_{(1)}\right)g^I_{(2)}$ it follows
$\varepsilon(g)\delta^i_h=\sum_I
\tau_k^i\left[S\left(g^I_{(1)}\right)\right]\tau^k_h\left(g^I_{(2)}\right)
$, and by (\ref{deftau})$_3$ in the sum the term with $k\!=\!m\!+\!1$ vanishes.
The latter relation implies that (\ref{deftau})$_2$ is the
transformation law needed to preserve
the relations $[\partial_j,x^h]=\1\delta^h_j$,
beside $[\partial_j,\partial_k]=0$ and  $\partial_j{}^*=-\partial_j$.}
\be
g\trc x^h=\tau^k_h(g)x^k, \qquad
g\trc\partial_i=\tau_j^i[S(g)]\partial_j, \qquad
\mbox{with }\tau_{m\!+\!1}^k(g)=\varepsilon(g)\delta_{m\!+\!1}^k,
                                                \label{deftau}
\ee
where repeated indices are summed over $1,2,...,m\!+\!1$.

For any Lie subalgebra $\g\subseteq igl(m)$
$\D_p$ is a $U\g$-module $*$-algebra with
a $U\g$-module $*$-subalgebra $\X_p$.
Taking $\A=\D_p$, $H=U\g$ and choosing a twist
$\F$, we can define $\hH$-module
$*$-algebras $\D_{p\star},\X_{p\star},...$ with
$\D_{p\star}\supset\X_{p\star}$,
as well as the map $\wedge:\D_{p}[[\lambda]]\mapsto\D_{p\star}$,
through the $\star$-deformation procedure
described in the previous subsections. In the appendix we show that
$\widehat{\D_{p}}\sim\D_{p\star}$ amounts to the algebra generated
by the unit $\1\!\equiv\!\hat x^{m\!+\!1}$
and $\hat x^1,...,\hat x^m,\hat \partial_1',...,\hat \partial_m'$,
 fulfilling
\be\ba{lll}
g\trc \hat x^h=\tau^k_h(g)\hat x^k, \qquad\qquad
&g\trc\hat \partial_a'=\tau_b^a[\hat S(g)]\hat \partial_b', &\\[8pt]
 \hat x^a{}^{\hat *} =\tau^k_a \left[ S(\beta)\right]\hat x^k,\qquad
\qquad &\hat \partial_a'{}^{\hat *} =-
\tau^a_k\left(\beta^{-1}\right)\hat \partial_k',&\\[8pt]
 \hat x^a\hat x^b= R^{kh}_{ab}\hat x^h\hat x^k,\qquad
&\hat \partial_a'\hat \partial_b'= R^{ab}_{kh}\hat \partial_h'\hat \partial_k',\qquad &\hat \partial_a'\hat x^b=\1\delta^b_a+
R_{bk}^{ha}\hat x^h\hat \partial_k'.
\ea                            \label{hatD}
\ee
where $R^{ab}_{hk}=(\tau^{a}_{h}\ot \tau^{b}_{k})(\R)$,  and
$R^{ab}_{k(m\!+\!1)}=\delta^a_k\delta^b_{m\!+\!1}=R^{ba}_{(m\!+\!1)k}$.
One can easily check that
\be
\sigma(g):=(g\trc x^h)\partial_h=\tau^k_h(g)x^k\partial_h,
\qquad\qquad g\in \g       \label{Msigma}
\ee
determines a map $\sigma:H\mapsto \D_p$ of the type described in
subsection \ref{defmaps},
so we can define also $D^\sigma_{\f}$ and therefore also
$\check  x^i,\check \partial_i$.
For $f\in\X_p$  (\ref{Df-hat})
becomes $D^\sigma_{\f}[f(x)]=\hat f(\check x)$
where $\check x^h:=D^\sigma_{\f}(x^h)$; applying the
pseudodifferential operator $D^\sigma_{\f}(f)$ to the constant
function $\1$ one finds [with the help of (\ref{altdefDf}),
$g\!\trc \!\1=\varepsilon(g)\1$ and
(\ref{twistcond})$_2$]
\be
\left[\wedge^{-1}\hat f\right](x)\equiv f(x)=D^\sigma_{\f}(f)\trc  \1=
\hat f(\check x)\trc  \1.            \label{inversehat}
\ee
Thus (\ref{inversehat}) gives an alternative way to compute
the restriction $\wedge^{-1}:\widehat{\X_{p}}\mapsto\X_p[[\lambda]]$.
Using the methods of the previous subsections one can similarly
$\star$-deform $\X_p^{\ot n}$, $\D_p^{\ot n}$ and find the corresponding
$\hH$-module $*$-algebra isomorphism
$\hat D^{\sigma_n}_{\f}:
\widehat{\D_p^{\ot n}}\mapsto\D_p^{\ot n}[[\lambda]]$ (with
$\sigma_n=\sigma^{\ot n}\circ\Delta^{(n)} $).
The existence of the latter is consistent with the well-known fact
\cite{Duc85} that all deformations of the Heisenberg algebra are ``trivial'',
i.e. can be reabsorbed into (formal) changes of generators.
Denoting by $x^{h}_1$, $x^{h}_2$,... the
generators $x^{h}\!\ot\!\1\!\ot...$, $\1\!\ot\!x^{h}\!\ot\!...$,...
of $\X_p^{\ot n}$ and by
$\partial_{x_1^a}=\partial/\partial x_1^a$,
$\partial_{x_2^a}=\partial/\partial x_2^a$,... the remaining generators
of $\D_p^{\ot n}$, the relations [generalizing (\ref{hatD})] which
characterize $\widehat{\D_p^{\ot n}}$ read
\be\ba{lll}
 \hat x_i^a{}^{\hat *} =\tau^k_a\! \left[ S(\beta)\right]\hat x_i^k,
\quad\: &\hat \partial_{x_i^a}'{}^{\hat *} =-
\tau^a_k\!\left(\beta^{-1}\right)\hat \partial_{x_i^k}',&\\[8pt]
 \hat x_i^a\hat x_j^b= R^{kh}_{ab}\hat x_j^h\hat x_i^k,\quad\:
&\hat \partial_{x_i^a}'\hat \partial_{x_j^b}'= R^{ab}_{kh}\hat \partial_{x_j^h}'\hat \partial_{x_i^k}',\quad\: &\hat \partial_{x_i^a}'
\hat x_j^b=\1\delta^b_a\delta^i_j\!+\!
R_{bk}^{ha}\hat x_j^h\hat \partial_{x_i^k}'.
\ea                            \label{hatDn}
\ee

Since working with $\X_p$ is not sufficient for Quantum Mechanics
(QM) and QFT purposes ($\X_p$ has e.g. no integrable functions) one has to extend
the $\star$-deformation to larger function spaces $\X$.
One could start with some subalgebra $\X$ of $C^{\infty}(\b{R}^m)$
closed under the action of derivatives and with the algebra
(and $\X$-bimodule) $\D$ of
smooth differential operators
consisting of polynomials in $\partial_1,...,\partial_m$
with (left, say) coefficients in $\X$;
its only nontrivial basic commutation relations are
$$
[\partial_h,f]=\partial_h\trc f\qquad \qquad\qquad
f\in \X.
$$
$\D$ is a $H$-module $*$-algebra, which we can take as our $\A$.
$\X$ is the $H$-module $*$-subalgebra
consisting of differential operators of order zero.
Clearly $\D\subset\En\!:=\!\mbox{End}(\X)$: each $D\in\D$ defines
an endomorphism $D:f\!\in\!\X\mapsto D\!\trc\! f\in\!\X$.
On $\X$  the $\g$-action is given by
$g\trc f=(g\trc x^h)(\partial_h\trc f)$.
Applying the $\star$-deformation procedure
to  $\A=\D$ one obtains $\hH$-module
$*$-algebras $\D_{\star},\X_{\star}$ with
$\D_{\star}\supset\X_{\star}$; these are clearly extensions of
$\D_{p\star},\X_{p\star}$ if $\X\supset\X_p$.
The same can be done with tensor powers $\X^{\ot n}$, $\D^{\ot n}$.

Riemann integral is defined using the volume form
$d\nu\!=\!d^m x$ associated to the Euclidean metric of $\b{R}^m$; both
$d\nu$ and   $\int_X\!\!\! d\nu({\rm x})$ are
invariant under the isometry group $G\!=\!ISO(m)\!\subset\! IGL(m)$,
or equivalently under $H\!=\!U\g$:
\be
\int_X\!\!\!
d\nu (g\trc f)= \epsilon(g)\!\!\int_X\!\!\! d\nu
f\qquad\Rightarrow \qquad \int_X\!\!\! d\nu f (g\trc h)=\!
\int_X\!\!\! d\nu [S(g)\trc f]h. \label{InvInt}
\ee
Fixed a
$\F\!\in\!(H\ot H)[[\lambda]]$, this implies the
$\hH$-invariance of $\int_X\!\!\! d\nu$, as well as
\be
\ba{l}
\displaystyle\int_X\!\!\! d\nu\, h\star h'=
\int_X\!\!\! d\nu\, h (\beta^{-1}\trc h')
=\int_X\!\!\! d\nu\, [S(\beta^{-1})\trc h] h'\\[10pt]
\displaystyle\int_X\!\!\! d\nu\, h^{*_\star}\star h'=
\int_X\!\!\! d\nu\,  h^* h'
\ea\qquad
\label{Intstarprop}
\ee
for the corresponding $\star$-product.  Hence
\be
\ba{l}
\overline{\displaystyle\int_X\!\!\! d\nu f}=
\displaystyle\int_X\!\!\! d\nu \, f^{*_\star},\qquad\qquad
\displaystyle\int_X\!\!\! d\nu \, h\star h'
=\displaystyle\int_X\!\!\! d\nu \, (w\trc h')\star h,\\[8pt]
\displaystyle\int_X\!\!\! d\nu\, h^{*_\star}\!\star\! h\ge 0,\qquad\qquad
\mbox{and }= 0\qquad\mbox{iff }h\!\equiv\! 0,\ea
\qquad \label{starintprop}
\ee
where $w\!:=\!S(\beta)\beta^{-1}$:
the Riemann integral fulfills  in general the {\it modified
trace property} (\ref{starintprop})$_2$ w.r.t. such $\star$-products. So
$w\neq \1$ is an obstruction for the star-product to be
strongly closed in the sense of Connes-Flato-Sterheimer
\cite{ConFlaSte}.
Of course, for the moment (\ref{InvInt}-\ref{starintprop}) make sense
(as formal power series in $\lambda$) if the derivatives of $f,h,h'$
of all orders are well-defined and integrable, e.g. if
$f,h,h'\!\in\! \X={\cal S}(\b{R}^m)$ (the Schwarz space).
[But, again, $\X={\cal S}(\b{R}^m)$
is not large enough for QM and QFT purposes].
Analogous relations hold for integration over $n$ independent
${\rm x}$-variables. In $(\X^{\ot n})_\star$
we define integration over the $j$-th set of coordinates
${\rm x}_j$ ($j=1,...,n$) of $\X^{\ot n}$ in the natural way, i.e. as
in the first equality of the formula
\be
\int_{X}\!\!\!\! d\nu_j [f({\rm x}_j) \star \omega]:=
\left[\!\int_{X}\!\!\!\!  d\nu_j f({\rm x}_j)\!\right] \star \omega
=\omega\star\!\int_{X}\!\!\!\! d\nu_j f({\rm x}_j),
\label{intstarcom}
\ee
for any $\omega$ depending only on the ${\rm x}_k$  ($k\neq j$); 
here $d\nu_jr\!:=r\!d^mx_j$. 
The second equality holds (with and without $\star$) trivially,
because the integral belongs to $\b{C}[[\lambda]]$. It follows
(see the appendix)
\be
\int_{X}\!\!\!\! d\nu_j \omega\star f({\rm x}_j)=
\omega\star\!\int_{X}\!\!\!\!  d\nu_j f({\rm x}_j),\label{intstarcom'}
\ee
namely integration $\int_{X}\! d\nu_j$ `$\star$-commutes' with
(i.e. may be moved beyond) $\omega$. More generally,
(\ref{intstarcom}-\ref{intstarcom'}) hold in
any $\hH$-module $*$-algebra $\Phi_\star\!\supseteq\!(\X^{\ot n})_\star$
(e.g. the field $\star$-algebras of the next sections)
if $\omega\!\in\!\Phi_\star$, does not depend on
${\rm x_j},\partial_{{\rm x_j}}$.

If we could define, in terms of generators
and relations only, $\widehat{\X}, \widehat{\D}$
isomorphic to $\X_\star,\D_\star$ for $\X,\D$ large enough, and
we could extend $\wedge$ to such $\X,\D$
\footnote{It is not known how to do this for general $\star$-deformation.
For entire functions $f$  one idea could be
 to express their power series expansion in
$x^1,...,x^m$ as formal power
series $\hat f$ in $x^1\star,...,x^m\star$, but even then
this seems in general a non-computable operation; the coefficients of $n$-th
degree $\star$-monomials may receive contributions from the
coefficients all monomials of degree $n'\ge n$.} then we could define a
$\hH$-invariant ``integration over $\hat X$''
$\int_{\hat X}\!\! d\hat\nu(\hat {\rm x})$ by the requirement
\be
\int_{\hat X}\!\!\! d\hat\nu \, \hat f(\hat {\rm x})
=\int_X\!\!\! d\nu\,  f({\rm x}).          \label{hatInt}
\ee
Then (\ref{InvInt}),
(\ref{starintprop}), the definition of $\int_{\hat X}\! d\hat\nu_j$
 and (\ref{intstarcom'})  would take the form
\be
\ba{l}
\displaystyle\int_{\hat X}\!\!\! d\hat\nu\,  [g\trc\hat f(\hat {\rm x})]
=\epsilon(g)\displaystyle\int_{\hat X}\!\!\! d\hat\nu \, \hat f(\hat {\rm x}),
\qquad\qquad
\displaystyle\overline{\int_{\hat X}\!\!\! d\hat\nu\,  \hat f(\hat {\rm x})}=
\displaystyle\int_{\hat X}\!\!\! d\hat\nu\, [\hat f(\hat {\rm x})]^{\hat *},\\[10pt]
\displaystyle\int_{\hat X}\!\!\!\! d\hat\nu\,  \hat h(\hat {\rm x})
\hat h'(\hat {\rm x})=\!
\displaystyle\int_{\hat X}\!\!\!\! d\hat\nu\,  [w\trc\!\hat h'(\hat {\rm x})]
\hat h(\hat {\rm x}), \qquad
\displaystyle\int_{\hat X}\!\!\!\! d\hat\nu\,  [\hat h(\hat {\rm x})]^{\hat *}
\hat h(\hat {\rm x})\ge 0,\quad
= 0\:\:\mbox{iff }\hat h\!\equiv\! 0,\\[10pt]
\displaystyle\int_{\hat X}\!\!\! d\hat\nu_j\,[\hat f(\hat {\rm x_j})
\hat \omega]:= \left[\displaystyle\int_{\hat X}\!\!\! d\hat\nu_j\,\hat f(\hat {\rm x_j})\right]\hat \omega
, \qquad \displaystyle\int_{\hat X}\!\!\! d\hat\nu_j\,
\hat \omega\hat f(\hat {\rm x_j})=\hat \omega
\displaystyle\int_{\hat X}\!\!\! d\hat\nu_j\,\hat f(\hat {\rm x_j}),
\ea \label{hatintprop}
\ee
(with any
$\hat \omega$ not depending on $\hat \partial_{\rm x_j},\hat {\rm x_j}$).
If we forget (\ref{hatInt}), requiring directly
(\ref{hatintprop})
and $\b{C}[[\lambda]]$-linearity should be sufficient to determine
up to a normalization constant the integration functional
$\int_{\hat X}\!\! d\hat\nu(\hat {\rm x}):
\hat f\in\widehat{\X}\mapsto \b{C}[[\lambda]]$.
As a matter of fact, such an approach has been used quite
successfully even on $\b{R}^m_q$ \cite{Fio93,Ste96},
 the deformation  \cite{FadResTak89} of
$\b{R}^m$ covariant under the
quantum group (i.e. proper {\it quasitriangular} Hopf algebra) $U_qso(m)$.

The formal procedure just sketched can be extended to a class of
$\g$-symmetric  (possibly Riemannian) algebraic variety
$X\!\subset\!\b{R}^m$, where $\g$ is a Lie subalgebra of $igl(m)$.
This will be developed elsewhere.

The relations/definitions presented in this subsection are formal,
both because of formal $\lambda$-power series and of
unspecified $\widehat{\X}$.
In the next subsection we present the specific examples of
$\star$-deformations induced by Moyal twists and
sketch why they should allow to go beyond the
formal level (this will be studied more in detail elsewhere).

\subsubsection{Application to Moyal deformations}
\label{moyal}

In the last fifteen years Moyal spaces\footnote{In the literature these
are also called {\it canonical}, or more often denoted
by some combination of the
names of Weyl, Wigner, Gr\"onewold, Moyal. This is due to the
relation between canonical commutation relations and the
$\star$-product (or twisted product) of Weyl and Von Neumann, which
in turn was used by Wigner to introduce the Wigner transform;
Wigner's work led Moyal to define the socalled Moyal bracket
$[f\stackrel{\star}{,} g]\!=\! f \star g \!-\! g \star f$; the
$\star$-product in position space [in the form of the asymptotic
expansion of (\ref{explstarprod}) with $x_i\!=\!x_j\!\equiv\! x$] first
appeared in a paper by Gr\"onewold.} have been the subject of intensive
investigations for their  potential physical relevance. Because of
the simplicity of their twist $\F$ they are also particularly
pedagogical models; moreover, 
the associated $\star$-products admit (section \ref{moyal})
non-perturbative (in $\lambda$) definitions
in terms of Fourier transforms. Here we give a detailed description of them.

One chooses a real Lie subalgebra $\g$
of $igl(m)$  containing all the generators $iP_h$ ($h=1,...,m$) of
translations, and as a twist
 \be \ba{l}
\F\equiv\sum_I\F^{(1)}_I\ot\F^{(2)}_I:= \mbox{exp}\left(\frac
i2\lambda\vartheta^{hk}P_h\ot
P_k\right)=\mbox{exp}\left(\frac i2\theta^{hk}P_{h}\ot
P_{k}\right). \label{twist} \ea
\ee
Here  $\vartheta^{hk}$ is
a fixed real antisymmetric matrix, $\sum_I$ includes the $\lambda$-power
series arising from the expansion of the
exponential, and as conventional we absorb the deformation parameter
$\lambda$ in $\theta^{hk}\!:=\!\lambda\vartheta^{hk}$.
Straightforward computations  give $\beta\!:=\!\sum_I
\F^{(1)}_I S\left(\F^{(2)}_I\right)\!=\!\1$, whence $*_\star\!=\!*$, and
$\hat\Delta (P_h)=\Delta (P_h)$, so that  the Hopf $P$-subalgebra remains
undeformed and equivalent to the abelian translation group  $\b{R}^m$.

Since $iP_h$ acts on $\X_p$ as
 $\partial_h=\partial/\partial x^{h}$, it is $iP_h\trc x^k=\delta^k_h$ and
$\tau^l_k(iP_h)=\delta^k_h\delta^l_{m\!+\!1}$.
The $\star$-product
 (\ref{starprod})+(\ref{twist}) on $\X_p^{\ot n}$ becomes
\be
\ba{l}
a(x_1\!,\!...,\!x_n)\star b(x_1\!,\!...,\!x_n)\!=\!\left. \exp\!\left[\frac
i2\!\left(\!\sum_i\!\partial_{x_i}\!\right)\!\theta\!
\left(\!\sum_j\!\partial_{y_j}\!\!\right)\!\right]\!\trc
a(x_1\!,\!...,\!x_n) b(y_1\!,\!...,\!y_n)\right\vert_{y\!=\!x}
\ea
\label{explstarprod}
\ee
(we abbreviate $p\theta q\!:=\!p_{h}\theta^{hk}q_{k}$ for any m-vectors $p,q$),
in particular $x^{h}_i\!\star\! x^{k}_j\!=\!x^{h}_i x^{k}_j\!+\!i\theta^{hk}/2$,
whereas $a\star\partial_{x^h_j}=a\partial_{x^h_j}$,
$\partial_{x^h_j}\star a=\partial_{x^h_j}a$ for any $a\in\D_p^{\ot n}$
(the differential calculus is not deformed). Hence the
relations (\ref{hatDn}) for $\widehat{\D_p^{\ot n}}\sim(\D_p^{\ot n})_\star$ take the form
\be\ba{lll}
\hat x^{h}_i{}^{\hat *}=\hat x^{h}_i ,\qquad
\qquad &\hat \partial_{x_i^h}{}^{\hat *}=-\hat \partial_{x_i^h},&\\[8pt]
[\hat x^{h}_i,\hat x^{k}_j] =\1 i\theta^{hk},\qquad\qquad
&[\hat \partial_{x_i^h},\hat x^{k}_j]=\1\delta^k_h\delta_j^i,\qquad\qquad
&[\hat \partial_{x_i^h},\hat \partial_{x^{k}_j}]=0.
\ea                            \label{hatDmoyal}
\ee
Using (\ref{Msigma}), (\ref{defDfsAB}) one constructs the isomorphism
$\hat D^{ \sigma_n }_{\f}:
\widehat{\D_p^{\ot n}}\mapsto\D_p^{\ot n}[[\lambda]]$; one finds
\be
\ba{l}
\check x^h_i\!=\!\hat D^{\sigma_n }_{\f}(\hat x^h_i)
\!=\!x^h_i\!+\!i\theta^{hk}\left[\frac 12\partial_{x^k_i}+
\sum_{j\!>\!i}^n\partial_{x^k_j}\right],\qquad\quad
\check \partial_{x^h_i}\!=\!\hat D^{\sigma_n }_{\f}
(\hat \partial_{x^h_i})\!=\!\partial_{x^h_i}
\ea                                  \label{defmapmoyal}
\ee
where $\sigma_n=\sigma^{\ot n}\circ\Delta^{(n)} $,
in particular $\check x^h\!=\!x^h\!+\!i\theta^{hk}\partial_k/2$,
$\check \partial_h\!=\!\partial_h$ for $n\!=\!1$ (the latter is
a result appeared several times in the literature).
Formally, (\ref{Intstarprop})$_1$ and $\beta=\1$ imply also
\be
\int\!\!
d^mx  \, a\star b=\int\!\! d^mx \, a\,b=\int\!\! d^mx \, b \star a,
\label{intprop}
\ee
and similarly for multiple integrations.

For any subspace  $\X\!\subset\!C^{\infty}(\b{R}^m)$,
the $\lambda$-power series involved  in (\ref{explstarprod})
is termwise well-defined if $a,b\!\in\!\X^{\ot n}$ and reduces
to a finite sum if at least one out of $a,b$ is in
$\X_p^{\ot n}$ (a polynomial); the determination of the largest
$\X$ such that it has a positive radius
of convergence $r$ is a delicate issue,
about which little is known \cite{EstGraVar89} (but again such a $\X$
would be too small for QM and QFT purposes).
On the other hand, exponentials certainly belong to
such a $\X$, e.g.
$e^{ih\cdot x_i}\star e^{ik\cdot x_j}=
e^{i\left(h\cdot x_i+k\cdot x_j-\frac{h\theta k}2\right)}$
for all matrices $\theta=\lambda\vartheta$ ($r=\infty$), and their $\star$-product
is associative as a consequence of the cocycle condition (\ref{cocycle}).
 Therefore
if $a,b\!\in\!\X$ admit Fourier transforms $\tilde
a,\tilde b$ then \footnote{
(\ref{IntForm}) has the series (\ref{explstarprod}) as a formal power expansion;
see \cite{EstGraVar89} for the conditions under which the latter is in fact an asymptotic expansion.}
 \be
a(x_i)\star b(x_j)= \int\! d^mh\!\int\!  d^mk\:\,
e^{i\left(h\cdot x_i+k\cdot x_j-\frac{h\theta k}2\right)}\tilde a(h) \tilde b(k).
                                                       \label{IntForm}
\ee
As a matter of fact, (\ref{IntForm}) and
its generalization to $a,b$ depending on all the $x_i$ can be used as a {\it
more general definition} of a (associative) $\star$-product. If $n\!=\!1$
($i\!=\!j\!=\!1$) it is a well-defined function for $a,b\!\in\!
L^1(\b{R}^m)\cap \widetilde{L^1(\b{R}^m)}$; it makes sense also for
$b\!\in\!\X\!=\!{\cal S}(\b{R}^m)$  and $a\in \X'$
(the space of tempered distributions; $\tilde a$ is then the Fourier transform
in the distribution sense) as the (symbolic) integrand in the definition of
the distribution
$a_\star:b\!\in\!\X\mapsto\int\!\!d^mx  \,a\star b\in\b{C}$,
as the latter integral is well-defined
[and actually equal to $\int\!\!d^mx  \, a\, b$], but if $a\!\in\!\X_p$
or $a\!\in\!\X$ this integrand
$a\star b$ is also a true [actually $C^{\infty}(\b{R}^m)$] function;
and similarly with $a,b$ interchanged; and so on.
If $n\!=\!2$ and $i\!=\!1$, $j\!=\!2$, it makes sense even
for $a,b\in \X'$, as defining a new distribution
$a\ot_\star b$ on $c(x_1,x_2)\in\X\ot \X$,
showing $\X'\!\ot_\star\!\X'\!=\! \X'\!\ot\!\X'$\footnote{Actually,
for $i\!=\!j$
and some $a,b\in \X'$ it may
even happen that (\ref{IntForm}) is ill-defined for
$\theta^{\mu\nu}\!=\!0$, but well-defined \cite{GraVar88} (and thus
``regularized'') for $\theta^{hk}\!\neq\! 0$. For
instance, for $a(x)=\delta^m(x)=b(x)$ and invertible $\theta$ one
easily finds $a(x_i)\!\star\! b(x_j)=(\pi^m\det
\theta)^{-1}\exp[2ix_j\theta^{-1}x_i]$; in particular for $i\!=\!j$
the exponential becomes 1 by the antisymmetry of $\theta^{-1}$, and
one finds a diverging constant as $\det \theta\to 0$, cf.
\cite{GraVar88,EstGraVar89}. In \cite{GraVar88} the largest algebra
of distributions for which the $\star$-product is well-defined and
associative was determined.}. In other words, using Fourier
transforms the $\star$-products $a\star b$ become well-defined
in the same sense as the products $ab$; this holds also
for $a,b$ depending on all the $x_i$. So these $\star$-products are
enough to replace all the products used in ordinary QFT,
with results reducing to the commutative ones for
$\theta^{hk}\!=\!0$.


Morally, to extend $\widehat{\X_p^{\ot n}},\widehat{\D_p^{\ot n}}$ we could
consider the Weyl form of (\ref{hatDmoyal}),
\be
\hat U_i^p \hat U_j^q=\hat U_j^q\hat U_i^pe^{-ip\theta q},\qquad \hat V_i^y\hat U_j^q=\hat U_j^q\hat V_i^y
e^{iy\cdot q\,\delta_{ij}},\qquad  \hat V_i^y \hat V_j^z=\hat V_j^z\hat V_i^y
\ee
(with $p,q,y,z\in\b{R}^m$) obtained by setting
$\hat U_i^p\!:=\!e^{ip\cdot \hat x_i}$,
$ \hat V_i^y\!:=\!e^{y\cdot \hat \partial_{x_i}}$.
The  $\hat U_i^p,\hat V_i^y$ generate a $C^*$-algebra
isomorphic to the canonical one generated by
$U_i^p\!:=\!e^{ip\cdot  x_i}$,
$ V_i^y\!:=\!e^{y\cdot  \partial_{x_i}}$: the isomorphism is given by the
extension of (\ref{defmapmoyal}) to these exponentials:
\be
\hat D^{\sigma_n }_{\f}\big(
\hat V_i^y\big)\!=\!e^{y\cdot \check \partial_{x_i}}=V_i^y,\qquad\qquad
\hat D^{\sigma_n }_{\f}\big(\hat U_i^p\big)\!=\!e^{ip\cdot \check x_i}
=U_i^p V_i^{\theta p/2}\prod\limits_{j>i}V_j^{\theta p}
\ee
The spaces of functions on $\b{R}^m$ that one needs for QM and QFT
[space of test functions like $\X\!=\!{\cal S}(\b{R}^m)$, of square integrable functions
$\L^2\!\equiv\!\L^2(\b{R}^m)$, space of distributions $\X'$, etc.,
and their tensor powers] all admit
suitably generalized notions of Fourier transformation
(Fourier, Fourier-Plancherel, Fourier for distributions), and
the generic element $a$ of each of them
can be expressed in terms of an anti-Fourier transform
\be
a(x_1,...,x_n) =\int\!\!\! d^mq_1...\!\!\int\!\!\!
d^mq_n\:\, e^{iq_1\cdot x_1}\!...e^{iq_n\cdot x_n}\tilde
a(q_1\!,\!...,q_n),
\ee
where the symbols $\tilde a$ respectively belong to
$\widetilde{\X^{\ot n}}\!=\!\X^{\ot n}$,
$\widetilde{\L^2{}^{\ot n}}\!=\!\L^2{}^{\ot n}$,
$\widetilde{\X'{}^{\ot n}}\!=\!\X'{}^{\ot n}$.
We correspondingly define
$\widehat{\X^{\ot n}},\widehat{\L^2{}^{\ot n}},\widehat{\X'{}^{\ot n}}$
as the spaces of objects of the form
\be
\hat a(\hat x_1,...,\hat x_n)=\int\!\!\! d^mq_1...\!\!\int\!\!\!
d^mq_n\:\, e^{iq_1\cdot \hat x_1}\!...e^{iq_n\cdot \hat x_n}
\tilde a(q_1\!,\!...,q_n).
\ee
By formally applying (\ref{inversehat}) to
$\hat U^p(\hat x)\!=\!e^{ip\cdot \hat x}$
one finds
$$
\left[\wedge^{-1}\hat U^q\right](x)=
e^{iq\cdot \check x}\trc  \1=e^{iq\cdot (x\!+\!\frac i2\theta\partial)}\trc  \1
=e^{iq\cdot x}=U^q(x)
$$
(here $(\theta\partial)^h\!=\!\theta^{hk}\partial_k$),
or equivalently $\wedge\left(U^q \right)=\hat U^q $,
and, by iterated application of (\ref{hatprop}),
\be
\wedge^n\left( U_1^{q_1} ...U_n^{q_n} \right)=\hat U_1^{q_1} ...\hat U_n^{q_n}
e^{\frac i2\sum_{i< j} q_i\theta q_j}.
\ee
We can {\it define} the extensions of $\wedge^n$ as maps from
$\X^{\ot n},\L^2{}^{\ot n},\X'{}^{\ot n}$ respectively
to $\widehat{\X^{\ot n}},\widehat{\L^2{}^{\ot n}},\widehat{\X'{}^{\ot n}}$
 by linearity w.r.t. the ``generalized basis''
$\{\hat U_1^{q_1} ...\hat U_n^{q_n}\}$:
\be
\left[\wedge^n(a)\right] (\hat x_1,...,\hat x_n):=\int\!\!\! d^mq_1...\!\!\int\!\!\!
d^mq_n\:\, e^{iq_1\cdot \hat x_1}\!...e^{iq_n\cdot \hat x_n}e^{\frac i2\sum_{i< j} q_i\theta q_j}\tilde
a(q_1\!,\!...,q_n);                        \label{wedgen}
\ee
in the latter case this extends the linear map $\wedge^n\!:\!\X_p^{\ot
n}\!\mapsto\! \widehat{\X_p^{\ot n}}$ defined in terms of the polynomial equation
(\ref{defhat}). For $n\!=\!1$ (\ref{wedgen}) is nothing but the well-known Weyl transformation.
We can trivially extend $\wedge^n$ to the rest of $\D$ by
setting $\wedge^n\left(V_i^p\right)=\hat V_i^p$,
$\wedge^n\left(\partial_{x^h_i}\right)=\hat \partial_{x^h_i}$.

 Another peculiar feature of Moyal spaces is that in
$\widehat{\X^{\ot n}}$ there is essentially {\it only one} set of
noncommuting coordinates $(\hat X^h)$. Let $w_i\!\in\!\b{R}$ with
$\sum_{i=1}^n\!w_i\!=\!1$. An alternative set of real generators of
both $\X^{\ot n}$ and $(\X^{\ot n})_{\star}$ is:
\be \ba{l}
\xi^{h}_i\!:=\! x^{h}_{i\!+\!1}\!-\! x^{h}_i,\quad
i\!=\!1,...,n\!-\!1, \qquad\quad
 X^{h}\!:=\!\sum_{i=1}^nw_i x^{h}_i \quad
\ea                             \label{defXxi}
\ee
(one simplest choice is $X^h=x^h_1$). All $\xi^{h}_i$ are translation invariant
and therefore $\star$-commute with all the $x^{k}_j$, whereas $X^{h}$ are not
 and fulfill $[X^{h}\!\stackrel{\star},\! X^{k}]=\1
i\theta^{hk}$, whence
\be
[\hat \xi^{h}_i, \hat x^{k}_j]=0, \qquad\qquad\qquad[\hat X^h,\hat X^k]=\1
i\theta^{hk}.
\ee
Therefore the $\hat \xi^{h}_i$ generate a polynomial algebra
$\X_{p}^{n\!-\!1}$ isomorphic
to $\X_p^{\ot(n\!-\!1)}$, whereas
$\hat X^h$  generate a polynomial algebra
$\widehat{\X}_{pX}$ isomorphic to $\widehat{\X_p}$, and
$\widehat{\X_p^{\ot n}}\sim\!\X_{p}^{n\!-\!1}\! \ot\widehat{\X}_{pX}$.
Reasoning as above one can extend such a ``decoupling'' to
$\widehat{\X^{\ot n}},\widehat{(\L^2)^{\ot n}}, \widehat{\X^{\prime\ot n}}$
It is immediate to check that the map $\wedge^n:\X^{\prime\ot n}\mapsto
\widehat{\X^{\prime\ot n}}$ is drastically simplified if we express the
generic $f\!\in\!\X^{\prime\ot n}$ as a function of $X,\xi_i$,
$f(x_1,\!...,\!x_n)=\sum_If^I_X(X)f^I_\xi(\xi_1,\!...,\!\xi_{n\!-\!1})$:
\be
\widehat{f^I_Xf^I_\xi}=\widehat{f^I_X}\,f^I_\xi \label{simplehat}
\ee

\subsection{Twisting Heisenberg/Clifford algebras $\A^\pm$}
\label{Hei-Cli}

Here we $\star$-deform the Heisenberg algebra $\A^+$ (resp. Clifford
algebra $\A^-$) associated to a species of bosons (resp. fermions)
and its transformation properties.

We start fixing the notation while recalling the construction of $\A^\pm$
and how the transformation properties of $\A^\pm$ are induced from the
1-particle ones. We describe the quantum system
abstractly (i.e. in terms of bra, kets, abstract operators) and assume
that a Lie group $G$ (the ``active
transformations'')  is unitarily implemented on the Hilbert space
of the system. The action of $U\g$ will be defined on a dense
subspace; in particular, on the one-particle sector the action of
$U\g$ will be defined on a pre-Hilbert space $\H$. Its closure $\bH$
is the one-particle Hilbert space.
We call $\rho:H \hookrightarrow \O\!:=\!\mbox{End}(\H)$ the $*$-algebra map
such that $g \trc s = \rho(g)s$
for $s\in\H$. The compatibility condition $g \trc (Os)=\!\sum_I\!
\left(g^I_{(1)}\!\trc\! O\right)\!
 g^I_{(2)}\!\trc\!s$ induces on $\O$ a
$H$-module $*$-algebra structure (that of the adoint action).
The actions on $\H,\O$ are thus defined by
\be
g \trc s = \rho(g)s,\qquad \qquad g\trc O=
\sum_I \rho\left(g^I _{(1)}\right) O\, \rho\!\left[S\left(
g^I _{( 2)}\right)\right].  \label{optransf}
\ee
The transformation of multi-particle systems, i.e. of $s\!\in\!
\H^{\ot n}$ and $O\! \in\! \O^{\ot n}$, is obtained  replacing
$\rho$ in (\ref{optransf})
by $\rho^{(n)}\!:=\!\rho^{\otimes n} \!\circ\! \Delta^{(n)}$.
For all $ s,v\in\H^{\ot n}$ $\la v,g \trc s\ra=\la g^* \trc v,
s\ra$.
The pre-Hilbert space $\H^{\ot n}_+$ of
$n$-boson (resp. $\H^{\ot n}_-$ of $n$-fermion) states
is a $H$-$*$-submodule of $\H^{\ot n}$, whereas 
$\O^{\ot n}_+$ is a $H$-module $*$-subalgebra of $\O^{\ot n}$;  its
elements map each of $\H^{\ot n}_+, \H^{\ot n}_-$ into itself.
$\rho^{(n)}(H)$ is a module $*$-subalgebra of $\O^{\ot n}_+$.

Let $\{ e_i\}_{i\in\b{N}}\!\subset\!\H$ be an  orthonormal basis of
$\bH$. For any $i_1\!,\!i_2,...\!,\!i_n\!\in\!\b{N}$ we denote
\be
e_{i_1,i_2,...,i_n}^{\pm}:= N\,
\P^n_{\pm}{}_{i_1i_2...i_n}^{j_1j_2...j_n}\,(e_{j_1}\ot e_{j_2}\ot
...\ot e_{j_n})\in\!\H^{\ot n}_{\pm},  \label{twistnbasis}
\ee
where $N$
is a normalization factor and $\P^n_{\pm}$ is the completely
(anti)symmetric projector on $\H^{\ot n}$. An orthonormal basis
${\cal B}^{n}_+$ (resp. $\B^n_-$) of $\bH^{n}_+$ (resp. $\bH^n_-$)
is the set of the vectors (\ref{twistnbasis}) with $i_1\le
i_2\le...\le i_n$ (resp. $i_1\!<\! i_2\!<\!...\!<\! i_n$). As known,
each of the latter is more conveniently characterized by the
sequence of occupation numbers $n_j\ge 0$; the integer $n_j$ counts
for how many $h$ it occurs $i_h=j$:
$$
\vert
n_1,n_2,...\ra:=e_{i_1,i_2,...,i_n}^{\pm}. \label{occnbasis}
$$
Clearly $\sum_{j=1}^{\infty}\!n_j\!=\!n$. Up to a phase
$N\!=\!\sqrt{n!/\prod_{j=1}^{\infty}\! n_j!}$ (with $0!\!=\!1$).
For $n$ identical
fermions it can be only $n_j\!=\!0,1$, and $N\!=\!\sqrt{n!}$. Denoting
as $\Psi_0$ the vacuum state, let
$$
\H^{\infty}_{\pm}:=\left\{\mbox{finite sequences }(s_0,s_1,s_2,...)
\in\b{C}\Psi_0\oplus\H\oplus \H^{2}_{\pm}\oplus...\right\}
$$
(finite sequence means that there exists an integer $l\!\ge\!
0$ such that $s_n=0$ for all $n\!\ge\! l$).
$\H^{\infty}_{\pm}$ is itself a $H$-$*$-module
(we assume the vacuum to be $H$-invariant,
$g\trc\Psi_0=\varepsilon(g)\Psi_0$). The Fock space is
defined as the closure $\bH^{\infty}_{\pm}$ of $\H^{\infty}_{\pm}$;
it consists of sequences $(s_0,s_1,s_2,...) \in\b{C}
\Psi_0\oplus\bH\oplus \bH^{2}_{\pm}\oplus...$ with finite norm.
Let ${\cal B}^0_{\pm}:=\{\Psi_0\}$. The set ${\cal
B}^{\infty}_{\pm}:=\cup_{n=0}^{\infty}{\cal B}^{n}_{\pm}$ is an
orthonormal basis of the Fock space. It is the set of $\vert
n_1,n_2,...\ra$ with $\sum_{j\in\b{N}}n_j<\infty$. Creation,
annihilation operators for bosons are defined by
$$
\ba{l} a^+_i\vert
...,n_i,...\ra:=\sqrt{n_i\!+\!1}\:\vert ...,n_i\!+\!1,...\ra,\qquad
a^i\vert n_1,n_2,...\ra:=\sqrt{n_i}\:\vert ...,n_i\!-\!1,...\ra, \ea
\label{defabose}
$$
and for fermions by
$$
\ba{l} a^+_i\vert
...,n_i,...\ra \!:=\!(-1)^{s_i}(1\!-\!n_i)\:\vert
...,n_i\!+\!1,...\ra,\qquad a^i\vert
...,n_i,...\ra\!:=\!(-1)^{s_i}n_i\:\vert ...,n_i\!-\!1,...\ra,
\nonumber \ea \label{defafermi}
$$
where $s_i:=\sum_{j=1}^{i\!-\!1}n_j$. They fulfill the Canonical
(anti)Commutation Relations (CCR)
\be
[a^i,a^j]_{\mp}=0,\qquad\quad [a^+_i,a^+_j]_{\mp}=0,
\qquad\quad[a^i,a^+_j]_{\mp}=\delta^i_j\1. \label{ccr}
\ee
The ``number-of-particles'' $\mbox{\bf n}\!:=\! a^+_ia^i$ (an infinite
sum over $i$) is a densely defined operator on  $\bH^\infty_{\pm}$
(a dense domain being $\H^\infty_{\pm}$) with nonnegative discrete
spectrum (actually $\{0,1,2,...\}$, $\overline{H^{\ot n}_{\pm}}$ being the
eigenspace with eigenvalue $n$). As known, this property, or
alternatively the property $a^i\Psi_0=0$, uniquely characterize
(up to unitary equivalences) the Fock space representation among all
irreducible representations of the Heisenberg (resp. Clifford)
$*$-algebra $\A^{\pm}$ generated by $a^i,a^+_i,\1$
fulfilling (\ref{ccr}). The $H$-$*$-module and the $\A$-module structures
of $\H^{\infty}_{\pm}$ induce a $H$-module $*$-algebra
structure on $\A^{\pm}$ (that of the adoint action), 
through the compatibility requirement
\be
g\trc(c\, s)=\sum_I(g^I_{(1)}\trc c)\, g^I_{(2)}\trc 
s, \qquad\qquad \forall g\in H,\:\:\: c\!\in\!\A^{\pm},\:\:\:
s\in\H^{\infty}_{\pm}.  \label{comp}
\ee
Since $g\trc\Psi_0=\varepsilon(g)\Psi_0$,
$a^+_i,a^i$ transform as $e_i=a^+_i\Psi_0$
and $\la e_i,\cdot\ra=\la \Psi_0, a^i\cdot\ra $ respectively:
\be
g\trc e_i=\rho_i^j(g)e_j\qquad \Rightarrow\qquad
g\trc a^+_i=\rho_i^j(g) a^+_j,\qquad
g\trc a^i=\rho^\vee{}_i^j(g)a^j = \rho^i_j\big[S(g)\big] a^j
                                       \label{lineartransf}
\ee
($\rho^\vee\!=\!\rho^T\!\circ\! S$ is the contragredient
 of $\rho$). So $\{a^+_i\}$ and $\{a^i\}$ respectively
span carrier spaces of the representations $\rho,\rho^\vee$ of $H$.
$\A^{\pm}$ is a $H$-module $*$-algebra because (\ref{ccr}) generate
a $H$-invariant $*$-ideal [$\trc$ is extended to products using
(\ref{leibniz})].

\bigskip
After this warm-up, we apply the $\star$-product deformation
procedure and obtain a
$*_\star$-algebra $\A^{\pm}_\star$ with generators $a^+_i,a^i$. It is
convenient to replace the $a^i$ by the $a^{\prime
i}\!:=\!\rho^i_j(\beta )a^j\!=\!a^+_i{}^{*_\star}$, as the latter
transform after the twisted contragredient representation, $g\trc
a^{\prime i}=\hat \rho^\vee{}_i^j(g)a^{\prime j} = \rho^i_j\big[\hat
S(g)\big] a^{\prime j}$, and, with the $\hat a^+_i$, fulfill the
commutation relations at the lhs of \be \ba{l}
 a^{\prime i} \star a^{\prime j} =
\pm R^{ij}_{vu} a^{\prime u}\star  a^{\prime v} ,\\[8pt]
 a^+_i \star a^+_j = \pm R_{ij}^{vu} a^+_u \star a^+_v,\\[8pt]
 a^{\prime i}\star  a^+_j     = \delta^i_j\1_{\scriptscriptstyle{\cal A}}
\pm R^{ui}_{jv} a^+_u \star a^{\prime v},
\ea
\qquad \quad\Leftrightarrow\qquad \quad
\ba{l}
\hat a^{\prime i}\hat a^{\prime j} = \pm R^{ij}_{vu}\hat a^{\prime u}\hat
a^{\prime v} ,\\[8pt]
\hat a^+_i\hat a^+_j = \pm R_{ij}^{vu}\hat a^+_u\hat a^+_v,\\[8pt]
\hat a^{\prime i}\hat a^+_j     = \delta^i_j\1_{\scriptscriptstyle
\hat{\cal A}} \pm R^{ui}_{jv}\hat a^+_u\hat a^{\prime v}; \ea
\label{hqccr} \ee here $R\!:=\!(\rho\ot \rho)(\R)$.
Omitting $\star$-product symbols and putting a $\hat{}$ over $*$ and all
generators we obtain the  isomorphic $\hH$-module $*$-algebra
$\widehat{\A^\pm}$ generated by $\hat a^+_i,\hat a^{\prime i}$ fulfilling the
relations at rhs(\ref{hqccr}), the $\hat
*$-conjugation relations $\hat a^+_i{}^{\hat *}=\hat a^{\prime
i}\!=\!\rho^i_j(\beta )\hat a^j$, and  \be g\tr \hat a^+_i=
\rho_i^j(g) \hat a^+_j,\qquad \qquad g\tr \hat a^{\prime i}=\hat
\rho^\vee{}_i^j(g)\hat a^{\prime j} = \rho^i_j\big(\hat S(g)\big)
\hat a^{\prime j}. \label{twistedlineartransf}
\ee
Such a general class of equivariantly deformed Heisenberg/Clifford
algebras was introduced in Ref. \cite{Fio96}\footnote{In
\cite{Fio96} $\hat a^+_i,\hat a^{\prime i},\beta,\hat
S,\tr,\hat\rho,\check a_i^+, \check a^{\prime i}$ were respectively
denoted as  $\tilde A^+_i,\tilde
A^i,\gamma^{-1},S_h,\trc_h,\tilde\rho,A_i^+,A^i$.
}. Up to normalization of $R$ the
relations at rhs(\ref{hqccr}) are actually identical to
the ones defining the (older) $q$-deformed Heisenberg algebras of
\cite{PusWor89,Pus89,WesZum90}, based on a quasitriangular $\R$ in (only) the
{\it fundamental} representation of $H=U_qsu(N)$ (i.e.
 $i,j,u,v\!\in\!\{1,...,N\}$).

On the other hand, following \cite{Fio96}, it is immediate to check
that setting
\be
\sigma(g):=(g\trc a^+_j)a^j=\rho^i_j(g)a^+_i a^j,
\qquad\qquad g\in \g,      \label{jordan}
\ee
defines a Lie $*$-algebra homomorphism $\sigma: \g\rightarrow \A^{\pm}$.
This is extended as a $*$-algebra homomorphism $\sigma:
\hH=H[[\lambda]]\rightarrow \A^{\pm}[[\lambda]]$ over
$\b{C}[[\lambda]]$ by setting $\sigma(\1_{\scriptscriptstyle
H})\!:=\!\1$ [for $\g=su(2)$ $\sigma$ is the well-known
Jordan-Schwinger realization of $Usu(2)$]; moreover $\sigma$
fulfills (\ref{gag}). It is also immediately checked that $\sigma(g)
\Psi_0=\varepsilon(g) \Psi_0$. It follows 
\be 
\sigma(g)s= g\trc s            \label{gonfock} 
\ee 
for any
$g\!\in\!\hH=H[[\lambda]]$, $
s\!\in\!\H^{\infty}_{\pm}[[\lambda]]$ (see the appendix). Setting
\be 
g\tr s:=g\trc s=\sigma(g) s        \label{deftronfock} 
\ee 
makes $\H^{\infty}_{\pm}[[\lambda]]$ also
into a $\hH$-$*$-module.
$\A^{\pm}[[\lambda]]$ endowed with a $\tr$ defined as in (\ref{gog})
is a compatible $\hH$-module $*$-algebra,  in the sense
\be
g\tr(c\, s)=\sum_I(g^I_{(\hat 1)}\tr c)\,g^I_{(\hat 2)}\tr
 s, \qquad\qquad \forall g\in \hH,\:\:
c\!\in\!\A^{\pm}[[\lambda]]  \label{hatcomp}
\ee
(see the appendix). Under $\tr$ $a^+_i,a^i$ {\it do not} transform as
$\hat a^+_i,\hat a^i$ in (\ref{twistedlineartransf}), but the
elements
\be
\check a_i^+=D_{\f}^{\sigma}(a^+_i)=:\widehat{D}_{\f}^{\sigma}(\hat a^+_i), \qquad
\qquad \check a^{\prime i} = D_{\f}^{\sigma}\!\big(a^{\prime
i}\big)= D_{\f}^{\sigma}\!\big[\rho^i_j\big(\beta
\big)a^j\big]=:\widehat{D}_{\f}^{\sigma}(\hat a^{\prime
i}) \label{defA}
\ee
{\it do} \cite{Fio96}. Moreover, the latter fulfill
rhs(\ref{hqccr}) and
$\check a^{\prime i}=\check a^+_i{}^*$. In other words, the ``dressed
operators'' $\check a^+_i,\check a^i$ provide a {\it realization of
$\hat a^+_i,\hat a^i$ within $\A^{\pm}[[\lambda]]$}.
 Summing up, (\ref{defA}) define a
$\hH$-module $*$-algebra isomorphism (over $\b{C}[[\lambda]]$)
$\widehat{D}_{\f}^{\sigma}:\widehat{\A^\pm}\leftrightarrow\A^{\pm}[[\lambda]]$\footnote{The
``triviality'' of any deformation $\widehat{\A^\pm}$ of any
Heisenberg algebra $\A^{\pm}$, i.e. the existence of a isomorphism
$\widehat{D}_{\f}^{\sigma}:\widehat{\A^\pm}\leftrightarrow\A^{\pm}[[\lambda]]$
of $*$-algebras in the form of a formal power series in $\lambda$
which is the identity at order $\lambda^0$, is also a consequence
\cite{Duc85} of the vanishing of its second cohomological group;
this however does not give any such isomorphism explicitly.}:
\be
\widehat{D}_{\f}^{\sigma}(\hat c\hat
d)=\widehat{D}_{\f}^{\sigma}(\hat c) \widehat{D}_{\f}^{\sigma}(\hat
d),\quad\widehat{D}_{\f}^{\sigma}(\hat c^{\hat *})
=\widehat{D}_{\f}^{\sigma}(\hat c)^*,\quad
g\tr\widehat{D}_{\f}^{\sigma}(\hat c)=\widehat{D}_{\f}^{\sigma}(g\tr
\hat c). \quad          \label{hatDfsigma}
\ee
Then the $\A^{\pm}[[\lambda]]$-$*$-module
$\H^{\infty}_{\pm}[[\lambda]]$ becomes also a
$\widehat{\A^\pm}$-$*$-module  if one sets
\be
\hat c  s:=\widehat{D}_{\f}^{\sigma}(\hat c) s\label{FockhHmod}
\ee
for any $\hat c\!\in\! \widehat{\A^\pm}$, 
$ s\!\in\!\H^{\infty}_{\pm}[[\lambda]]$. As $\hat a^{\prime i}
\Psi_0=\check a^{\prime i}\Psi_0=0$, one finds that this is a
Fock-space type  representation of $\widehat{\A^\pm}$ on
$\H^{\infty}_{\pm}[[\lambda]]$. $\hat *$ plays also the role of Hermitean
conjugation w.r.t. scalar product of $\H^{\infty}_{\pm}$, since
\be
\la \hat c\, s_1,s_2\ra\stackrel{(\ref{FockhHmod})}{=}
\la \widehat{D}_{\f}^{\sigma}(\hat c) s_1,s_2\ra=
\la s_1,\widehat{D}_{\f}^{\sigma}(\hat c)^* s_2\ra
\stackrel{(\ref{hatDfsigma})}{=}
\la s_1,\widehat{D}_{\f}^{\sigma}(\hat c^{\hat *}) s_2\ra
\stackrel{(\ref{FockhHmod})}{=}\la s_1,\hat c^{\hat *} s_2\ra
\label{hatHermConj}
\ee
for any $\hat c\!\in\! \widehat{\A^\pm}$,
$s_1,s_2\!\in\!\H^{\infty}_{\pm}[[\lambda]]$.  
Note also that $\hat a^+_i\Psi_0=\check a^+_i\Psi_0=a^+_i\Psi_0=e_i$. 
Finally, the
analog of the compatibility (\ref{hatcomp}) (with $\hat c$ replacing $c$) holds.

In terms of $\star$-products the previous property
becomes the first equality in
\be
\la c\, s_1,s_2\ra=\la s_1,c^{*_\star}\star  s_2\ra=\la s_1,c^* s_2\ra,\label{starHermConj}
\ee
valid for any $c\!\in\! \A^{\pm}[[\lambda]]$: the
Hermitean conjugation is $*$ if we multiply operators by the original
product, $*_\star$ if we multiply them  by the
$\star$-product. An independent proof of this (see the appendix)
relies on the relation
\be
\la \Psi_0,r^*t\,\Psi_0\ra=\la \Psi_0,
r^{*_\star}\!\star\! t\,\Psi_0\ra, \qquad\qquad\forall
r,t\!\in\! \A^\pm[[\lambda]] \label{*starstar}
\ee
[this is the analog of (\ref{Intstarprop})$_2$] ,
which  in turn is proved using the $H$-invariance of $\Psi_0$.

\bigskip
Choosing $\hat c\!=\!\hat a^+_{i_1}...\hat a^+_{i_n}$ in
definition (\ref{FockhHmod}) one obtains the first equality in the formula
\be
\hat a^+_{i_1}...\hat a^+_{i_n}\Psi_0=
\check a^+_{i_1}...\check a^+_{i_n}\Psi_0=
\overline{F}^n{}_{i_1,...,i_n}^{j_1,...,j_n}\: a^+_{j_1}... a^+_{j_n}\Psi_0
=a^+_{i_1}\star...\star a^+_{i_n}\Psi_0
\in\H^n_\pm[[\lambda]].
\label{Aavec} \ee
The second is easily proved using
(\ref{leibniz})$_3$ and the inverse of definition
(\ref{iter-n})$_1$. The third follows applying to the vacuum
both sides of the identity $a^+_{i_1}\!\star\!...\!\star\! a^+_{i_n}\!=\!
\overline{F}^n{}_{i_1,...,i_n}^{j_1,...,j_n} \: a^+_{j_1}... a^+_{j_n}$, where
$\overline{F}^n\!:=\!\rho^{\ot n}[(\F^n)^{-1}]$
(a unitary transformation of $\H^{\ot n}$); this is an identity in
$V(\A^{\pm})[[\lambda]]$ following  from (\ref{starprod}).
A straightforward computation gives
\be
\ba{l}
 \sigma(g) a^+_{i_1}... a^+_{i_n}\Psi_0=
[\rho^{(n)}(g)]^{j_1...j_n}_{i_1...i_n}\, a^+_{j_1}... a^+_{j_n}\Psi_0,\\[8pt]
\sigma(g) a^+_{i_1}\star...\star a^+_{i_n}\Psi_0=
[\hat\rho^{(n)}(g)]^{j_1...j_n}_{i_1...i_n}\, a^+_{j_1}\star...\star a^+_{j_n}\Psi_0,\ea
\ee
where $\hat\rho^{(n)}(g)\!=\! \rho^{\otimes n} \!\circ\! \hat\Delta^{(n)}(g)\!=\!
F^n \rho^{(n)}(g) \overline{F}^n$. Note also that
$a^+_i\star a^{\prime i}\!=\! \check a^+_i\check a^{\prime i}\!=\!a^+_ia^i\!=\!\mbox{\bf n}$. Let
$e_{i_1,...,i_n}^{\prime\pm}$ be the vectors (\ref{Aavec})
multiplied by $\sqrt{n!}/N$;
a (non-orthonormal) basis $\B^{\prime n}_+$ of $\H^{\ot
n}_+[[\lambda]]$ (resp. $\B^{\prime n}_-$ of $\H^n_-[[\lambda]]$)
over $\b{C}[[\lambda]]$ consists of those with $i_1\!\le\!
i_2\!\le\!...\!\le\! i_n$ (resp. $i_1\!<\! i_2\!<\!...\!<\! i_n$).

The vectors (\ref{Aavec}) are up to now formal $\lambda$-power
series (arising from the $\lambda$-power series expansions of
the $\overline{F}^n{}_{i_1,...,i_n}^{j_1,...,j_n}$)
with coefficients in $\H^{\ot n}_\pm$. If
one can really define a 1-parameter continuous family of
unitary operators $\lambda\!\in \b{R}\mapsto\overline{F}^n(\lambda)$
(at least in a neighbourhood of $\lambda\!=\!0$)
such that the mentioned $\lambda$-power series are the asymptotic expansions of
the $\overline{F}^n{}_{i_1,...,i_n}^{j_1,...,j_n}$, then
(\ref{Aavec}) are true vectors of $\H^{\ot n}_\pm$ allowing a
{\it representation of $\widehat{\A^\pm}$ on the ordinary (Bose/Fermi) Fock
space $\bH^{\infty}_{\pm}$}; as a consequence, {\bf no ``change of
statistics'' is needed to represent the deformed Heisenberg (resp.
Clifford) algebra $\widehat{\A^\pm}$}. 
This occurs e.g. in all deformations based on  `Reshetikhin twists', i.e.
twists of the form 
$\F=e^{i \lambda f}$, where $f\in \h\wedge\h$ and $\h$ is a real
Cartan subalgebra of $\g$; then choosing a basis $e_i$ of $\H$ 
consisting of eigenvectors of $\h$ the unitary matrix
$\overline{F}^n(\lambda)$  becomes diagonal with elements
of modulus 1 (phases). This applies
in particular to the Moyal twists (see sections \ref{moyalE}, \ref{MMQFT}), where
one uses a generalized basis of $\H$ consisting of eigenvectors 
of the translation Lie subalgebra $\h$.  
More generally, this will occur whenever $\hH$ 
admits a non-pertubative (in $\lambda$) representation on $\H$
in one-to-one correspondence with that of $H$;
then the matrix elements of $\overline{F}^n(\lambda)$ will be 
given by contractions of deformed and undeformed
Clebsch-Gordan coefficients \cite{Blo05}. This works
even for some {\it quasitriangular} (but not triangular) Hopf algebras 
like the semi-simple Drinfel'd-Jimbo
quantum groups (in {\it rational} form), where the dependence on $\lambda$ 
of the deformed Clebsch-Gordan coefficients of the can be concentrated
in  a {\it rational} dependence on the new parameter $q=e^{\lambda}\in\b{C}$, 
for generic $q$.

Does $\widehat{\A^\pm}$ admit other $*$-representations of Fock type which are
not unitarily equivalent to $*$-representations of $\A^{\pm}$? Using
standard arguments it is easy to see that the answer is negative in
either characterization of the Fock representation. In fact, from
the first characterization (the existence of a unique
vacuum $\hat\Psi_0$: $\hat a^{\prime i}\hat\Psi_0=0$ for
all $i$) and the commutation relations (\ref{hqccr}) it follows that
the map
$$
P(\hat a^{\prime},\hat a^+)\hat\Psi_0\mapsto
P(\check a^{\prime},\check a^+)\Psi_0
$$
for all polynomial $P(\hat a^{\prime},\hat a^+)$ preserves
the scalar products, and therefore is unitary. As for the second,
(\ref{hqccr}) imply that the
nonnegative-definite, infinite sum over $i$ $\hat{\mbox{\bf
n}}\!:=\!\hat a^+_i\hat a^{\prime i}= (\hat a^{\prime i})^{\hat
*}\hat a^{\prime i}$ fulfills $[\hat{\mbox{\bf n}},\hat a^{\prime
i}]\!=\!-\!\hat a^{\prime i}$, $[\hat{\mbox{\bf n}},\hat
a^+_i]\!=\!\hat a^+_i$; requiring that $\hat{\mbox{\bf n}}$ is densely defined
and has a
nonnegative discrete spectrum, in particular an eigenvector
$\nu$ with eigenvalue $\nu\!\ge\! 0$, implies that a vacuum
$ \hat \Psi_0\neq 0$ arises applying to $\nu$ a suitable
monomial in the $\hat a^{\prime i}$.

\section{Nonrelativistic second quantization}
\label{N2ndQ}

\subsection{Twisting quantum mechanics in configuration space}
\label{Config}

Dealing with the wave-mechanical description of a system of quantum
particles in space $X$ means that the state vectors $s$'s are described by
wavefunctions $\psi$'s on $X$ and the abstract operators by
differential or more generally integral operators on the $\psi$'s.
For simplicity we stick to\footnote{Morally, one could
apply such an approach to any Riemannian manifold $X$ (playing the role
of ``space'') for which first quantization is well-defined and has
a covariance Lie group $G'$ containing as a subgroup the isometries of the spacetime $X'\!=\!\b{R}\times X$.} $X=\b{R}^3$, consider spinless
particles and derive consequences from the
covariance of the description first under the Euclidean group $G$
(thought as a group of {\it active}  space-symmetry transformations), then
under the whole Galilei group $G'$. We shall call
${\rm x}^a $ a set of Cartesian coordinates of $\b{R}^3$.
Going to the infinitesimal form,
all elements $H=U\g$ will be well-defined differential operators e.g. on the
pre-Hilbert space ${\cal S}(\b{R}^3)$, so we can choose $\X$
as a dense subspace $\X\subseteq {\cal S}(\b{R}^3)$ (to be specified later)
and tailor the 1-particle pre-Hilbert space $\H$
and the algebra of endomorphisms
$\O\!:=\!\mbox{End}(\H)$ as respectively
isomorphic to $\X$ and $\En\!:=\!\mbox{End}(\X)$, by definition;
we shall call the isomorphisms $\kappa,\tilde\kappa$.
[One reason why we do not identify $\H$ with $\X$ is that we wish to introduce a
{\it realization} (i.e. representation) of the {\it same} state $s\in\H$ of the quantum system also by a noncommutative wavefunction.]
Summarizing, there exists a (reference frame dependent) {\it $H=U\g$-equivariant configuration space realization of $\{\H,\O\}$ on
$\{\X,\En\}$}, i.e.

\begin{enumerate}

\item there exists a $H$-equivariant, unitary transformation
$\kappa:s\!\in\!\H\leftrightarrow\psi_s\!\in\!\X$, i.e.
\be
g\trc
\psi_s\!=\!\psi_{g\trc s},\qquad\quad \la s |v \ra=\! \int_X\!\!\!
d\nu\: [\psi_s({\rm x})]^*\psi_v({\rm x}).
\label{scalprod1}
\ee

\item $\kappa(Os)=\tilde\kappa(O)\kappa(s)$ for any
$s\!\in\!\H$ defines a $H$-equivariant map
$\tilde\kappa\!:\!\O\!\leftrightarrow\!\En$
[equivariance means $g\trc\tilde\kappa(O)=\tilde\kappa(g\trc O)]$,
and $\D\subset\En$
($\D$ was defined in section \ref{twistdifcal}).
In particular  $\tilde\kappa(q^a)\!=\!{\rm x}^a\cdot$,
$\tilde\kappa(p^a)\!=\!-i\hbar\frac{\partial}{\partial {\rm x}^a}$
on the canonical variables $\{q^a,p^a\}$.

\end{enumerate}

This implies for a system of $n$ distinct particles (resp. $n$ bosons/fermions)
on $X$:

\begin{enumerate}
\item $\kappa^{\ot n}\!:\!\H^{\ot n}\!\leftrightarrow\!\X^{\ot n}$
(resp. their restrictions $\kappa^n_\pm\!:\!\H^{\ot
n}_{\pm}\!\leftrightarrow\!\X^{\ot n}_{\pm}$) are $H$-equivariant
unitary transformations.

\item  $\tilde\kappa^{\ot n}\!:\!\O^{\ot
n}\!\leftrightarrow\! \En^{\ot n}$ (resp. its restriction
$\tilde\kappa^n_+\!:\!\O^{\ot n}_+\!\leftrightarrow\! \En^{\ot
n}_+$) are $H$-equivariant maps.
\end{enumerate}

\noindent The maps $\kappa^n_{\pm},\tilde\kappa^n_+$ define a
(frame-dependent) {\it $U\g$-equivariant, commutative configuration
space realization} of $\{\H^{\ot n}_{\pm},\O^{\ot n}_+\}$ (the
Hilbert space and algebra of observables of a system of $n$
bosons/fermions) on $\{\X^{\ot n}_{\pm},\En^{\ot n}_+\}$.

\medskip
For any twist $\F\!\in\!(H\ot H)[[\lambda]]$ and the associated
$\star$-deformation one finds
\be \ba{lll}
\la s, v\ra &=&
\int_X\! d\nu({\rm x}_1)...\!\!\int_X\! d\nu({\rm x}_n) [\psi_s({\rm
x}_1,...,{\rm x}_n)]^*
\psi_v({\rm x}_1,...,{\rm x}_n)\\[8pt]
&\stackrel{(\ref{Intstarprop})}{=}& \int_X\! d\nu({\rm x}_1)
...\!\!\int_X\! d\nu({\rm x}_n) [\psi_s({\rm x}_1,...,{\rm x}_n)]^{*_\star}
\star\psi_v({\rm x}_1,...,{\rm x}_n). \ea
\ee
If in addition $\F$ is such that one can define $\widehat{\X},\widehat{\D}$ and
the map $\wedge$ (see section \ref{twistdifcal}),
we introduce noncommutative
wavefunctions $\hat\psi\!=\!\wedge^n(\psi)$.
Then the previous equation becomes
\be
\la s, v\ra =\int_{\hat X}\!
d\hat\nu(\hat{\rm x}_1)...\!\!\int_{\hat X}\! d\hat\nu(\hat{\rm
x}_n) [\hat\psi_s(\hat{\rm x}_1,...,\hat{\rm x}_n)]^{\hat
*}\hat\psi_v(\hat{\rm x}_1,...,\hat{\rm x}_n).   \label{scalprodn}
\ee
The map $\wedge^n:\psi_s\!\in\!\X^{\otimes
n}\!\mapsto\!\hat\psi_s\!\in\!\widehat{\X^{\otimes n}}$ is thus unitary
and $\hH$-equivariant, $\widehat{g\trc
\psi_s}\!=\!g\trc \hat\psi_s$; using (\ref{hatprop}) one finds
$\wedge^n\!=\!\wedge^{\ot n}\!\circ\!\big(\F^n\trc^{\ot n}\!\big)$.

The action of the symmetric group $S_n$ on
$\widehat{\X^{\otimes n}},\widehat{\En^{\otimes n}}$
is obtained by ``pull-back'' from that on
$\X^{\otimes n},\En^{\otimes n}$. A permutation
$\tau\!\in\!S_n$ is represented on $\X^{\otimes n},\widehat{\X^{\otimes n}}$
 respectively by the permutation operator $\P_\tau$ and
the ``twisted permutation operator'' $\P^F_\tau=\wedge^n \P_\tau
[\wedge^n]^{-1}$ (cf. \cite{FioSch96}). The completely
(anti)symmetric projector $\P_{\pm}^n\!=\!(1/{n!})\sum_{\tau\in
S_n}\eta_\tau P_\tau $ ($\eta_{\tau}= -1$ if $\tau$ is an odd
permutation and we consider $\P^n_-$, $\eta_{\tau}=1$ otherwise) and
its twisted counterpart $\P_{\pm}^{nF}:=\wedge^n \P_{\pm}^n
\left[\wedge^n\right]^{-1}$ respectively project $\X^{\ot n}$ and
$\widehat{\X^{\otimes n}}$ onto their subspaces $\X^{\ot n}_{\pm}$ and
$\wedge^n(\X^{\otimes n}_{\pm})$, which are
eigenspaces of $\P_\tau$ and $\P_\tau^{F}$ with eigenvalue
$\eta_{\tau}$\footnote{One can easily show that
$\P^F_\tau,\P_{\pm}^{nF}$ depend on $\F$ through the $\R$-matrix
only. If $\P_\tau=\prod_{\{(hk)\}} \P_{hk}$ is a decomposition of
$\P_\tau$ as a product of transpositions ($\P_{hk}$ stands for the
transposition of the $h$-th and $k$-th tensor factor), $\P^F_\tau$
can be decomposed as $\P^F_\tau=\prod_{\{(hk)\}}
\P_{hk}\R_{hk}\trc^{\ot n}$, where
$\R_{hk}=\sum_I\1^{\ot(h\!-\!1)}\ot\!\R^{(1)}_I\!\ot\!
\1^{\ot(k\!-\!h\!-\!1)}\!\ot\!\R^{(2)}_I\!\ot\!\1^{\ot(n\!-\!k)}$.
If e.g. $n=2$ and $\tau=(21)$, we find $\P_\tau[\psi_s({\rm
x}_1)\psi_v({\rm x}_2)]=\psi_v({\rm x}_1)\psi_s({\rm x}_2)$, whereas
\be \P^F_\tau[\hat\psi_s(\hat{\rm x}_1)\hat\psi_v(\hat{\rm
x}_2)]=\sum_{I,I'}\big[\Fu_I\bF^{(2')}_{I'}\trc\hat\psi_v(\hat{\rm
x}_1)\big] \big[\Fd_I\bF^{(1')}_{I'}\trc\hat\psi_s(\hat{\rm
x}_2)\big]= \sum_I\big[\R^{(1)}_I\trc\hat\psi_v(\hat{\rm
x}_1)\big]\,\big[\R^{(2)}_I\trc\hat\psi_s(\hat{\rm x}_2)\big].
\label{PFtau} \ee}. Thus, $\widehat{\X^{\otimes n}},\widehat{\En^{\otimes n}}$
 are (anti)symmetric  up to the similarity transformation $\wedge^n$ (cf.
\cite{FioSch96}). As an example, in section \ref{moyalE} we
exhibit $\P_{\pm}^{2,F}$ on the Moyal space.

\medskip
Let $\hat \kappa^n:=\wedge^n \kappa^{\ot n}$,
$\widehat{\tilde\kappa}^n(\cdot):=\wedge^n [\tilde\kappa^{\ot n}(\cdot)][\wedge^n]^{-1}$.
The restrictions $\hat\kappa^n_{\pm}:=\hat \kappa^n\!\!\upharpoonright\!_{\!\H^{\ot
n}_{\scriptscriptstyle\pm}}$,
$\widehat{\tilde\kappa}^n_+:=\widehat{\tilde\kappa}^n\!\!\upharpoonright\!_{\!\O^{\ot
n}_{\scriptscriptstyle +}}$ define a (frame-dependent) {\it
$\hH$-equivariant, noncommutative configuration space realization}
of $\{\H^{\ot n}_{\pm},\O^{\ot n}_+\}$ on
$\{\widehat{\X^{\ot n}_{\pm}},\widehat{\En^{\ot n}_{+}}\}$.

\subsection{Twisting quantum fields in the Schr\"odinger picture}
\label{Schrpic}

Let $\{ \varphi_i\}_{i\!\in\!\b{N}}$ be an orthonormal basis of $\X$,
$\{ e_i\}_{i\!\in\!\b{N}}$ the corresponding basis of $\H$
($\varphi_i\!=\!\kappa(e_i)$),  $a^+_i,a^i$ the associated wavefunction,
creation, annihilation operators.
We adopt the dual $\X'$ of $\X$ as space of distributions.
The nonrelativistic field
operator $\varphi$ in the {\it Schr\"odinger picture} and its hermitean conjugate
$\varphi^*$ defined in (\ref{Schfield})
are operator-valued distributions
fulfilling the canonical (anti)commutation relations
\be
[\varphi({\rm x}),\varphi({\rm y})]_{\mp}=h.c.=0, \qquad \qquad
[\varphi({\rm x}),\varphi^*({\rm y})]_{\mp}= \varphi_i({\rm
x})\varphi_i^*({\rm y})=\delta({\rm x}\!-\!{\rm
y}) \label{fccr}
\ee
($\mp$ for bosons/fermions; infinite sum over $i$;
$\varphi_i^*\!\equiv\!\overline{\varphi_i}$).  One needs to work
 with polynomials of arbitrarily high degree $n$
in $\varphi,\varphi^*$ evaluated at independent points
${\rm x}_1,...,{\rm x}_n$; the generic normal ordered monomial of
degree $n$ will read
\be
\varphi^*({\rm x}_1)....\varphi^*({\rm x}_m)
\varphi({\rm x}_{m\!+\!1})...\varphi({\rm x}_n) \label{fieldmon}
\ee
with $0\le m\le n$. We shall call the linear span of all such monomials
(for all $m,n$) the {\it field $*$-algebra} $\Phi$. This is a subalgebra
of the tensor product algebra
$\Phi^e\!=\!\A^{\pm}\ot\left(\bigotimes_{i=1}^{\infty}\! \X'\right)$,
where the first, second,...
tensor factor $\X'$ refers to the dependence of the distribution
on ${\rm x}_1,{\rm x}_2,...$ [by definition, the dependence
of the monomial (\ref{fieldmon}) on ${\rm x}_{h}$ is trivial for $h> n$].
The CCR (\ref{ccr}) of $\A^\pm$ are the only nontrivial commutation
relations in $\Phi^e$. Relations  (\ref{fccr}) hold with
any ${\rm x},{\rm y}\in\{{\rm x}_1,{\rm x}_2,...\}$.

The key property is that $\varphi,\varphi^*$ are basis-independent,
i.e. {\bf invariant under the
group $U(\infty)$ of unitary transformations of $\H$ applied to $\{
e_i\}_{i\!\in\!\b{N}}$}, in particular under the subgroup
$G$ of active space-symmetry transformations (transformations
of the states $e_i$ obtained by translations or rotations of the
1-particle system), or (in infinitesimal form)
{\bf under $U\g$}:
 \be
g\trc \varphi({\rm x})=\epsilon(g)\varphi({\rm x}),\qquad\quad g\trc \varphi^*({\rm x})=\epsilon(g)\varphi^*({\rm x}).
\label{fieldinv}
\ee
Instead, $\varphi_i,a^+_i,e_i$ transform after the
same non-trivial representation $\rho$ of $U(\infty)$,
$\varphi^*_i,a^i,\la e_i,\cdot\ra$ after the contragredient $\rho^\vee$.
Altogether, $\Phi^e$ is a huge $U\g$-module [and also $Uu(\infty)$-module] $*$-algebra.

\medskip
As a consequence, if we extend the $\star$-deformation of the previous
sections to  $Uu(\infty)$ and $\Phi^e$ we obtain the
Hopf algebra $\widehat{Uu(\infty)}$
and the $\widehat{U\g}$-module [and also $\widehat{Uu(\infty)}$-module]
$*$-algebra $\Phi^e_\star$.
We find $\varphi^{*_\star}({\rm x})=\varphi^*({\rm x})$
by (\ref{star'}), (\ref{fieldinv}) and $\epsilon(\beta )=1$.
Moreover, by (\ref{Trivstar}) the $\star$ has no effect on the
product of $\varphi({\rm x})$ with any
$\omega\!\in\!V\big(\Phi^e\big)$:
\be
\varphi({\rm x})\star \omega=\varphi({\rm
x}) \omega, \qquad \qquad \omega\star\varphi({\rm
x})=\omega\varphi({\rm x}), \qquad \qquad \&\quad  \mbox{herm.
conj.}      \label{utilissima}
\ee
(as usual, these products are well-defined only if $\omega$
is ${\rm x}$-independent). Consequently, for any ${\rm
x},{\rm y}\!\in\!\{{\rm x}_1,{\rm x}_2,...\}$ the CCR (\ref{fccr})
can be rewritten in the form
\be
[\varphi({\rm x})\stackrel{\star}{,}\varphi({\rm y})]_{\mp}=h.c.=0,
\qquad\qquad\qquad [\varphi({\rm
x})\stackrel{\star}{,}\varphi^{*_\star}({\rm
y})]_{\mp}=\varphi_i({\rm x})\star\varphi_i^{*_\star}({\rm
y})\qquad\quad \label{qfccr}
\ee
(here and below
$[A\!\stackrel{\star}{,}\!B]_{\mp}\!:=\!A\star B\!\mp\! B\star A$).
By Lemma \ref{lemmainv} in the appendix, in
$V(\Phi^e_\star)=V(\Phi^e)[[\lambda]]$ one can re-express the field
itself in terms of $\star$-products:
\be
\varphi({\rm x})
=\varphi_i({\rm x}) \star a^{\prime i}, \qquad\qquad\varphi^*({\rm
x})=\varphi^{*_\star}({\rm x}) =a^+_i\star \varphi_i^{*_\star}({\rm
x}).        \label{fieldeco'}
\ee
The $\star$-commutation rules
among $a^{\prime i}, a^+_j,\varphi_i({\rm x}),\varphi_i({\rm y}),
\varphi^{*_\star}_i({\rm x}), \varphi^{\hat
*}_i({\rm y})$ beside (\ref{hqccr}) are reported in formula (\ref{earel})
in the appendix. In particular, $a^{\prime i}, a^+_j$ do not
$\star$-commute with functions of ${\rm x}$, (for all ${\rm
x}\in\{{\rm x}_1,{\rm x}_2,....\}$). If $\F$ is such that
one can define the map $\wedge$ on $\X$, in the
``hat notation'' (\ref{fieldeco'}), (\ref{qfccr}) become the first
two lines in (\ref{hatqfccr}).

\medskip
\begin{figure}
\begin{center}
\begin{picture}(20000,18000)
\put(0,15700){$\H^{\ot n} $}
\put(16000,15700){$\X^{\ot n} $}
\drawline\fermion[\E\REG](3000,16200)[12000]
\drawline\fermion[\E\REG](3000,15500)[12000]
\drawarrow[\E\ATBASE](15000,16200)
\drawarrow[\W\ATBASE](3000,15500)
\put(7700,16800){$\kappa^{\ot n}$}
\put(7500,14400){$(\kappa^{\ot n})^{-1}$}
\put(0,1300){$\H^{\ot n}_{\pm} $}
\put(16000,1300){$\X^{\ot n}_{\pm} $}
\drawline\fermion[\E\REG](3000,1800)[12000]
\drawline\fermion[\E\REG](3000,1100)[12000]
\drawarrow[\E\ATBASE](15000,1800)
\drawarrow[\W\ATBASE](3000,1100)
\put(7700,2400){$\kappa^n_{\pm}$}
\put(7500,0){$(\kappa^n_{\pm})^{-1}$}
\drawline\fermion[\NE\REG](3000,2800)[16700]
\drawline\fermion[\NW\REG](15000,2800)[16700]
\drawarrow[\SW\ATBASE](3000,2800)
\drawarrow[\SE\ATBASE](15000,2800)
\put(5800,4500){$\Pi^n_{\pm}$}
\put(11000,4500){$\pi^n_{\pm}$}
\drawline\fermion[\N\REG](1000,3300)[12000]
\drawline\fermion[\N\REG](17000,3300)[12000]
\drawarrow[\S\ATBASE](1000,3300)
\drawarrow[\S\ATBASE](17000,3300)
\put(1500,8500){$\P^n_{\pm}$}
\put(17500,8500){$\P^{\prime n}_{\pm}$}
\end{picture}
\end{center}
\caption{\label{diag} Commutative diagram.}
\end{figure}
The undeformed fields are characterized by the `kinematical' property that setting
\be\ba{l}
\left[\pi^n_\pm(s)\right]({\rm x}_1,\!...,{\rm x}_n)
:=\frac 1{\sqrt{n!}}
\left\la\varphi^{ *}({\rm x}_1\!)...
\varphi^{ *}({\rm x}_n)\Psi_0,s\right\ra,\\[10pt]
\Pi^n_\pm(\psi) :=\frac 1 {\sqrt{n!}}\! \displaystyle\int_X \!\! \!\!d\nu({\rm
x}_1\!)...\!\! \displaystyle\int_X \!\!\!\! d\nu({\rm x}_n)
\varphi^{ *}({\rm x}_1\!)...
\varphi^{*}({\rm x}_n)\Psi_0 \: \psi({\rm
x}_1,\!...,{\rm x}_n)\ea             \label{wfket}
\ee
(the scalar products in the first, second line are in $\H^{\ot n}$, $\X^{\ot n}$
respectively) defines `crossed' (anti)symmetric projectors, i.e. linear maps
$\pi^n_\pm\!:\!\H^{\ot n}\!\mapsto\! \X^{\ot n}_{\pm}$ and
$\Pi^n_\pm\!:\!\X^{\ot n}\!\mapsto\!\H^{\ot n}_{\pm}$ making
diagram \ref{diag} commutative (see the appendix).
If $\psi\!=\!f_1\!\ot\!...\!\ot\! f_n$ one finds
$$
\Pi^n_\pm(f_1\!\ot\!...\!\ot\! f_n)=\frac 1 {\sqrt{n!}}
\varphi^{*}(f_1)... \varphi^{*}(f_n)\Psi_0,
\qquad\qquad \varphi^{ *}(f):=\displaystyle\int_X \!\! \!\!d\nu({\rm
x})\varphi^{*}({\rm x})f({\rm x});
$$
since for $f_l=\varphi_{i_l}$ the lhs is proportional to
$e^{\pm}_{i_1,...,i_n}\!\in\!\B^{n}_{\pm}$, it follows the
well-known statement that polynomials in $\varphi^{ *}(f)$ (fields
smeared with test functions $f$) applied to the vacuum make up a
subspace dense in the Fock space. The restrictions of
$\pi^n_\pm,\Pi^n_\pm$ to $\H^{\ot n}_{\pm},\X^{\ot n}_{\pm}$
respectively reduce to $\kappa^n_\pm,\big(\kappa^n_\pm\big)^{-1}$,
in particular give $\kappa,\kappa^{-1}$ for $n\!=\!1$.

$\star$-Deformation preserves these properties.
After linearly extending $\pi^n_\pm,\Pi^n_\pm$ resp. to
$\H^{\ot n}[[\lambda]],\X^{\ot n}[[\lambda]]$, by
(\ref{utilissima}), (\ref{fieldeco'}) the rhs(\ref{wfket})'s do not
change as elements resp. of $V\big(\X^{\ot
n}_\pm\big)[[\lambda]]\!=\!V\left[\big(\X^{\ot
n}_\pm\big)_\star\right]$ and $\H^{\ot n}_\pm[[\lambda]]$ if we
replace all products by  $\star$-products and $*$'s by $*_\star$'s.
Hence, if $\wedge\!:\!\X[[\lambda]]\!\mapsto\!\hat\X$ exists, $\hat{\pi}^n_\pm\!:=\!\wedge^n\!\circ\!\pi^n_\pm$,
$\hat{\Pi}^n_\pm\!:=\!\Pi^n_\pm\!\circ\!(\wedge^n)^{-1}$ fulfill
\be\ba{l} \hat{\pi}^n_\pm\!:\!\H^{\ot
n}[[\lambda]]\!\mapsto\!\widehat{\X^{\ot n}_{\pm}}\qquad
\left[\hat{\pi}^n_\pm(s)\right]\!(\hat{\rm x}_1,\!...,\hat{\rm
x}_n)=\frac 1 {\sqrt{n!}} \left\la \hat\varphi^{\hat *}(\hat{\rm
x}_1\!)...\hat\varphi^{\hat *}(\hat{\rm x}_n)\Psi_0,s\right\ra,\\[10pt]
\hat\Pi^n_\pm\!:\!\widehat{\X^{\ot n}}\!\!\mapsto\!\H^{\ot
n}_{\pm}\![[\lambda]]\qquad\hat\Pi^n_\pm\!(\hat\psi) \!=\!\frac 1
{\sqrt{n!}}\! \!\!\displaystyle\int_{\hat X} \!\!\!
\!\!d\hat\nu(\hat{\rm x}_1\!)...\!\!\! \displaystyle\int_{\hat X}
\!\!\!\!\! d\hat\nu(\hat{\rm x}_n\!) \hat\psi(\!\hat{\rm
x}_1\!,\!...,\!\hat{\rm x}_n\!)\hat\varphi^{\hat *}(\hat{\rm
x}_1\!)...\hat\varphi^{\hat
*}(\hat{\rm x}_n\!)\Psi_0 \ea\label{hatwfket}
\ee
and
analogous properties. Polynomials in  $\hat\varphi^{ *}(\hat
f)\!:=\!\int\!\!d\hat\nu\hat\varphi^{\hat *}(\hat{\rm x})\hat
f(\hat{\rm x})$ applied to the vacuum make up a subspace dense in
the Fock space. $s=e_{i_1}\ot...\ot e_{i_n}$ is mapped into
\bea
&&\pi^n_\pm(s)({\rm x}_1,...,{\rm x}_n)=
\varphi_{(i_1}({\rm x}_1\!)...\varphi_{i_n]}({\rm x}_n\!), \label{n-wf}\\
&&\hat{\pi}^n_\pm(s)(\hat {\rm x}_1,\!...,\hat {\rm x}_n)=
F^n{}^{j_1...j_n}_{(i_1...i_n]} \hat\varphi_{j_1}(\hat{\rm
x}_1\!)...\hat\varphi_{j_n}(\hat{\rm x}_n\!),\label{n-wfh}
\eea
where $(...]$ means (anti)symmetrization of the indices. The same
result is obtained if
$Ns\!=\!e^{\pm}_{i_1,...,i_n}\!\in\!\B^{n}_{\pm}$. If
$Ns\!=\!e^{\prime\pm}_{i_1,...,i_n}\!\in\!\B^{\prime n}_{\pm}$
[as defined after (\ref{Aavec})] we find instead
\be
\left[\hat{\pi}^n_\pm(s)\right](\hat {\rm x}_1,\!...,\hat {\rm
x}_n)=\P^{n,F}_{\pm}{}^{j_1...j_n}_{i_1...i_n}
\hat\varphi_{j_1}(\hat{\rm x}_1\!)...\hat\varphi_{j_n}(\hat{\rm
x}_n\!) = \P^n_{\pm}{}^{h_1...h_n}_{1...n}
\hat\varphi_{i_1}(\hat{\rm x}_{h_1}\!)...\hat\varphi_{i_n}(\hat{\rm
x}_{h_n}\!)\label{n-wf'}
\ee
($h_l\!\in\!\{1,...,n\}$, whereas $i_l,j_l\!\in\!\b{N}$), see the
appendix. In particular, if $n\!=\!1,2$, $i_1\!\neq\! i_2$
\bea
&&[\hat\kappa(e_i)](\hat{\rm
x})\equiv[\hat{\pi}(e_i)](\hat{\rm x})=\left\la \hat\varphi^{\hat
*}(\hat{\rm x})\Psi_0, e_i\right\ra =\left\la\Psi_0,
\hat\varphi(\hat{\rm x})\hat a^+_i \Psi_0\right\ra
=\hat\varphi_i(\hat{\rm x}),\\
&&\left[\hat\kappa^2_{\pm}\!\left(\!{\scriptscriptstyle\frac
1{\sqrt{2}}} e^{\prime\pm}_{i_1{i_2}}\right)\!\right]\!(\hat{\rm
x},\hat{\rm
y})\equiv\left[\hat{\pi}^2_{\pm}\!\left(\!{\scriptscriptstyle\frac
1{\sqrt{2}}} e^{\prime\pm}_{i_1{i_2}}\right)\!\right]\!(\hat{\rm
x},\hat{\rm y}) \nn
&&\qquad \qquad=\frac 12\left\la\Psi_0,\hat\varphi(\hat{\rm
y})\hat\varphi(\hat{\rm x})\hat a^+_{i_1}\hat a^+_{i_2}\Psi_0\right\ra
={\cal
P}^{2,F}_{\pm}{}^{{j_1}{j_2}}_{{i_1}{i_2}}\hat\varphi_{j_1}(\hat{\rm
x})\hat\varphi_{j_2}(\hat{\rm y})\\ &&\qquad \qquad=\frac
12[\hat\varphi_{i_1}(\hat{\rm x}) \hat\varphi_{i_2}(\hat{\rm y})\!\pm\!
R^{{j_2}{j_1}}_{{i_1}{i_2}}\hat\varphi_h(\hat{\rm x}) \hat\varphi_{j_2}(\hat{\rm
y})]=\frac 12[\hat\varphi_{i_1}(\hat{\rm x}) \hat\varphi_{i_2}(\hat{\rm
y})\!\pm\!\hat\varphi_{i_1}(\hat{\rm y}) \hat\varphi_{i_2}(\hat{\rm x})].
\nonumber
\eea

\subsection{Field equations of motion and Heisenberg picture}
\label{Heispic}

Assume the $n$-particle wavefunction $\psin$ fulfills the
 Schr\"odinger equation (\ref{1Schr}) if $n\!=\!1$,
and
\be \ba{l}
i\hbar \frac{\partial}{\partial t}\psin=\Hans\psin,\qquad \quad
\Hans\!:= \sum\limits_{h=1}^{n}\Haus({\rm x}_h,\partial_{{\rm x}_h},t)
+\sum\limits_{h<k} W(\rho_{hk})\star\ea \label{Schr}
 \ee
if $n\!\ge\! 2$; here the time coordinate $t$  remains ``commuting''.
The class  is sufficently general to include $\star$-deformations of
the Schr\"odinger equations of many physically relevant models: it
includes possible $\star$-local interactions with external
background potential $V({\rm x},t)$ and $U(1)$ gauge potential ${\bf
A}({\rm x},t)$ [$q\!\equiv$electrical charge of the particle] as
well as internal interactions through a two-body potential $W$
depending only on the (invariant) distance
\ $\rho_{hk}\!=\!|{\rm x}_h\!-\!{\rm x}_k|$ \ between
${\rm x}_h,{\rm x}_k$. $\Hans$ will be
hermitean provided $\Hau$ is and $\beta\!\trc\! \Hau\!=\!\Hau$,
as we shall assume. In general (\ref{Schr}) is a
$\star$-differential, pseudodifferential equation, preserving the
(anti)symmetry of $\psin$.
The Fock space Hamiltonian
$$
\Has\!(\varphi)= \int_X\!\!\!\! d\nu({\rm x}) \varphi^{\hat *}({\rm
x})\star\Haus\! \varphi({\rm x})\!\star  + \!\int_X \!\!
\!\!d\nu({\rm x})\!\!\! \int_X \!\!\!\! d\nu({\rm y})
\varphi^{\hat *}({\rm y})\!\star\!\varphi^{\hat *}({\rm x})\!\star\!W(\rho_{{\rm x}{\rm y}})\!\star\!
\varphi({\rm x})\!\star\!\varphi({\rm y})\star
$$
annihilates the vacuum, commutes with the number-of-particles
operator $\mbox{\bf n}\!:=\!a^+_i\!\star a^i$ and its restriction to
$\H^{\ot n}_{\pm}$ coincides with $\Hans$ up to the unitary
transformation $\tilde\kappa^{\ot n}$ [this can be easily checked
using (\ref{wfket})]. As in the undeformed case, in Fock space one
can consider also Hamiltonians not commuting with $\mbox{\bf n}$;
consequently, the latter will no longer be a constant of motion.

We now introduce the evolution operator $U(t)\!:=\!T\left[e^{-\frac
i{\hbar}\! \int_0^t\!\!dt\,\Has}\right]$ and go from the Schr\"odinger to
the {\it Heisenberg picture}. The Heisenberg field $\varphihs({\rm
x},t):= [U(t)]^{*_\star}\varphi({\rm x})U(t)$ fulfills the equal
time commutation relations
\be
[\varphihs({\rm
x},t)\stackrel{\star}{,}\varphihs({\rm y},t)]_{\mp}=h.c.=0, \qquad
\quad [\varphihs({\rm x},t)\stackrel{\star}{,}
\varphihs{}^{*_\star}({\rm y},t)]_{\mp}=\varphi_i({\rm
x})\!\star\!\varphi_i^{*_\star}({\rm y}) \label{Hfccrstar}
\ee
and the evolution equation
\be
i\hbar \frac{\partial}{\partial
t}\varphihs= [\varphihs,\Has].\label{feqstar}
\ee
If  $\Haus$ is
$t$-independent, so is $\Has$, then $\Has(\varphihs)=\Has(\varphi)$,
and (\ref{Hfccrstar}-\ref{feqstar}) can be formulated directly in
the Heisenberg picture as equations in the unknown $\varphihs(t)$.
If $W=0$ (\ref{feqstar}) formally coincides with (\ref{1Schr}),
\be
 i\hbar \frac{\partial\varphihs}{\partial t}= \Haus\varphihs,\label{2ndquant}
\ee
a $\star$-differential equation; following the conventional
terminology we shall call ``second quantization'' the replacement of
the wavefunction $\psi$ by the field operator $\varphihs$. If in
addition $\Haus$ is $t$-independent (in particular, in the free case
$V\!\equiv\! 0\!\equiv\!A^a$), setting $x\!:=\!(t,{\rm x})$ we can
express the Heisenberg field solution of (\ref{2ndquant}) in the form
\be
\varphihs(x)=\varphi_i(x) \star a^{\prime i}
\:\:\quad\mbox{(sum over $i$)}, \qquad\qquad [\kappa^H(t)](e_i)=
\varphi_i({\rm x})e^{-i\omega_i t}=:\varphi_i(x)
\label{Heisfield}
\ee
in terms of an orthormal basis
$\{\varphi_i({\rm x})\}$ of eigenfunctions of $\Haus$ with
eigenvalues $\hbar\omega_i$ (the sum over $i$ actually becomes an
integral over the continuous part of the spectrum).
Now we assume $\F$ is such that $\widehat{\X},\widehat{\D}$,
$\wedge: \X\mapsto\widehat{\X}$ exist and
$V,A^a,W\!\in\!\X$, so that $\hat V\!=\!\wedge(V)$,
$\hat A^a\!=\!\wedge(A^a)$, $\hat W\!=\!\wedge(W)$ are well-defined
(actually, $\hat W\!=\!W$ by the $U\g$-invariance of $\rho_{hk}$).
The map $\hat\kappa^{n,{\scriptscriptstyle H}}_\pm(t)=
\hat\kappa^{n}_\pm\!\circ\! U(t) :\H^{\ot n}_\pm\mapsto\widehat{\X^{\ot
n}_\pm}$ is a $t$-dependent $\widehat{U\g}$-equivariant noncommutative
configuration space realization of $\H^{\ot n}_{\pm}$
(for $n=1$ we denote it by $\hat\kappa^{\scriptscriptstyle H}$).
We shall denote $\hat\kappa^{\scriptscriptstyle H}(e_i)=\hat\varphi_i(\hat
x)$. The analog of (\ref{hatwfket}) are now
$$
\ba{l} \hat\psi_s(\hat{\rm x}_1,\!...,\hat{\rm x}_n,t)\!:=\!
\left\{\hat\pi^n_\pm\left[U(t) s\right]\right\} \!(\hat{\rm
x}_1,\!...,\hat{\rm x}_n)=\frac 1 {\sqrt{n!}} \left\la
\hat\varphi^{\scriptscriptstyle H\hat *}(\hat{\rm x}_n,t)...
\hat\varphi^{\scriptscriptstyle H\hat *}(\hat{\rm x}_1,t)\Psi_0,s\right\ra,\\[10pt]
s\!=\! \hat\Pi^n_\pm[\hat\psi_s(t)] \!:=\!\frac 1 {\sqrt{n!}}\!
\displaystyle\int_{\hat X} \!\! \!\!d\hat\nu(\hat{\rm x}_1\!)...\!\!
\displaystyle\int_{\hat X} \!\!\!\! d\hat\nu(\hat{\rm x}_n)
\hat\varphi^{\scriptscriptstyle H\hat
*}(\hat{\rm x}_1,t)...\hat\varphi^{\scriptscriptstyle H\hat *}
(\hat{\rm x}_n,t)\Psi_0\hat\psi_s(\hat{\rm x}_1,\!...,\hat{\rm
x}_n,t); \ea 
$$
$\hat\psi_s$ fulfills the Schr\"odinger eq.
(\ref{hatqfccr})$_4$ and
$\hat\psi_s(t\!=\!0)\!=\!\hat\pi^n_\pm(s)$.
Replacing
$\hat V,\hat {\bf A},\hat\varphi_i,\hat\psi^{(n)},\int_{\hat X}\!\! d\hat\nu(\hat {\rm x})$  in the previous equations we can
reformulate the latter using only ``hatted'' objects:
\be
\ba{l}\hat\varphi(\hat {\rm x}) =\hat\varphi_i(\hat {\rm x})
\hat a^{\prime i},\qquad \qquad\qquad\qquad\qquad\:\:\varphi^{\hat
*}(\hat {\rm
x}) =\hat a^+_i \hat \varphi_i^{\hat *}(\hat {\rm x}) \\[10pt]
[\hat \varphi(\hat {\rm x}),\hat \varphi(\hat {\rm
y})]_{\mp}=\mbox{h.c.}=0, \qquad \qquad\qquad\quad [\hat \varphi(\hat
{\rm x}),\hat \varphi^{\hat
*}(\hat {\rm y})]_{\mp}=\hat\varphi_i(\hat{\rm
x})\hat\varphi_i^{\hat *}(\hat{\rm y}),\\[10pt]
\hat\Ha^{\scriptscriptstyle(1)}\!=\!\!\frac{-\hbar^2}{2m}\hat D^a\!
\hat D_a\!+\!\hat
V\!,\quad \hat D_a\!=\!\hat \partial_a\!+\! iq\hat A_a,\qquad\hat
\Ha^{(n)}\!=\! \sum\limits_{h=1}^{n}\!\!\hat\Ha^{\scriptscriptstyle(1)}\!
(\hat{\rm x}_h\!,\!\hat\partial_{{\rm x}_h}\!,\!t\!)\!+\!
\!\sum\limits_{h<k}\!\! \hat W(\hat\rho_{hk}),\\[12pt]
i\hbar
\frac{\partial}{\partial t}\hat\psi^{(n)}=\hat\Ha^{(n)}\hat\psi^{(n)}\\[12pt]
\hat\Ha=\!\displaystyle\int_{\hat X} \!\!\!\! d\hat\nu(\hat {\rm x})
\hat \varphi^{\hat *}(\hat {\rm x})\hat\Ha^{\scriptscriptstyle(1)}\!(\hat{\rm
x},\hat\partial_{{\rm x}},t)\hat\varphi(\hat{\rm x}) \! + \!\!\int_{\hat X} \!\! \!\! d\hat\nu(\hat
{\rm x})\!\!\! \int_{\hat X} \!\!\!\!\! d\hat\nu(\hat {\rm y})
\hat\varphi^{\hat *}\!(\hat {\rm y})\hat\varphi^{\hat *}\!(\hat {\rm x})
W(\hat\rho_{{\rm x}{\rm y}}) \hat\varphi({\rm x})\hat\varphi({\rm y}),\\[12pt]
[\hat\varphi^{\scriptscriptstyle H}(\hat{\rm
x},t),\hat\varphi^{\scriptscriptstyle H}(\hat{\rm
y},t)]_{\mp}=\mbox{h.c.}=0, \qquad \qquad
[\hat\varphi^{\scriptscriptstyle H}(\hat{\rm
x},t),\hat\varphi^{\scriptscriptstyle H\hat
*}(\hat{\rm y},t)]_{\mp}= \hat\varphi_i(\hat{\rm
x})\hat\varphi_i^{\hat *}(\hat{\rm y}), \\[10pt]
i\hbar \frac{\partial}{\partial t}\hat\varphi_{\scriptscriptstyle
H}= [\hat\varphi^{\scriptscriptstyle H},\hat\Ha]\\[10pt]
\hat\varphi^{\scriptscriptstyle H}(\hat x) =\hat\varphi_i(\hat x)
\hat a^{\prime i},\qquad \quad\:\:\varphi_{\scriptscriptstyle
H}^{\hat
*}(\hat x) =\hat a^+_i \hat \varphi_i^{\hat *}(\hat  x)
\qquad\qquad\mbox{if $\hat W\!=\!0$ and
$\partial_t\Ha^{(1)}\!=\!0$}\ea\label{hatqfccr}
\ee
[$\hat x,\hat y\!\in\!\{\hat x_1,\hat x_2,...\}$ are two sets 
of noncommutative coordinates fulfilling (\ref{hatDn})]. Summing up: at
least formally, {\bf we can formulate the same quantum theory on
either the commutative or the noncommutative space}: (\ref{hatDn})  \&
(\ref{hatintprop}) \& (\ref{hatqfccr}) summarize a candidate framework for
nonrelativistic {\bf Field Quantization on the noncommutative spacetime
$\b{R}\!\times\!\hat X$ compatible with QM axioms and Bose/Fermi
statistics}, in that it has been obtained by a {\bf Second
Quantization} procedure. Note that the field
commutation relations, in both the Schr\"odinger and Heisenberg
picture, are of the type ``field (anti)commutator=a distribution''.
The framework is not only $\widehat{U\g}$-, but also
$\widehat{U\g'}$-covariant ($\g'$ is the Galilei Lie algebra); to account for
the $t$-dependence $C^1(\b{R},\H),C^1(\b{R},\X)$,... must replace
$\H,\X$,... as carrier spaces of the representations. Now one can
forget how we have got it and investigate case by case
its consistency beyond the level
of formal $\lambda$-power series, using only ``noncommutative
mathematics''.

\medskip
{\bf Remark 1.} In general
(\ref{Schr}) are pseudodifferential (and therefore highly non-local)
equations, but second order as $\star$-differential equations. Similarly
(\ref{feqstar}). However, by (\ref{Trivstar}) the Hamiltonians
$\Haus,\Hans,\Has$ coincide with the local $\Hau,\Han,\Ha$
[defined by (\ref{Schr})
{\it without} $\star$-products] if $V,{\bf A}$ are $H_s$-invariant,
where $H_s$ is the smallest Hopf $*$-subalgebra $H_s\!\subseteq\!U\g$
such that $\F\!\in\!(H_s\ot H_s)[[\lambda]]$ [the
Laplacian $\Delta\!=\!|g|^{-\frac 12}\partial^a\partial_a$ and $W(\rho)$ are
already $U\g$-invariant]; then the dynamics remains undeformed.
If so there is no simplification in treating the dynamics in
the noncommutative setting, although this is possible.
This may become convenient if $\hat V\!\neq\! V$ or $\hat{\bf A}\!\neq\!{\bf A}$
and the dependence of $\hat V,\hat{\bf A}$ on $\hat {\rm x}$ is simpler than
the  dependence of $V,{\bf A}$ on ${\rm x}$, e.g.
are solutions themselves of $\star$-differential equations which are
truly pseudodifferential. See
section \ref{moyalE} and section 3 in \cite{FioWes07} for examples.

\section{Nonrelativistic QM on Moyal-Euclidean space}
\label{moyalE}

$X=\b{R}^3$, $G=ISO(3)$, the twist is \ref{twist}, and we use
the results of subsection \ref{moyal}. It is instructive to see
explicitly how $\wedge^n$, acting on (anti)symmetric
wavefunctions, ``hides'' their (anti)symmetry.
Sticking to $n\!=\!2$, we find e.g.
on the generalized basis of (anti)symmetrized plane waves
$$
\wedge^2\left(e^{iq_1\cdot {\rm x}_1} e^{iq_2\cdot {\rm x}_2} \pm e^{iq_2\cdot {\rm x}_1}
e^{iq_1\cdot {\rm x}_2}\right)=
e^{iq_1\cdot \hat {\rm x}_1} e^{iq_2\cdot \hat {\rm x}_2}e^{\frac i2 q_1\theta q_2}
\pm e^{iq_2\cdot \hat {\rm x}_1} e^{iq_1\cdot \hat {\rm x}_2}e^{-\frac i2 q_1\theta q_2}.
$$
As noted in section 3 of \cite{FioWes07}, the (anti)symmetry remains
manifest if we use coordinates $\xi^a_i,X^a$ of the type
(\ref{defXxi}) with $X^a\!=\!\sum_{i=1}^n{\rm x}_i^a/n$ the coordinates of
the center-of-mass of the system (which are completely symmetric).
By (\ref{simplehat}), the map $\wedge^n$ deforms only the $X$ part
of the wavefunction, leaving unchanged and completely
(anti)symmetric the $\xi$-part. For instance, the previous equation
becomes
$$
\wedge^2\left[e^{i(q_1\!+\!q_2)\cdot X} \left(e^{i(q_2\!-\!q_1)\cdot \xi_1} \pm
e^{-i(q_2\!-\!q_1)\cdot \xi_1}\right)\right] =e^{i(q_1\!+\!q_2)\cdot \hat X}
\left(e^{i(q_2\!-\!q_1)\cdot \hat\xi_1} \pm
e^{-i(q_2\!-\!q_1)\cdot \hat\xi_1}\right)
$$

By Remark 1, if $V,{\bf A}$ are translation invariant (i.e.
constant, in particular vanish) the nonlocality of the Hamiltonians
disappears\footnote{Actually, to this end it is sufficient that
$V,{\bf A}$ are constant along each plane perpendicular to the
vector of components
$\theta^a\!:=\!\varepsilon^{abc}\theta^{bc}/2$.}, and the dynamics
reduces to the undeformed one. Then
formulating the dynamics in the $\star$-deformed, noncommutative
setting brings no formal advantage in solving the equation of
motion. Therefore we consider two very simple choices of
non-constant vector potential (and $A^0\!\equiv\!V\!=\!0$). They
fulfill the Coulomb gauge condition and the free field equation not
only in the standard differential version
$A^0\!=\!\partial_iA^i\!=\!0$, $\partial_\mu F^{\mu i}\!=\!\Box
A^i\!=\!0$, $F^{\mu\nu}\!:=\!\partial^{[\mu} A^{\nu]}$, but also in
the $\star$-differential version \be \ba{l}
A^0\!=\!\partial_i\!\star\! A^i\!=\!0,\quad
\partial_\mu F_\star^{\mu i}\!\!+\!ie
[A_\mu\!\stackrel{\star}{,}\!F_\star^{\mu i}]\!=\!0 ,\quad
F_\star^{\mu\nu}\!:=\!\partial^{[\mu} A^{\nu]}
\!-\!\frac{e^2}2[A^\mu\!\stackrel{\star}{,}\!A^\nu].
\ea\label{stareq}
\ee

\bigskip
{\bf 1. Charged particle in a constant magnetic field ${\bf B}$.} The
simplest gauge choice is $A^i({\rm x})\!=\!\epsilon^{ijk}B^j {\rm x}^k/2$. One
finds
\bea
&&\Haus=-\frac{\hbar^2}{2m}\!\left\{\Delta\!-\!\frac{iq}{\hbar c}
\varepsilon^{abc}B^a{\rm x}^b\partial^c\!-\!\frac{q^2}{4\hbar^2 c^2}
\left[{\bf B}^2 {\rm x}^2\!-\!({\bf B}\cdot{\rm x})^2\right]
\right.\nn&&\left.-\frac{q}{2\hbar c}\varepsilon^{abc}B^a\partial^b
\theta^{cd}\partial^d\!-\!\frac{q^2}{16\hbar^2 c^2}\left[{\bf B}^2
\big[4i({\rm x}\theta\partial)\!+\!(\theta^2)^{ab}\partial^a\partial^b\!
\big]\!-\!4i({\bf B}\!\cdot\!{\rm x})({\rm x}\theta\partial)\!+\!4({\bf B}
\theta\partial)^2\right]
\right\}\quad~\nonumber
\eea
(we have displayed the undeformed Hamiltonian in the first line,
the corrections in the second line),
so in $\X$ the Schr\"odinger equation is still differential of
second order, but more complicated than in the commutative case. We
show that in terms of ``hatted'' objects it can be formulated and
solved as in the undeformed case.
We choose the ${\rm x}^3$-axis so that  $q{\bf B}=qB\vec{k}$ with $qB>0$; this gives
$$
\ba{l}
\hat D^3\!=\!\hat\partial^3\!,\:\:\:\hat D^1\!=\!\hat\partial^1\!-\!i\beta
\hat {\rm x}^2\! ,\:\:\:\hat D^2\!=\!\hat\partial^2\!+\!i\beta\hat {\rm x}^1
\qquad\Rightarrow\qquad[\hat\partial^3\!,\!\hat D^a]\!=\!0,\:\:\:
[\hat D^1\!,\!\hat
D^2]\!=\!i2\beta[1\!-\!\frac{\beta\theta^{12}}{2}],
\ea
$$
where $a\!=\!1,2$ and $\beta\!:=\!qB/2\hbar c$. We assume
$qB\theta^{12}\!<\!4\hbar c$. Then one finds \bea &&\ba{l}
a\!:=\!\alpha[\hat D^1\!\!-\!i\hat D^2], \:\:
\alpha\!:=\!\!\left[\!4\beta(1\!-\!\frac{\beta\theta^{12}}2)\!\right]^{\frac{-1}2}
\qquad\Rightarrow\qquad a^*\!=\!-\!\alpha[\hat D^1\!\!+\!i\hat D^2],
\:\:[a,a^*]=1,
\ea \qquad\\[8pt]
&&\ba{l}
\hat\Hau\!=\!\frac{-\hbar^2}{2m}\hat D^i\! \hat D^i
\!=\!\frac{-\hbar^2}{2m}\left[(\hat\partial^3)^2\!-\!\frac 1{2\alpha^2}
(aa^*+a^*a)\right]\!=\!\hat\Hau_\parallel\!+\!\hat\Hau_\perp\\[8pt]
\hat\Hau_\parallel\!:=\!\frac{(-i\hbar\hat\partial^3)^2}{2m},\qquad
\hat\Hau_\perp\!:=\!\hbar\omega\left(a^*a\!+\!\frac 12\right),\qquad
\omega\!:=\!\frac {qB}{mc}\left(1\!-\!\frac{\beta\theta^{12}}2\right)
\ea\eea
As $\hat\partial^3$ commutes with $a,a^*$,
the operators $\hat\Hau_\parallel,\hat\Hau_\perp$
commute with each other. The first is as on the commutative space,
and has continuous spectrum $[0\!,\!\infty[$; the generalized
eigenfuntions are the eigenfuntions $e^{ik\hat {\rm x}^3}$ of
$p^3=-i\hbar\hat\partial^3$ with eigenvalue $\hbar k$. The second is
an harmonic oscillator Hamiltonian with $\omega$ modified
by the noncommutativity through $\theta^{12}$
(but not $\theta^{13},\theta^{23}$).
So the spectrum of $\hat\Hau$
is the set of  $E_{n,k^3}\!=\!\hbar\omega (n\!+\!1/2)+(\hbar k^3)^2/2m $.

To find a basis of eigenfunctions we define in analogy with the undeformed case
$$
\ba{l} \hat z\!:=\!\sqrt{\!\frac
{\beta}2}(\hat {\rm x}^1\!\!+\!i\hat {\rm x}^2),\quad
\partial_{\hat z}\!:=\!\frac1{\sqrt{2\beta}} (\hat\partial_1\!-\!i
\hat\partial_2),\qquad\Rightarrow\qquad
\hat z^*\!=\!\sqrt{\!\frac
{\beta}2}(\hat {\rm x}^1\!-\!i\hat {\rm x}^2),
\quad\partial_{\hat z}^* \!=\!-\partial_{\hat z^*} \ea
$$
and find that the only nontrivial commutators among $\hat z,\hat z^*,
\partial_{\hat z},\partial_{\hat z^*}$ are
$$
\ba{l} [\partial_{\hat z},\hat z]=1,\qquad [\partial_{\hat z^*},\hat
z^*]=1,\qquad[\hat z,\hat z^*]\!=\!\beta\theta^{12}.
\ea
$$
We can thus re-express $a,a^*$ in the form
$$
a= \alpha\sqrt{2\beta}(\hat
z^*\!+\!\partial_{\hat z}), \qquad \quad  a^*= \alpha\sqrt{2\beta}
(\hat z\!-\!\partial_{\hat z^*}).
$$
Setting $\hat l^3\!:=\!\hat z\partial_{\hat z}\!-\!\hat z^*\partial_{\hat z^*}
\!-\!\beta\theta^{12}\partial_{\hat z}\partial_{\hat z^*}$
and ${\rm n}\!:=\!a^*a$ we also find
\be
[\hat l^3,\hat z^*]=-\hat z^*
,\qquad [ \hat l^3,\hat z]=\hat z,
\qquad [\hat l^3,a^*]= a^*,\qquad [ \hat l^3,a]=- a,
\qquad [ \hat l^3,{\rm n}]=0.
\ee
The existence of
``ground state'' eigenfunctions $\hat\psi_0$ characterized by the
condition $a\hat\psi_0=0$ is proved as in the commutative case. As
$[a,\hat z^*]=0$, if
$\hat\psi_0$ fulfills this condition, so does $(\hat z^*)^r\hat\psi_0$
for all $r\!\in\!\b{N}$; moreover, if $\hat l^3\hat\psi_0=m\hat\psi_0$,
then $\hat l^3(\hat z^*)^r\hat\psi_0=(m\!-\!r)(\hat z^*)^r\hat\psi_0$. In particular we find
\be
\hat\psi_{0,0}(\hat z^*,\hat z)\!:=\!\int\!\!dk\,dk^*e^{ik\hat z^*}
e^{ik^*\hat z} e^{-kk^*} \qquad\Rightarrow\qquad
a\hat\psi_{0,0}=0=\hat l^3\hat\psi_{0,0}.
\ee
In analogy with the undeformed case we choose as a complete set of commuting
observables $\{p^3,{\rm n} ,\hat l^3\}$. The deformed Landau eigenfunctions
\be
\hat\psi_{k^3,n,m}(\hat {\rm x})=(a^*)^n(\hat z^*)^{n\!-\!m}\hat\psi_{0,0}
(\hat z^*,\hat z)e^{ik^3\hat {\rm x}^3}
\ee
are generalized eigenfunctions with eigenvalues $p^3\!=\!\hbar k^3\!\in\!\b{R}$,
$,{\rm n}\!=\!n\!=\!0,1,...$, $\hat l^3\!=\!m\!=n,n\!-\!1,...$ and build up
an orthogonal basis of of $\L^2(\b{R}^3)$.
They are also eigenfunctions of $\hat\Hau$ with eigenvalues
$E_{n,k^3}\!=\!\hbar\omega (n\!+\!1/2)+(\hbar k^3)^2/2m $.
Replacing $\hat {\rm x}^a\!\to\! {\rm x}^a\star$ and performing all the $\star$-products
one finds the corresponding eigenfuntions
$\psi_{k^3,n,m}({\rm x})$ of $\Haus$. To contain the size of this paper
we don't consider multi-particle systems here.

\bigskip
 {\bf 2. Charged particle in a plane wave electromagnetic field.}
$A^a({\rm x})\!=\!\varepsilon^a({\rm p}) \exp[-ip\cdot {\rm x}]
\!\equiv\!\varepsilon^a({\rm p}) \exp[i({\rm p}\cdot{\rm
x}\!-\!|{\rm p}| t)]$, (the amplitude vector fulfilling
$\varepsilon^a({\rm p}){\rm p}^a\!=\!0$). To check (\ref{stareq})
it is useful to use the properties
$$
 e^{i{\rm p}\cdot{\rm x}}\star
f({\rm x})\!=\! e^{i{\rm p}\cdot{\rm x}} f({\rm x}\!+\!\theta{\rm
p}/2) \qquad\Rightarrow\qquad e^{i{\rm p}\cdot{\rm x}}\star
e^{ia{\rm p}\cdot{\rm x}}= e^{i{\rm p}\cdot{\rm x}} e^{ia{\rm
p}\cdot{\rm x}}
$$
where $(\theta{\rm
p})^a\!:=\!\theta^{ab}{\rm p}^b$,  as ${\rm p}\theta{\rm p}\!=\!0$.
The Schr\"odinger equation (\ref{Schr}) for $n\!=\!1$ particle
becomes
$$
i\hbar \partial_t\psi\!({\rm x},t)
=\frac{-\hbar^2}{2m}\!\left[\!\Delta\psi\!({\rm x},t)\!+\!2iq
e^{-ip\cdot x}\varepsilon^a\hat\partial_a\psi\!\left(\!{\rm x}\!+\!
\frac{\theta{\rm p}}2,t\!\right)\!-\!q^2e^{-2ip\cdot x}
|\varepsilon|^2\psi\!({\rm x}\!+\!\theta{\rm p},t)\right];
$$
the nonlocality induced by the $\star$-product is here particularly
simple, in that it involves the wavefunction at points ${\rm x},{\rm
x}\!+\!\theta{\rm p}/2,{\rm x}\!+\!\theta{\rm p}$ related by the
constant shift $\theta{\rm p}/2$. As in the undeformed case, one can
approximate the dynamics by perturbative methods.

\section{Relativistic second quantization}
\label{R2ndQ}

Now we adopt as a starting point
Minkowski spacetime (i.e. $\b{R}^4$ endowed with
Minkowski metric) as a pseudo-Riemannian manifold $X$; we shall
denote as $x^\mu$ ($\mu=0,1,2,3$) the coordinates w.r.t. a fixed
inertial frame. The isometry group is the Poincar\'e Lie group,
whose Lie algebra we denote as $\P$; we denote as $P_\mu,M_{\mu\nu}$
the generators of spacetime translations and Lorentz transformations
respectively. A relativistic particle is described choosing as the
algebra of observables $\O=H=U{\cal P}$ and as the Hilbert space
$\bH$ the completion of a pre-Hilbert space $\H$ carrying an
irreducible $*$-representation of $U{\cal P}$ charaterized by a
nonnegative eigenvalue $m^2$ of the Casimir $P^\mu P_\mu$ and a
nonnegative spectrum for $P^0$. For simplicity we stick to the case
of a scalar particle (i.e. the Pauli-Lubanski Casimir vanishes) of
positive mass $m$. Normalizable states $s$, in particular the
vectors of an orthonormal basis $\{e_i\}_{i\!\in\!\b{N}}$ of $\H$,
can be decomposed as combinations (integrals) of generalized
eigenstates $e_{{\rm p}}$ of the operators $P_\mu$ with eigenvalues
$p_\mu$ [$(p^a)\!\equiv\!{\rm p}\!\in\!\b{R}^3$ and
$p^0\!\equiv\!\sqrt{{\rm p}^2+m^2}\!>\!0$]: $s\!=\!\int\!\!
d\mu(p)e_{{\rm p}}\tilde\psi_s({\rm p})$, where $d\mu({\rm p})=d^3{\rm
p}/2p^0$ is the Poincar\'e invariant measure. Fixing the
normalization of $e_{{\rm p}}$ by setting as usual $(e_{\rm
p},e_{\rm q})\!=\!2p^0\delta^3({\rm p}\!-\!{\rm q})$, the scalar
product is expressed as
$$
(s,v)=\int\!\! d\mu({\rm p})\tilde\psi_s^*({\rm p})\tilde\psi_v({\rm p}).
$$
The $\P$-action on $\H$ reads
$$
P\!_\mu\trc s\!=\!\!\!\int\!\!\! d\mu({\rm p})\, e_{{\rm p}}\,
p_\mu\tilde\psi_s({\rm p}),\qquad M_{\mu\nu}\trc
s\!=\!\!\!\int\!\!\! d\mu({\rm p})\, e_{{\rm p}}\,
i\!\left[p_\mu\partial_{p^\nu}\!- p_\nu\partial_{
p^\mu}\!\right]\!\tilde\psi_s({\rm p}),
$$
(where we have to replace $\partial_{p^0}\!\to\!0$, as $p^0$ is no
more and independent variable). As a $\H$ we take the space of
vectors $s$ for which $\tilde\psi_s\!\in\!{\cal S}(\b{R}^3)$;
clearly $s$ are in the domain of all elements of $H$.
 We denote by $a_i,a^+_i$ the creation and annihilation operators
corresponding to $e_i$ and $a^+_{\rm p},a^{\rm p}$ the generalized
ones corresponding to $e_{{\rm p}}$\footnote{In order to uniquely
fix the signs in the definitions of the ``generalized bases'' ${\cal
B}^n_{\pm}$ and of the creation/annihilation operators one needs to
choose a complete ordering within the set of labels ${\rm
p}\!\in\!\b{R}^3$, since no natural one is available: one can
declare e.g. ${\rm p}\!>\!{\rm p}'$ if  $p_1\!>\!p'_1$, or
$p_1\!=\!p'_1$ and $p_2\!>\!p'_2$, or $p_1\!=\!p'_1$ and
$p_2\!=\!p'_2$ and $p_3\!>\!p'_3$. As known, the physical results
are independent of this choice.}; the former fulfill
(\ref{lineartransf}), (\ref{ccr}), the latter  $P_\mu\trc a^+_{\rm
p}\!=\!p_\mu a^+_{\rm p}$, $P_\mu\trc a^{\rm p}\!=\!-p_\mu a^{\rm
p}$ and
\be
[a^{\rm p},a^{\rm q}]=0,\qquad\quad [a^+_{\rm
p},a^+_{\rm q}]=0,\qquad\quad [a^{\rm p},a^+_{\rm q}]
=2p^0\delta^3({\rm p}-{\rm q}). \label{ccr'}
\ee
A $t$-dependent
configuration space realization (Heisenberg picture) is obtained
setting $\kappa^{\scriptscriptstyle H}(e_{{\rm p}})\!=\! e^{-ip\cdot
x}$, $\tilde\kappa^{\scriptscriptstyle H}(P_\mu)\!=\!i\partial_\mu$,
$\tilde\kappa^{\scriptscriptstyle H}(M_{\mu\nu})\!=\!
i(x_\mu\partial_\nu\!-\!x_\nu\partial_\mu)$, and the ``on-shell''
condition $\big(P^\mu P_\mu\!-\!m^2\big)s\!=\!0$ becomes on
$\psi_s\!=\!\kappa^{\scriptscriptstyle H}(s)\!=\!\int\!\!
d\mu(p)e^{-ip\cdot x}\tilde\psi_s({\rm p})$ the Klein-Gordon
equation $(\Box\!+\!m^2)\psi_s\!=\!0$. So
$\X\!:=\!\kappa^{\scriptscriptstyle H}(\H)$
is the pre-Hilbert space of normalizable, smooth,
positive energy
solutions of the K-G equation (these functions depend both on space
and time). We denote $\kappa^{\scriptscriptstyle H}(e_i)=\varphi_i$
and $\Phi^e=\A^{\pm}\!\ot\!\left(\bigotimes_{i=1}^{\infty}\! \X'\right)$.
The hermitean relativistic free field (in the Heisenberg picture) is
$\varphi(x)\!=\!\varphi_i(x)a^i
\!+\!a_i^+\!\varphi_i^*(x)=\!\int\!\!d\mu(p)\, [e^{-ip\cdot x} a^{
{\rm p}} \!+\!a_{\rm p}^+ e^{ip\cdot x}]$; it is basis-independent,
$H$-invariant and fulfills the Klein-Gordon equation. The
`crossed' (anti)symmetric projectors
$\pi^n_\pm,\Pi^n_\pm$ of fig. \ref{diag} are defined replacing in
(\ref{wfket}) the scalar product integral (\ref{scalprod1}) by the one
(\ref{relscalprod}) and $\varphi^*$ by the creation part of $\varphi$.

Fixed a twist $\F\!\in\!(U\P\ot U\P)[[\lambda]]$ (examples are in
next subsection and in \cite{BorLukTol}), we can apply the
associated deforming procedure to the whole setting. The deformed
annihilation/creation operators associated to
$\{e_i\}_{i\!\in\!\b{N}}$ will fulfill the commutation relations
(\ref{hqccr}). By (\ref{Trivstar}), $\Box\star\omega=\Box\omega$ for
all $\omega\in\Phi^e$ (as the d'Alembertian is $U\P$-invariant), so
the free field equation and its solutions are not deformed. By Lemma
\ref{lemmainv} one can express $\Box$ in the form
$\Box\!=\!\partial_\mu\star\partial^{\prime \mu}$ ($\partial^{\prime
\mu}\!:=\!S(\beta )\trc\partial^\mu$) and $\varphi$ in the form
\be
\varphi(x)=\varphi_i(x)\!\star\!a^{\prime i}
\!+\!a_i^+\!\star\!\varphi_i^{\hat
*}(x)=\!\! \int \!\!\!d\mu(p) [e^{-ip\cdot x}\!\star\! a^{\prime {\rm p}}
\!+\!a_{\rm p}^+ \!\star\! e^{ip\cdot x}\,]  \label{fielddeco}
\ee
and the free field commutation relation in the form
$$
[\varphi(x)\stackrel{\star},\varphi(y)]=
\varphi_i(x)\star\varphi_i^{*_\star}(y)\!-\!\varphi_i(y)\star
\varphi_i^{*_\star}(x)= \int\!
d\mu(p)\left[e^{-ip\!\cdot \!(x\!-\!y)}\!-\!e^{ip\!\cdot
\!(x\!-\!y)}\right]
$$
for any $x,y\!\in\!\{x_1,x_2,...\}$. As known, this vanishes if
$x\!-\!y$ is space-like (microcausality). Provided $\F$ allows the
definition of $\widehat{\X},\widehat{\D}$ and
$\wedge\!:\!\D\!\mapsto\!\widehat{\D}$,
and going to the ``hat notation'' we find
for any two sets $\hat x,\hat y\!\in\!\{\hat x_1,\hat x_2,...\}$
of noncommutative coordinates fulfilling (\ref{hatDn})
\be
\ba{l} \hat\varphi(\hat x)=\hat\varphi_i(\hat x)\hat
a^{\prime i} \!+\!\hat a_i^+\hat\varphi_i^{\hat
*}(\hat x)=\! \displaystyle\int \!\!\!d\mu(p) \left[\wedge(e^{-ip\cdot x})
\,\hat a^{\prime {\rm p}}
\!+\!\hat a_{\rm p}^+  \wedge\!(e^{ip\cdot x})\right] ,\\[8pt]
 [\hat\varphi(\hat x),\hat\varphi(\hat y)]= \hat\varphi_i(\hat
x)\hat\varphi_i^{\hat
*}(\hat y)\!-\!\hat\varphi_i(\hat y)\hat\varphi_i^{\hat
*}(\hat x),\\[8pt]
(\hat\Box+m^2)\hat\varphi(\hat x)=0.      \ea \label{freecomm}
\ee
All $\hat\varphi(\hat x_h)$ belong to the $\hH$-module $*$-algebra $\hat\Phi^e$.
$\hat\Phi^e$ is generated by the $\hat a^+_i,\hat a^{\prime i}$ and the sets
of noncommutative coordinates $\hat x_1,\hat x_2,...$
fulfilling (\ref{hatDn}), (\ref{hqccr}), (\ref{earel}); the latter
relations provide our answer to the issues {\it a), b), c)} mentioned in the introduction.

\subsection{Relativistic QFT on Moyal-Minkowski space}
\label{MMQFT}

Taking (\ref{twist}) as a twist (with Latin letters
replaced by Greek ones as indices) one obtains Moyal-Minkowski
noncommutative spacetime and the twisted Poincar\'e Hopf algebra
$\hH=\widehat{U\P}$ of
\cite{ChaKulNisTur04,Wes04,Oec00}\footnote{In section 4.4.1
of \cite{Oec00} this was formulated in terms of the dual Hopf
algebra of $\hH$.},
$$
\hat\Delta (P\!_\mu)=P\!_\mu\!\ot\!\1\!+\!\1\!\ot\! P\!_\mu=
\Delta(P\!_\mu),\qquad\quad\hat\Delta (M_\omega)=
M_\omega\!\ot\!\1\!+\!\1\!\ot\! M_\omega\!+\!
[\omega,\!\theta]^{\mu\nu}P\!_\mu \!\ot\!
P\!_\nu\neq\Delta (M_\omega);
$$
here we have set $M_\omega\!:=\!\omega^{\mu\nu}M_{\mu\nu}$.
It is convenient to
write formulae (\ref{hqccr}), (\ref{defA}), (\ref{Aavec}) for
generalized creation \& annihilation operators:
 \bea
&&\ba{l}
 a^{+}_{\rm p} \!\star\! a^{+}_{\rm q}= e^{-i
 p\theta q}\,     a^{+}_{\rm q} \!\star\! a^{+}_{\rm p}, \\[8pt]
a^{\rm p}\!\star\!  a^{\rm q}=
e^{-i p\theta q} \,    a^{\rm q} \!\star\!  a^{\rm p}, \\[8pt]
a^{\rm p}\!\star\!  a^{+}_{\rm q}=e^{i p\theta q} \,  a^{+}_{\rm q}
\!\star\! a^{\rm p}\!+\!
2p^0\delta^3({\rm p}\!-\!{\rm q})\\[8pt]
a^{\rm p}\!\star\! e^{iq\cdot x} =e^{-ip\theta q} \,e^{iq\cdot x}\!\star\!
a^{\rm p},
\quad \& \mbox{ h.c.},
\ea
\qquad \Leftrightarrow\qquad
\ba{l}\hat a^+_{\rm p} \! \hat a^+_{\rm q}= e^{-i
p\theta\! q}\,     \hat a^+_{\rm q} \! \hat a^+_{\rm p},\\[8pt]
\hat a^{\rm p}  \hat a^{\rm q}\!=\! e^{-i p\theta\! q}     \hat a^{\rm q}
\! \hat a^{\rm p},\\[8pt]
\hat a^{\rm p}\! \hat a^+_{\rm q}\!=\!
e^{i  p\theta\! q}   \hat a^+_{\rm q} \! \hat a^{\rm p}\!+\!2
p^0\delta^3\!({\rm p}\!-\!{\rm q}),\\[8pt]
\hat a^{\rm p} e^{iq\cdot \hat x} =e^{-ip\theta q} \,e^{iq\cdot \hat x}
\hat a^{\rm p},\quad \& \mbox{ h.c.} ;
\ea\quad\label{aa+cr}\\[8pt]
&& \check a^+_{\rm p}\equiv D_{\f}^{\sigma}\left(a^+_{\rm p}\right)=
a^+_{\rm p}e^{-\frac i2 p\theta\sigma(P)}, \qquad \qquad \check a^{\rm p}\equiv
D_{\f}^{\sigma}\left(a^{\rm p}\right)=a^{\rm p}e^{\frac i2 p\theta\sigma(P)}
\label{defA"}\\
&&\hat a^+_{{\rm p}_1}...\hat  a^+_{{\rm p}_n}\Psi_0=
a^+_{{\rm p}_1}\star...\star a^+_{{\rm p}_n}\Psi_0=
\check a^+_{{\rm p}_1}...\check a^+_{{\rm p}_n}\Psi_0=
\exp\!\left[-\frac i2\!\sum\limits_{j,k=1 \atop j<k}^n p_j\theta p_k\right]\!
a^+_{{\rm p}_1}...a^+_{{\rm p}_n}\Psi_0 \label{genstates}
\eea
where, according to definition (\ref{jordan}),
$\sigma(P_\mu)=\int \!d\mu(p)\,p_\mu  a^+_{\rm p} a^{\rm p}$.
By (\ref{genstates}) generalized states differ from their undeformed
counterparts only by multiplication by a  phase factor.
As $\check a^+_{\rm p}\check a^{\rm p}\!=\!a^+_{\rm p}a^{\rm p}$,
$\sigma(P_\mu)\!=\!\int \!d\mu(p)\,p_\mu \check a^+_{\rm p}\check
a^{\rm p}$, from (\ref{defA"}) the inverse of $D_{\f}^{\sigma}$ is
readily obtained. Formulae (\ref{freecomm}) become
\be
\ba{l} \hat\varphi(\hat x)=\hat\varphi_i(\hat x)\hat
a^{\prime i} \!+\!\hat a_i^+\hat\varphi_i^{\hat
*}(\hat x)=\! \displaystyle\int \!\!\!d\mu(p) \left[e^{-ip\cdot \hat x}
\,\hat a^{\prime {\rm p}}
\!+\!\hat a_{\rm p}^+  e^{ip\cdot \hat x}\right] ,\\[8pt]
[\hat\varphi(\hat x),\hat\varphi(\hat y)]
=2\displaystyle\int \!\!\! d\mu(p)\:
\sin\left[p\!\cdot \!(\hat x\!-\!\hat y)\right],\\[8pt]
(\hat\Box+m^2)\hat\varphi(\hat x)=0.      \ea \label{freecommmoyal}
\ee
The rhs(\ref{freecommmoyal})$_2$ is like the undeformed one.

Incidentally, by an explicit computation one can easily show the analog of
(\ref{scalprod1}), i.e. that the realization of the scalar product in
($t$-dependent) configuration space holds also on the
Moyal-Minkowski space:
\be
(s,v)=i\!\!\int\!\!\!d^3\!x
\,[\psi_s^*\partial^0\psi_v\!-\!\big(\partial^0\psi_s^*\big)\psi_v]=
i\!\!\int\!\!\!d^3\!x \,\left[\psi_s^{*_\star}\star\partial^0\psi_v
\!-\!\big(\partial^0\psi_s^{*_\star}\big)\star\psi_v\right];
\label{relscalprod}
\ee
one just needs to recall that $\beta\!=\!\1$
implies $*_\star\!=\!*$, replace the plane-wave expansions of
$\psi_s,\psi_v$ in (\ref{explstarprod}) and note that integrating in
$d^3x$ gives a $\delta^3({\rm h}\!-\!{\rm k})$ and hence $h\theta k\!=\!0$.

As seen, in terms of generalized wavefunctions, creation, annihilation operators
the relations (\ref{hatDn}), (\ref{hqccr}), (\ref{earel})
characterizing $\hat\Phi^e$ become (\ref{hatDmoyal}), (\ref{aa+cr});
the latter provide our answer on the Moyal-Minkowski space to
the issues {\it a), b), c)} mentioned in the introduction.
It is remarkable that the free field  (\ref{fielddeco}) \&
(\ref{aa+cr}) coincides with the one found in formulae (37) \& (46)
of \cite{FioWes07} [see also formulae (32) \& (36) of
\cite{Fio08Proc}] imposing just the free field equation and Wightman
axioms (modified only by the requirement of {\it twisted} Poincar\'e
covariance). The present construction shows that
such a field is compatible with ordinary Bose/Fermi statistics - a
point only briefly mentioned in \cite{FioWes07,Fio08Proc}. In
\cite{FioWes07} it has been also shown that the $n$-point functions
of a (at least scalar) field theory, when expressed as functions of
coordinates differences $\xi$, coincide with the undeformed ones. To
a large extent this is due to (\ref{simplehat}). This result holds
in time-ordered perturbation theory also for interacting
fields with interaction $\varphi^{\star n}$, due to the translation
invariance of the latter.

In our notation essentially all sets of relations recently appeared in
\cite{ChaPreTur05,Tur06,BalManPinVai05,AkoBalJos08,
BuKimLeeVacYee06,Zah06,LizVaiVit06,Abe06,FioWes07,Fio08Proc,
RicSza07,AscLizVit07,GroLec07} can be
summarized as $\hat x^{h}_i{}^{\hat *}\!=\!\hat x^{h}_i$,
$\hat \partial_{x_i^h}{}^{\hat *}\!=\!-\hat \partial_{x_i^h}$ and
$$
\ba{l}
[\hat x^{h}_i,\hat x^{k}_j] =\1 i\theta^{hk}(\eta_1\!+\!\eta_2\delta_j^i),
\qquad\qquad
[\hat \partial_{x_i^h},\hat x^{k}_j]=\1\delta^k_h\delta_j^i,\qquad\qquad
[\hat \partial_{x_i^h},\hat \partial_{x^{k}_j}]=0,\\[9pt]
\hat a^+_{\rm p} \! \hat a^+_{\rm q}= e^{-i
p\tilde	\theta\! q}\,     \hat a^+_{\rm q} \! \hat a^+_{\rm p},\qquad\quad
\hat a^{\rm p}  \hat a^{\rm q}\!=\! e^{-i p\tilde\theta\! q}     \hat a^{\rm q}
\! \hat a^{\rm p},\qquad\quad
\hat a^{\rm p}\! \hat a^+_{\rm q}\!=\!
e^{i  p\tilde\theta\! q}   \hat a^+_{\rm q} \! \hat a^{\rm p}\!+\!2
p^0\delta^3\!({\rm p}\!-\!{\rm q}),\\[9pt]
\hat a^{\rm p} e^{iq\cdot \hat x} =e^{-i\eta_3 p\theta q} \,e^{iq\cdot \hat x}
\hat a^{\rm p},\qquad\qquad \hat a^+_{\rm p} e^{iq\cdot \hat x} =e^{i\eta_3 p\theta q} \,e^{iq\cdot \hat x}
\hat a^+_{\rm p},
\ea
$$
the choices of the parameters $\tilde\theta{}^{\mu\nu},\eta_1,\eta_2,\eta_3$
specifying the differences. Our relations (\ref{hatDmoyal}), (\ref{aa+cr}),
like (46) of \cite{FioWes07}  and (36) of \cite{Fio08Proc},
correspond to $\tilde\theta\!=\!\theta, \eta_1\!=\!\eta_3\!=\!1,\eta_2\!=\!0$.
In all other cases $\eta_3\!=\!0$, implying an answer to
the issues {\it a), b), c)} in the introduction different
from the present one.
The choice $\tilde\theta=0$ gives the canonical (anti)commutation relations
for the creation and annihilation operators and is assumed in most
papers which do not twist the Poincar\'e covariance group, both
in operator (see e.g.  \cite{DopFreRob95} and \cite{GroLec07}) 
and implicitly in
path-integral approach to quantization (see e.g. \cite{Fil96}),
together with $\eta_1\!=\!\eta_3\!=\!0,\eta_2\!=\!1$.
Chaichian and coworkers (see e.g. \cite{ChaPreTur05,Tur06}) do
twist the Poincar\'e covariance group and all the spacetime coordinates
adopting $\eta_1\!=\!1,\eta_2\!=\!\eta_3\!=\!0$, but not
the creation and annihilation operators ($\tilde\theta\!=\!0$).
Balachandran and coworkers (see \cite{BalManPinVai05} and e.g. the
review \cite{AkoBalJos08}) adopt $\tilde\theta\!=\!-\theta$ and
$\eta_1\!=\!\eta_3\!=\!0,\eta_2\!=\!1$
(equivalently, they do not perform the $\star$-product between
functions of different sets $x,y$ of coordinates).
Ref. \cite{LizVaiVit06,Abe06,RicSza07} adopt $\tilde\theta\!=\!-\theta$  and
$\eta_1\!=\!1,\eta_2\!=\!\eta_3\!=\!0$ in scalar field theories
respectively in 1+1 and in arbitrary dimension (\cite{Abe06}
only with $\theta^{0i}=0$). This is the only other choice leading
to the ``local'' free field commutation relation (\ref{freecommmoyal}),
and appears also in an alternative proposal
[formulae (44) instead of (46)] contained in
\cite{FioWes07}\footnote{However in \cite{Fio08Proc}
we have put aside this alternative because scalar products
cannot be expressed in terms of Wightman functions defined as vacuum
expectation values of $\star$-products of fields.}. The other choices lead to
``non-local'' free field commutation relation.
Ref. \cite{BuKimLeeVacYee06} adopts $\tilde\theta\!=\!\theta$ and
$\eta_1\!=\!\eta_2\!=\!\eta_3\!=\!0$ (only creation and
annihilation operators are twisted).
In the prescription of \cite{AscLizVit07}  creation and
annihilation are  $\star$-multiplied as in the first three left equations
of (\ref{aa+cr}), what corresponds again to $\tilde\theta\!=\!\theta$ 
(it is not clear what the remaining commutation relations are, 
as they consider a different kind of $\star$-commutator 
$[\cdot,\cdot]_\star:=[\cdot,\cdot]\circ \bF\trc^{\ot 2}$).
Ref. \cite{Zah06} considers
$\eta_1\!=\!1,\eta_2\!=\!\eta_3\!=\!0$, $\tilde\theta\!=\!0$
(the creation and annihilation operators are not twisted).

A realization in the form (\ref{defA"}) of generalized creation \&
annihilation operators fulfilling (\ref{aa+cr})$_2$ has  appeared in
\cite{BalManPinVai05}. It is also reminiscent of the Fock space
realization \cite{Kul81} of the Zamolodchikov-Faddeev \cite{ZamZam79}
algebra, which is generated by deformed creation/annihilation
operators of scattering states of some completeley integrable
1+1-dimensional QFT.


\section{Appendix}

In this appendix we collect the proofs of several statements made
in the previous
sections and miscellaneuous formulae needed for that.

We start by
writing in compact notation (\ref{cocycle}) and its consequences
\be\ba{ll}
\F_{12}\F_{(12)3}=\F_{23}\F_{1(23)} \qquad\qquad &
\F_{(12)3}\F_{(123)4}=\F_{34}\F_{(12)(34)},\\[8pt]
\F_{1(23)}\F_{(123)4}=\F_{(23)4}\F_{1(234)}\qquad\qquad &
\F_{12}\F_{(12)(34)}=\F_{2(34)}\F_{1(234)},\ea \label{bla0}
\ee
as well as the  inverse of (\ref{cocycle}) and its consequences
\be\ba{ll}
\bF_{(12)3}\bF_{12}=\bF_{1(23)}\bF_{23}, \qquad\qquad &
\bF_{(123)4}\bF_{(12)3}=\bF_{(12)(34)}\bF_{34},\\[8pt]
\bF_{(123)4}\bF_{1(23)}=\bF_{1(234)}\bF_{(23)4},\qquad\qquad &
\bF_{(12)(34)}\bF_{12}=\bF_{1(234)}\bF_{2(34)},\ea \label{bla}
\ee
obtained applying $\Delta$ on the first, second, third tensor factor
and taking into account the cocommutativity of $\Delta$;
the bracket encloses tensor factors
obtained from one by application of $\Delta$. To denote the
decomposition of $\F_{(12)3}$ we use a Sweedler-type notation
$$
\F_{(12)3}\equiv (\Delta\ot\id)(\F)=\sum_I\F^{(1)}_{(1)I}\ot \F^{(1)}_{(2)I}
\ot\F^{(2)}_I,
$$
and similarly for $\F_{1(23)},\bF_{(12)3}...$.
We denote as
$S_i$ the antipode on the $i^{th}$ tensor factor,
as $\tau_{ij}$ and $m_{ij}$ respectively the flip and the
multiplication of the $i^{th},j^{th}$ tensor factors.

Applying $m_{23}\!\circ\!S_3$,
$m_{12}\!\circ\!S_2\!\circ\!\tau_{12}$ to (\ref{bla0})$_1$,
$m_{12}\!\circ\!S_1$, $m_{23}\!\circ\!S_2\!\circ\!\tau_{23}$ to
(\ref{bla})$_1$, and recalling (\ref{defbeta}), (\ref{defbeta'}) we
respectively obtain the important relations (see e.g. also Lemma 1
in \cite{Fio97}) \bea
\F^{-1}&=& \sum_I\F^{(1)}_{(1)I}\ot \F^{(1)}_{(2)I} S
\left(\F^{(2)}_I\right)\beta^{-1} = \F^{(2)}_{(1)I} S
\left(\F^{(1)}_I\right)S(\beta^{-1})\ot \F^{(2)}_{(2)I}, \label{dritto4}\\
\F &=& \sum_I\beta\, S \!\left(\bF^{(1)}_I\right) \bF^{(2)}_{(1)I}\ot
\bF^{(2)}_{(2)I}= \bF^{(1)}_{(1)I}\ot S(\beta)\, S\!
\left(\bF^{(2)}_I\right)
\bF^{(1)}_{(1)I}. \label{dritto2}
\eea

Applying $m_{14}\!\circ\!m_{23}\!\circ\!S_3 \circ\!S_4$ to (\ref{bla0})$_2$
and recalling (\ref{defbeta}), (\ref{defbeta'}) we obtain
$\Delta(\beta)[(S\ot S)\F_{21}]=\F^{-1}(\beta \!\ot\! \beta )$,
implying the first equality in
\be \Delta(\beta)=\F^{-1}(\beta \ot \beta )[(S\ot S)\F^{-1}_{21}]
=\F^{-1}_{21}(\beta \ot \beta )[(S\ot S)\F^{-1}]    \label{deltabeta}
\ee
The second equality follows from the first and the cocommutativity of $\Delta$.

\medskip\noindent {\bf Proof of (\ref{*comp}), i.e. that (\ref{star'}) is an antihomomorphism for
$\A_\star$}: \bea &&(a\!\star\! b)^{*_\star}=\sum_I S(\beta
)\trc\left[\!\left(\bF^{(1)}_I\!\trc a\right)\! \left(\bF^{(2)}_I\!\trc
b\right)\!\right]^*\nn
&&=\sum_I S(\beta )\trc
\left\{\!\left[S\!\left(\bF^{(2)}_I\right)^*\trc b^*\right]\!
\left[S\!\left(\bF^{(1)}_I\right)^* \trc a^*\right]\!\right\}\nn
&&=\sum_I \left[S(\beta_{(1)})S\left(\F^{(2)}_I\right)S(\beta^{-1})\trc
b^{*_\star}\right]\! \left[S(\beta_{(2)})S\left(\F^{(1)}_I\right) S(\beta^{-1})
\trc a^{*_\star}\right]\nn
&&\stackrel{(\ref{deltabeta})}{=}\sum_I \left[\bF^{(1)}_IS(\beta)S(\beta^{-1})\trc
b^{*_\star}\right]\! \left[\bF^{(2)}_IS(\beta) S(\beta^{-1}) \trc
a^{\hat
*}\right]= b^{*_\star}\star a^{*_\star}.\qquad\qquad \Box\nonumber
\eea

\medskip\noindent {\bf Proof that $*_\star$ fulfills (\ref{defleibniz})$_2$}:
Since $S(\beta)\beta\in\mbox{Centre}(H)[[\lambda]]$ and 
$\beta^*=S\!\left(\beta^{-1}\right)$,
%
we find
\bea (g\trc
a)^{*_\star} &\stackrel{(\ref{star'})}{=}&S (\beta )\trc (g\trc a)^*= S(\beta
)\trc [S(g)]^*\trc a^*= S(\beta ) [S(g)]^* S(\beta^{-1})\trc a^{
*_\star}\nn &= &\beta^{-1}{}^* [S(g)]^* \beta
{}^*\trc a^{*_\star}= [\beta S(g)\beta^{-1}]^* \trc a^{*_\star}= [\hat
S(g)]^* \trc a^{*_\star}. \qquad\qquad \Box\nonumber
\eea

\medskip\noindent
{\bf Proof of (\ref{braid})}:  Using (\ref{bla}) and
$\R_{32}\!=\!\F_{23}\bF_{32}$ we find \bea
&&\bF_{(12)(34)}\bF_{34}\bF_{12}\R_{32}=\bF_{(123)4}\bF_{(12)3}\bF_{12}\R_{32}=
\bF_{(123)4}\bF_{1(23)}\bF_{23}\R_{32}\nn &&
=\bF_{(123)4}\bF_{1(23)}\bF_{32}=
\tau_{23}\!\left[\bF_{(123)4}\bF_{1(23)}\bF_{23}\right]\!
=\tau_{23}\!\left[\bF_{1(234)}\bF_{(23)4}\bF_{23}\right]\nn &&
=\tau_{23}\!\left[\bF_{1(234)}\bF_{2(34)}\bF_{34}\right]\!=
\tau_{23}\!\left[\bF_{(12)(34)}\bF_{12}\bF_{34}\right]\nonumber \eea
whence \bea
 \mbox{rhs}(\ref{braid})_1&=&m_{12} m_{34}\left\{
\left[\bF_{(12)(34)}\bF_{34}\bF_{12}\R_{32}\right]\trc^{\ot 4}(a\ot a'\ot b\ot b') \right\} \nn
&=&m_{12} m_{34}\tau_{23}\left\{\left[\bF_{(12)(34)}\bF_{12}\bF_{34}\right]
\trc^{\ot 4}(a\ot b\ot a'\ot b') \right\} \nn
&=&m_{13} m_{24}\left\{\left[\bF_{(12)(34)}\bF_{12}\bF_{34}\right]
\trc^{\ot 4}(a\ot b\ot a'\ot b') \right\}=\mbox{lhs}(\ref{braid})_1
\qquad\quad \Box\nonumber
\eea
$(a\ot_\star b)^{*_\star}=b_2^{*_\star}\star a_1^{*_\star}$ is just
an application of (\ref{*comp}) to $(\A\ot\B)_\star$; it takes the
form (\ref{braid})$_2$ upon using
(\ref{braid})$_1$. One can also directly check (\ref{braid})$_2$
using (\ref{deltabeta}). \hfill $\Box$

\medskip\noindent
{\bf Proof of (\ref{altdefDf})}:
\bea
&& rhs(\ref{altdefDf})\stackrel{(\ref{gag})}{=}
\sum_I \sigma\left(\bF^{(1)}_{(1)I} \right)\, a\,
\sigma\left[S\left(\bF^{(1)}_{(2)I} \right)\bF^{(2)}_I\right]=
\sum_I \sigma\left(\bF^{(1)}_{(1)I} \right)\, a\,
\sigma\left\{S\left[S\left(\bF^{(2)}_I\right)\bF^{(1)}_{(2)I}
\right]\right\}\nn
&&\stackrel{(\ref{dritto2})}{=}\sum_I \sigma\left(\Fu_{I} \right)\, a\,
\sigma\left\{S\left[S\left(\beta^{-1}\right)\Fd_{I} \right]\right\}
=\sum_I \sigma\left(\Fu_{I} \right)\, a\,
\sigma\left\{S\left[\Fd_{I} \right]\beta^{-1}\right\}
\stackrel{(\ref{defDf})}{=}D_{\f}^{\sigma}(a)\qquad\quad \Box\nonumber
\nonumber
\eea

\medskip\noindent
{\bf Proof of (\ref{*inter})}: \bea &&[D_{\f}^{\sigma}(a)]^*
\stackrel{(\ref{altdefDf})}{=} \left[\sum_I \left(\bF^{(1)}_I \trc
a\right)\, \sigma\left(\bF^{(2)}_I \right)\right]^*
\stackrel{\F^{*_2}=\F^{-1}}{=} \sum_I
\sigma\left(\F^{(2)}_I \right)\left[S\left(\F^{(1)}_I \right)\trc
a^*\right]\nn && \stackrel{(\ref{gag})}{=} \sum_I
\left[\F^{(2)}_{(1)I}S\!\left(\F^{(1)}_I \right)\trc
a^*\right]\sigma\!\left(\F^{(2)}_{(2)I} \right)
\stackrel{(\ref{dritto4})}{=} \sum_I \left[\bF^{(1)}_I
S\!\left(\beta\right)\trc a^*\right]\sigma\!\left(\bF^{(2)}_I
\right) \stackrel{(\ref{altdefDf})}{=} [D_{\f}^{\sigma}(a^{
*_\star})]\nonumber\eea

\medskip\noindent
{\bf Proof of (\ref{prodinter})}:
\bea
&& D_{\f}^{\sigma}(a)D_{\f}^{\sigma}(a') \stackrel{(\ref{altdefDf})}{=}
\sum_I \left(\bF^{(1)}_I \trc a\right)\,\sigma\left(\bF^{(2)}_I \right)
\sum_{I'} \left(\bF^{(1)}_{I'} \trc a'\right)\,\sigma\left(\bF^{(2)}_{I'} \right)
\nn && \stackrel{(\ref{gag})}{=}\sum_{I,I'} \left(\bF^{(1)}_I \trc a\right)\,
\left(\bF^{(2)}_{(1)I}\bF^{(1)}_{I'} \trc a'\right)\,\sigma\left(\bF^{(2)}_{(2)I}\bF^{(2)}_{I'} \right)\nn &&\stackrel{(\ref{bla})_1}{=}\sum_{I,I'} \left(\bF^{(1)}_{(1)I}\bF^{(1)}_{I'} \trc a\right)\,
\left(\bF^{(1)}_{(2)I}\bF^{(2)}_{I'} \trc a'\right)\,\sigma\left(\bF^{(2)}_I \right)\nn &&\stackrel{(\ref{leibniz})}{=}
\sum_{I,I'} \bF^{(1)}_{I}\trc \left[\left(\bF^{(1)}_{I'} \trc a\right)
\left(\bF^{(2)}_{I'} \trc a'\right)\right]\,\sigma\left(\bF^{(2)}_I \right)\nn &&\stackrel{(\ref{starprod})}{=}\sum_{I} \bF^{(1)}_{I}\trc \left[ a\star  a'\right]\,\sigma\left(\bF^{(2)}_I \right)\stackrel{(\ref{altdefDf})}{=}D_{\f}^{\sigma}(a\star  a')\qquad\quad \Box\nonumber
\nonumber
\eea

\medskip\noindent
{\bf Proof that (\ref{defDfsAB}) defines a deforming map
$D_{\f}^{\hat\sigma_{\scriptscriptstyle \A\ot\B}}:(\A\ot\B)_\star\mapsto
(\A\ot\B)[[\lambda]]$}: \bea g\, \tr
\left[D_{\f}^{\hat\sigma\!_{\scriptscriptstyle \A\ot\B}}(c)\right]
&=&\sum_I \hat \sigma\!_{\scriptscriptstyle \A\ot\B} \!\left(g^I
_{(\hat 1)}\right)\: \left[D_{\f}^{\hat\sigma\!_{\scriptscriptstyle
\A\ot\B}}(c)\right]\: \hat \sigma\!_{\scriptscriptstyle
\A\ot\B}\!\left[\hat S \left(g^I _{(\hat 2)}\right)\right]\nn
&\stackrel{(\ref{defDfsAB})}{=}& \F\!_\sigma\sum_{I,I'}
\left[\sigma\!_{\scriptscriptstyle \A\ot\B} \!\left(g^I _{(\hat
1)}\right)\right]\: \left(\bF^{(1)}_{I'} \trc c\right)\,
\sigma\!_{\scriptscriptstyle \A\ot\B} \left[\bF^{(2)}_{I'} \hat S
\!\left(g^I _{(\hat 2)}\!\right)\! \right]\bF\!_{\sigma} \nn &
\stackrel{(\ref{gag})}{=} & \F\!_\sigma\sum_{I,I'}
\left[\left(g^I_{(\hat 1)}{}_{(1')}\bF^{(1)}_{I'} \right)\trc
c\right]\, \sigma\!_{\scriptscriptstyle \A\ot\B}\left[g^I_{(\hat
1)}{}_{(2')}\bF^{(2)}_{I'} \hat S \!\left(g^I _{(\hat 2)}\!\right)\!
\right]\bF\!_{\sigma} \nn & \stackrel{(\ref{inter-2})}{=} &
\F\!_\sigma\sum_{I,I'} \left[\left(\bF^{(1)}_{I'}g^I_{(\hat
1)}{}_{(\hat 1')} \right)\trc c\right]\,\sigma\!_{\scriptscriptstyle
\A\ot\B}\! \left[\bF^{(2)}_{I'}g^I_{(\hat 1)}{}_{(\hat 2')} \hat S
\!\left(g^I _{(\hat 2)}\!\right)\! \right]\bF\!_{\sigma} \nn & = &
\F\!_\sigma \sum_{I,I'} \left[\left(\bF^{(1)}_{I'}g \right)\trc
c\right]\, \sigma\!_{\scriptscriptstyle \A\ot\B}\!
\left(\bF^{(2)}_{I'}\right)\bF\!_{\sigma}\stackrel{(\ref{defDfsAB})}{=}
\left[D_{\f}^{\hat\sigma\!_{\scriptscriptstyle \A\ot\B}}(g\trc
c)\right], \nonumber\eea proving the intertwining property
(\ref{intertw}) in the present case.  (\ref{*inter}) is proved as
follows: \bea \left[D_{\f}^{\hat\sigma\!_{\scriptscriptstyle
\A\ot\B}}(c)\right]^{*\ot *} &\stackrel{(\ref{defDfsAB})}{=}&
\left[\F\!_\sigma\sum_I \left(\bF^{(1)}_I \trc c\right)\,
\sigma\!_{\scriptscriptstyle \A\ot\B}\!\left(\bF^{(2)}_I
\right)\bF\!_{\sigma}\right]^{*\ot *} \nn &=& \F\!_\sigma\sum_I
\sigma\!_{\scriptscriptstyle \A\ot\B}\!\left(\F^{(2)}_I
\right)\left[S\!\left(\F^{(1)}_I \right)\trc c^{*\ot
*}\right]\bF\!_\sigma\nn & \stackrel{(\ref{gag})}{=} & \F\!_\sigma\sum_I
\left[\F^{(2)}_{(1)I}S\left(\F^{(1)}_I \right)\trc c^{*\ot
*}\right]\sigma\!_{\scriptscriptstyle \A\ot\B}\!\left(\F^{(2)}_{(2)I} \right)
\bF\!_\sigma\nn &\stackrel{(\ref{dritto4})}{=}& \F\!_\sigma\sum_I
\left[\bF^{(1)}_I S\left(\beta\right)\trc c^{*\ot
*}\right]\sigma\!_{\scriptscriptstyle \A\ot\B}\left(\bF^{(2)}
\right)\bF\!_\sigma \stackrel{(\ref{defDfsAB})}{=}
[D_{\f}^{\sigma}(c^{\widehat{*\ot
*}})]\nonumber\eea
Finally, here is the proof of (\ref{prodinter}) in the present case:
\bea D_{\f}^{\hat\sigma\!_{\scriptscriptstyle \A\ot\B}}(c\star c')
&\stackrel{(\ref{defDfsAB})}{=}&\F\!_\sigma \sum_{I}
\left[\bF^{(1)}_{I} \trc (c\star c')\right]\,
\sigma_{\scriptscriptstyle\A\ot\B}\!\left(\bF^{(2)}_{I}  \right)
\bF\!_{\sigma} \nn & \stackrel{(\ref{starprod})}{=}&
\F\!_\sigma\sum_{I,I'} \left[\left(\bF^{(1)}_{(1)I}
\bF^{(1')}_{I'}\right)\trc c\right] \left[\left(\bF^{(1)}_{(2)I}
\bF^{(2')}_{I'}\right)\trc c'\right]\,
\sigma_{\scriptscriptstyle\A\ot\B}\!\left(\bF^{(2)}_{I}\right)\bF\!_{\sigma}
\nn & \stackrel{(\ref{cocycle})}{=}&\F\!_\sigma\sum_{I,I'}
\left(\bF^{(1)}_I \trc c\right) \left[\left(\bF^{(2)}_{(1)I}
\bF^{(1')}_{I'}\right)\trc c'\right]\,
\sigma_{\scriptscriptstyle\A\ot\B}\! \left(\bF^{(2)}_{(2)I}
\bF^{(2')}_{I'} \right) \bF\!_{\sigma} \nn &
\stackrel{(\ref{gag})}{=}&\F\!_\sigma\sum_{I,I'} \left(\bF^{(1)}_I
\trc c\right) \, \sigma_{\scriptscriptstyle
\A\ot\B}\!\left(\bF^{(2)}_I \right)\, \left[ \bF^{(1')}_{I'}\trc
c'\right]\, \sigma_{\scriptscriptstyle\A\ot\B}\! \left(
\bF^{(2')}_{I'} \right)  \bF\!_{\sigma} \nn &
\stackrel{(\ref{defDfsAB})}{=}&\left[D_{\f}^{\hat\sigma\!_{\scriptscriptstyle
\A\ot\B}}(c )\right] \left[D_{\f}^{\hat\sigma\!_{\scriptscriptstyle
\A\ot\B}}(c' )\right] \qquad\qquad \qquad\qquad\qquad\qquad
\Box\nonumber \eea

\medskip\noindent {\bf Proof of (\ref{utile})}: \bea
D_{\f}^{\hat\sigma\!_{\scriptscriptstyle \A\ot\B}}(\alpha\ot \1)
&\stackrel{(\ref{defDfsAB})}{=}& \F\!_\sigma\sum_I \left(\bF^{(1)}_I
\trc \alpha\ot \1\right) \, \sigma\!_{\scriptscriptstyle
\A\ot\B}\!\left(\bF^{(2)}_I \right)\bF\!_{\sigma} \nn &=&
\sum_{I,I',I''} \left(\F^{(1')}_{(1')I'}\bF^{(1)}_I \trc
\alpha\right)\sigma\!_{\scriptscriptstyle \A}\!
\left(\F^{(1')}_{(2')I'}\bF^{(2)}_{(1)I} \bF^{(1'')}_{I''}\right)\ot
\sigma\!_{\scriptscriptstyle
\B}\!\left(\F^{(2')}_{I'}\bF^{(2)}_{(2)I}\bF^{(2'')}_{I''} \right) \nn
&\stackrel{(\ref{cocycle})}{=}&  \sum_{I,I',I''}
\left(\F^{(1')}_{(1')I'}\bF^{(1)}_{(1)I}\bF^{(1'')}_{I''}  \trc
\alpha\right)\sigma\!_{\scriptscriptstyle \A}\!
\left(\F^{(1')}_{(2')I'}\bF^{(1)}_{(2)I}\bF^{(2'')}_{I''} \right)\ot
\sigma\!_{\scriptscriptstyle \B}\!\left(\F^{(2')}_{I'}\bF^{(2)}_{I}
\right) \nn &=& \sum_{I''} \left(\bF^{(1'')}_{I''}  \trc
\alpha\right)\sigma\!_{\scriptscriptstyle \A}\!
\left(\bF^{(2'')}_{I''} \right)\ot
\1=D_{\f}^{\sigma\!_{\scriptscriptstyle \A}}(\alpha)\ot
\1=\check\alpha\ot \1 \nonumber \eea

\bea D_{\f}^{\hat\sigma\!_{\scriptscriptstyle \A\ot\B}}(\1\ot b)
&\!\stackrel{(\ref{defDfsAB})}{=}\!& \!(\bR\F_{21})_\sigma\sum_I
\left(\1\ot\bF^{(1)}_I \trc b\right) \, \sigma\!_{\scriptscriptstyle
\A\ot\B}\!\left(\bF^{(2)}_I \right)(\bF_{21}\R)_{\sigma} \nn &\!=\!&
\!\bR\!_\sigma\sum_{I,I',I''} \!\!\left[\sigma\!_{\scriptscriptstyle
\A}\! \left(\F^{(2')}_{I'}\bF^{(2)}_{(2)I} \bF^{(2'')}_{I''}\right)\ot
\left(\F^{(1')}_{(1')I'}\bF^{(1)}_I \trc
b\right)\sigma\!_{\scriptscriptstyle
\B}\!\left(\F^{(1')}_{(2')I'}\bF^{(2)}_{(1)I}\bF^{(1'')}_{I''}
\right)\right]\R\!_\sigma \nn &\!\stackrel{(\ref{cocycle})}{=}\!&\!
\bR\!_\sigma\sum_{I,I',I''}\!\!\left[ \sigma\!_{\scriptscriptstyle
\A}\! \left(\F^{(2')}_{I'}\bF^{(2)}_{I} \right)\ot
\left(\F^{(1')}_{(1')I'}\bF^{(1)}_{(1)I} \bF^{(1'')}_{I''}\trc
b\right)\sigma\!_{\scriptscriptstyle
\B}\!\left(\F^{(1')}_{(2')I'}\bF^{(1)}_{(2)I}\bF^{(2'')}_{I''}
\right)\right]\R\!_\sigma
 \nn &\!=\!& \!\bR\!_\sigma\sum_{I''}\left[ \1\ot \left(\bF^{(1'')}_{I''}\trc
b\right)\sigma\!_{\scriptscriptstyle \B}\!\left(\bF^{(2'')}_{I''}
\right)\right]\R\!_\sigma \nn &\!=\!& \!\bR\!_\sigma\left[\1\ot
D_{\f}^{\sigma\!_{\scriptscriptstyle
\A}}(b)\right]\R\!_\sigma=\bR\!_\sigma\left[\1\ot \check
b\right]\R\!_\sigma\qquad\qquad\qquad\qquad\Box\nonumber\eea

\medskip\noindent {\bf Proof of (\ref{hatD}), (\ref{hatDn})}:
Let $\partial_i'\!:=\!S(\beta)\trc\partial_i \!=\!\tau^i_j(\beta)\partial_j$.
(\ref{deftau})$_2$ implies $g\trc \partial_i'=\tau^i_j\big[\hat S(g)\big]\partial_j'$, which
with (\ref{deftau})$_1$ proves the first line of (\ref{hatD}).
The second, third lines are proved as follows:
\bea
&&x^a{}^{*_\star} \stackrel{(\ref{star'})}{=}S(\beta)\trc x^a{}^{*}=
S(\beta)\trc x^a=\tau^k_a \left[ S(\beta)\right]x^k,\nn
&&\partial_a'{}^{*_\star} \stackrel{(\ref{star'})}{=}S(\beta)\trc\left[S(\beta)\trc\partial_a \right]^*=S(\beta)\beta^*\trc\partial_a^* =-
S(\beta)S\left(\beta^{-1}\right)\trc \partial_a=-\tau^a_k\left(\beta^{-1}\right)\partial_k',\nn
%
%
&&x^a\!\star\! x^b\stackrel{(\ref{braid})}{=}\sum_I\!\left[\R^{(2)}_I\trc x^b\right]\star  \left[\R^{(1)}_I\trc x^a\right]=\sum_I\!\tau_b^h\left[\R^{(2)}_I\right]  \tau_a^k\left[\R^{(1)}_I\right]x^h\star x^k=R^{kh}_{ab}x^h\!\star\! x^k,\nn
&&\partial_a'\star  \partial_b'\stackrel{(\ref{braid})}{=}\sum_I\!\left[\R^{(2)}_I\trc \partial_b'\right]\star  \left[\R^{(1)}_I\trc \partial_a'\right]=
\sum_I\!\tau^b_h\!\left[\hat S\!\left(\R^{(2)}_I\right)\!\right]  \tau^a_k\!
\left[\hat S\!\left(\R^{(1)}_I\right)\!\right]\partial_h'\star \partial_k'\nn &&
\qquad\quad \stackrel{(\ref{SR})}{=}\sum_I\!\tau^b_h\left[\R^{(2)}_I\right]  \tau^a_k \left[\R^{(1)}_I\right]\partial_h'\star \partial_k'
=R_{kh}^{ab}\partial_h'\star \partial_k',\nn &&
\partial_a'\star x^b\stackrel{(\ref{starprod})}{=}
\sum_I\!\left[\bF^{(1)}_I\trc\partial_a'\right]\star
 \left[\bF^{(2)}_I\trc x^b\right]=\sum_I\!\tau^a_k\!\left[\beta S\!\left(\bF^{(1)}_I\right)\!\right]
\tau_b^h\!\left[\bF^{(2)}_I\right]\partial_k x^h\nn &&
\quad =\sum_I\!\tau^a_k\!\left[\beta S\!\left(\bF^{(1)}_I\right)\!\right]
\tau_b^h\!\left[\bF^{(2)}_I\right](\1\delta^h_k+x^h\partial_k)=
\delta^a_b\1+\sum_I\!\left[\R^{(2)}_I\!\trc\! x^b\right]\!\star\!  \left[\R^{(1)}_I\!\trc\! \partial_a'\right]\nn &&
\quad =\delta^a_b\1+\sum_I\!\tau_b^h\!\left[\R^{(2)}_I\right]\tau^a_k\!
\left[\hat S\!\left(\R^{(1)}_I\right)\!\right] x^h\!\star\!\partial_k'
 \stackrel{(\ref{SR})}{=}\delta^a_b\1+\sum_I\!\tau_b^h\!\left[\R^{(1)}_I\right]\tau^a_k\!
\left[\R^{(2)}_I\right] x^h\!\star\!\partial_k'\nn &&\quad=
\delta^a_b\1+R_{bk}^{ha} x^h\star\partial_k',\nonumber
\eea
where we have used the following relation, valid for any triangular Hopf algebra:
\be
(\hat S \otimes \mbox{id})(\R)=\R_{21}= (\mbox{id}\otimes \hat S^{-1})(\R).
       \label{SR}
\ee
 The proof
of (\ref{hatDn}) is completely analogous. \hfill $\Box$

\medskip\noindent {\bf Proof of (\ref{intstarcom'})}:
\bea
&&\int_{X}\!\!\!\! d\nu_j [\omega\star f({\rm x}_j)]
\stackrel{(\ref{braid})}{=}
\int_{X}\!\!\!\! d\nu_j\sum_I
\left[\R^{(2)}_I\trc f({\rm x}_j)\right]\star
\left[\R^{(1)}_I\trc \omega\right]\stackrel{(\ref{InvInt})}{=}
\nn&& \int_{X}\!\!\!\! d\nu_j \!\left\{\!
 f({\rm x}_j)\star \!\sum_I\varepsilon\!\left(\R^{(2)}_I\right)
 \!\left[\R^{(1)}_I\trc \omega\right]\!\right\}\stackrel{(\ref{twistcond})_2}{=}
\int_{X}\!\!\!\! d\nu_j f({\rm x}_j) \star \omega
\stackrel{(\ref{intstarcom})_2}{=}
\omega\star\!\int_{X}\!\!\!\!  d\nu_j f({\rm x}_j).\nonumber
\qquad\qquad\Box
\eea

\begin{lemma}
Let $(\M,\rho)$ be a representation of $H$ and $(\M^\vee,\rho^\vee)$
its contragredient, $\{\eta_i\}$ a basis of $\M$ and $\{\eta^i\}$
the contragredient one of $\M^\vee$. Then \be
\eta_i\ot\eta^i=\eta_i\ot_\star\eta^{\prime i}\qquad \mbox{(sum over
$i$)}\qquad \qquad\eta^{\prime
i}\!:=\!S\big(\beta\big)\trc\eta^i \ee and this sum is
$H$-invariant. \label{lemmainv}
\end{lemma}

\noindent {\bf Proof}: By definition $g\trc \eta_i=\rho_i^j(g)\eta_j$,
$g\trc \eta^i=\rho^\vee{}_i^j(g)\eta^j = \rho^i_j\big[S(g)\big]
\eta^j$. Therefore, \bea \eta_i\ot_\star\eta^{\prime
i}=\sum_I\bF^{(1)}_I\trc \eta_i\ot
\bF^{(2)}_IS\!\left(\beta\right)\trc \eta^i
=\sum_I\eta_j\rho_i^j\!\left(\bF^{(1)}_I\right)\ot\rho^i_k\!\left[\beta
S\!\left(\bF^{(2)}_I\right)\right]\eta^k\nn
=\sum_I\eta_j\ot\rho^j_k\!\left[\bF^{(1)}_I\beta
S\!\left(\bF^{(2)}_I\right)\right]\eta^k
\stackrel{}{=}\sum_{I,I'}\eta_j\ot\rho^j_k\!\left[\bF^{(1)}_I
\F^{(1')}_{I'}S\!\left(\bF^{(2)}_I\F^{(2')}_{I'}\right)\right]\eta^k=
\eta_j\ot\eta^j.\quad\Box\nonumber \eea

\medskip\noindent
{\bf The Commutation rules among
$\hat a^{\prime i}, \hat a^+_j,\hat \varphi_i(\hat {\rm x}),\hat \varphi_i(\hat {\rm y}), \hat \varphi^{\hat *}_i(\hat {\rm x}), \hat \varphi^{\hat *}_i(\hat {\rm y})$}, beside (\ref{hqccr}),  are:
\be
\ba{ll}
\hat \varphi_i\!(\hat y)\hat \varphi_j\!(\hat x)\!=\!R^{kh}_{ij}\hat \varphi_h\!(\hat x)\hat \varphi_k\!(\hat y),\qquad\quad &\hat \varphi_i^{\hat *}\!(\hat y)\hat \varphi_j^{\hat *}\!(\hat x)\!=\!R^{ij}_{kh}\hat \varphi_h^{\hat *}\!(\hat x)\hat \varphi_k^{\hat *}\!(\hat y),\\[8pt]
\hat \varphi_i^{\hat *}\!(\hat y)\hat \varphi_j\!(\hat x)\!=\!R^{hi}_{jk}\hat \varphi_h\!(\hat x)\hat \varphi_k^{\hat *}\!(\hat y),
\qquad\quad &\\[8pt]
\hat \varphi_i(\hat x)\hat a^{+}_j=R^{kh}_{ij}\hat a^{+}_h\hat \varphi_k(\hat x),\qquad\quad &\hat \varphi_i^{\hat *}(\hat x)\hat a^{+}_j=R^{hi}_{jk}\hat a^{+}_h\hat \varphi_k^{\hat *}(\hat x),\\[8pt]
\hat a^{\prime i}\hat \varphi_j(\hat x)=R^{hi}_{jk}\hat \varphi_h(\hat x)
\hat a^{\prime k},\qquad\quad
&\hat a^{\prime i}\hat \varphi_j^{\hat *}(\hat x)=
R^{ij}_{kh}\hat \varphi_h^{\hat *}(\hat x)\hat a^{\prime k},
\ea                                              \label{earel}
\ee
where $\hat x,\hat y\!\in\!\{\hat x_1,\hat x_2,...\}$ are any two sets
of noncommutative coordinates fulfilling (\ref{hatDn}). Relations (\ref{earel})
hold also if $\hat y=\hat x$, and are a straightforward
consequence of (\ref{braid}). \ $\Box$

\medskip\noindent
{\bf Proof of (\ref{gonfock})}. Any $ s$ can be expressed in the form
$ s=\sum_Jc^J_s\Psi_0$, with some $c^J_s\!\in\!\A^{\pm}[[\lambda]]$.
Using the identities $(\id\!\ot\!\varepsilon)\!\circ\!\Delta=\id$,
$\varepsilon\!\circ\! S=\varepsilon$ we find
\bea
&&g\trc s=\sum_Jg\trc (c^J_s\Psi_0)\stackrel{(\ref{comp})}{=}
\sum_{J,I}(g^I_{(1)}\!\trc c^J_s) g^I_{(2)}\!\trc\!\Psi_0=
\sum_{J,I}(g^I_{(1)}\!\trc c^J_s) \varepsilon\big(g^I_{(2)}\big)\Psi_0=
\sum_J(g\trc c^J_s) \Psi_0\qquad\nn &&
\stackrel{(\ref{gag})}{=}\sum_{J,I} \sigma\!\left(g^I_{(1)}\!\right)\!
c^J_s \sigma\!\left[S \!\left(g^I_{(2)}\right)\!\right]\!
\Psi_0= \sigma\!\left(g^I_{(1)}\!\right) c^J_s
\varepsilon\!\left[S \!\left(g^I_{(2)}\!\right)\!\right]\!
\Psi_0=\sum_J\sigma\left(g\right)c^J_s\Psi_0=
\sigma(g)s .\quad\Box\nonumber
\eea

\medskip\noindent
{\bf Proof of (\ref{hatcomp})}. Using the identities
$(\id\!\ot\!\hat\varepsilon)\!\circ\!\hat\Delta=\id$ and
$\cdot\circ (\hat S\!\ot\!\id)\!\circ\!\hat\Delta=\hat\varepsilon$
($\cdot$ stands for the product in $\hH$),  we find
\bea
&& g\tr(c s)\stackrel{(\ref{deftronfock})}{=}
\sigma(g)c s=\sum_I\sigma\!\left(g^I_{(1)}\!\right) c\,
\varepsilon\!\left(g^I_{(2)}\right)\! s=
\sum_I\sigma\!\left(g^I_{(1)}\!\right) c\,
\sigma\!\left[\hat S \! \left(g^I_{(2)}\right)\!g^I_{(3)}\!\right]\!
 s\nn && \stackrel{(\ref{gog})}{=} \sum_I \left(g^I_{(1)}\tr c\right)
\sigma\!\left(g^I_{(2)}\!\right) s\stackrel{(\ref{deftronfock})}{=}
\sum_I(g^I_{(\hat 1)}\tr c)\,g^I_{(\hat 2)}\tr  s
\qquad\qquad\qquad\Box\
\nonumber\eea

\medskip\noindent {\bf Proof of (\ref{starHermConj})}:
Let $v_1,v_2\!\in\! \A^\pm[[\lambda]]$ be such that
$s_h=v_h\Psi\!_0$; then
\bea
&&\la  c  s_1,s_2\ra\!=\!\la  c  v_1\Psi\!_0,v_2\Psi\!_0\ra
\!=\!\la  \Psi\!_0,(c  v_1)^*v_2\Psi\!_0\ra \!\stackrel{(\ref{*starstar})}{=}\!
\la  \Psi\!_0,(c  v_1)^{*_\star}\!\star\! v_2\Psi\!_0\ra
\!\stackrel{(\ref{*comp})}{=}\!\la  \Psi\!_0,  v_1^{*_\star}\!\star\!
c^{*_\star}\!\star\! v_2\Psi\!_0\ra
\nn&& \stackrel{(\ref{*starstar})}{=}\la
\Psi\!_0,  v_1\,^* (c^{*_\star}\!\star\! v_2)\Psi\!_0\ra
=\la  v_1\,\Psi\!_0,  c^{*_\star}\!\star\! v_2\,\Psi\!_0\ra
=\la  s_1, c^{*_\star}\!\star\! s_2\ra.\qquad\qquad\Box   \nonumber
\eea

\medskip\noindent
{\bf Proof of the commutativity of diagram \ref{diag}}. We start by
noting that
\bea
\frac1{\sqrt{n!}}\varphi^*\!({\rm x}_1\!)...\varphi^*\!({\rm x}_n\!)\Psi_0
\!=\!a^+_{i_1}... a^+_{i_n}\Psi_0 \,\frac1{\sqrt{n!}}
\varphi_{i_n}^*\!({\rm x}_n\!)...
\varphi_{i_1}^*\!({\rm x}_1\!)\!=\! \vert n_1,n_2...\ra\, \frac 1N
\varphi_{i_n}^*\!({\rm x}_n\!)...\varphi_{i_1}^*\!({\rm x}_1\!)
\!\nn
\qquad \qquad\qquad\!=\!e_{i_1,...,i_n}^{\pm}\frac 1N
\varphi_{i_n}^*\!({\rm x}_n\!)...\varphi_{i_1}^*\!({\rm x}_1\!)
=(e_{j_1}\!\ot...\ot e_{j_n})\P^n_{\pm}{}_{i_1...i_n}^{j_1...j_n}
\varphi_{i_n}^*\!({\rm x}_n\!)...\varphi_{i_1}^*\!({\rm x}_1\!).\nonumber
\eea
As an example, we prove that $\pi^n_\pm\!=\!\kappa^{\ot n}\!\circ\!\P^n_\pm$ and
$\Pi^n_\pm\!=\!\P^n_{\pm}\!\circ\!(\kappa^{\ot n})^{-1}$. In fact,
for any $s=s^{j_1,...,j_n}(e_{j_1}\!\ot ..\ot e_{j_n})\!\in\!\H^{\ot n}$,
$\psi=\psi^{j_1,...,j_n}(\varphi_{j_1}\!\ot ..\ot \varphi_{j_n})\!\in\!\X^{\ot n}$ we find
\bea
[\pi^n_\pm(s)]({\rm x}_1,\!...,{\rm x}_n)\!=\!\frac 1{\sqrt{n!}}
\left\la\left[\varphi^{ *}({\rm x}_1\!)...
\varphi^{*}({\rm x}_n)\Psi_0\right],s\right\ra=
\varphi_{j_1}({\rm x}_1\!)...
\varphi_{j_n}({\rm x}_n\!)\P^n_{\pm}{}_{i_1...i_n}^{j_1...j_n}s^{i_1,...,i_n}
\qquad \nn=\P^n_{\pm}{}_{i_1...i_n}^{j_1...j_n}s^{i_1,...,i_n}
[\kappa^{\ot n}(e_{j_1}\ot ..\ot e_{j_n})]({\rm x}_1,\!...,{\rm x}_n)=
\{\kappa^{\ot n}[\P^n_\pm(s)]\}({\rm x}_1,\!...,{\rm x}_n)
\nn \Pi^n_\pm(\psi)
\!=\!\! \displaystyle\int_X \!\! \!\!
d\nu({\rm x}_1\!)...\!\! \displaystyle\int_X \!\!\!\! d\nu({\rm x}_n)\,
\frac 1 {\sqrt{n!}}
\varphi^*\!({\rm x}_1\!)...\varphi^*\!({\rm x}_n\!)\Psi_0
\: \psi({\rm x}_1,\!...,{\rm x}_n)\qquad\qquad\qquad\qquad\qquad\qquad
\nn \qquad
\!=\!(\!e_{j_1}\!\ot...\ot e_{j_n}\!)\P^n_{\pm}{}_{i_1...i_n}^{j_1...j_n}\!\! \displaystyle\int_X
\!\!\! \!\!d\nu({\rm x}_1\!)...\!\! \displaystyle\int_X \!\!\!\!\!
d\nu({\rm x}_n) \:
\varphi_{i_n}^*\!({\rm x}_n\!)...\varphi_{i_1}^*\!({\rm x}_1\!)
\varphi_{h_1}({\rm x}_1\!)...
\varphi_{h_n}({\rm x}_n\!)\psi^{h_1,...,h_n} \:\nn \qquad
\!=\!(e_{j_1}\!\ot...\ot e_{j_n})\P^n_{\pm}{}_{i_1...i_n}^{j_1...j_n}
\psi^{i_1,...,i_n}=\P^n_{\pm}(e_{i_1}\!\ot...\ot e_{i_n})\psi^{i_1,...,i_n}
\qquad\qquad\qquad\qquad\qquad\quad\nn
\qquad=\P^n_{\pm}[\left(\kappa^{\ot n}\right)^{-1}(\varphi_{i_1}\!\ot...\ot \varphi_{i_n})]\psi^{i_1,...,i_n}=\P^n_{\pm}[\left(\kappa^{\ot n}\right)^{-1}(\psi)]
\qquad\qquad\qquad\qquad\qquad\quad \Box\nonumber
\eea

\medskip\noindent
{\bf Proof of (\ref{n-wf'})}: By (\ref{Aavec}) $e^{\prime\pm}_{i_1,...,i_n}=\overline{F}^n{}^{j_1...j_n}_{i_1...i_n}e^{\pm}_{j_1,...,j_n}$, whence
$$
\mbox{lhs}(\ref{n-wf'})
\stackrel{(\ref{n-wfh})}{=}\overline{F}^n{}^{j_1...j_n}_{i_1...i_n}
F^n{}_{(j_1...j_n]}^{l_1...l_n} \hat\varphi_{l_1}(\hat{\rm
x}_1\!)...\hat\varphi_{l_n}(\hat{\rm x}_n\!)
=\P^{n,F}_{\pm}{}^{l_1...l_n}_{i_1...i_n}
\hat\varphi_{j_1}(\hat{\rm x}_1\!)...\hat\varphi_{j_n}(\hat{\rm x}_n\!)
$$
proving the first equality in (\ref{n-wf'}); to prove the second one
can use either (\ref{bcsym}) or (\ref{PFtau}) (together with their
generalizations to $n\!>\!2$). \hfill $\Box$

\end{document}